\documentclass[12pt]{article}
\usepackage{graphicx}
\usepackage{verbatim} 
\usepackage{amsmath}
\usepackage{authblk}
\def\hybrid{\topmargin 0pt      \oddsidemargin 0pt
        \headheight 0pt \headsep 0pt  
        \voffset=-0.5cm
        \textwidth 6.25in       
        \textheight 9.5in       
        \marginparwidth 0.0in
        \parskip 5pt plus 1pt   \jot = 1.5ex}
\catcode`\@=11
\def\marginnote#1{}

\newcount\hour
\newcount\minute
\newtoks\amorpm
\hour=\time\divide\hour by60
\minute=\time{\multiply\hour by60 \global\advance\minute by-\hour}
\edef\standardtime{{\ifnum\hour<12 \global\amorpm={am}%
        \else\global\amorpm={pm}\advance\hour by-12 \fi
        \ifnum\hour=0 \hour=12 \fi
        \number\hour:\ifnum\minute<10 0\fi\number\minute\the\amorpm}}
\edef\militarytime{\number\hour:\ifnum\minute<10 0\fi\number\minute}

\def\draftlabel#1{{\@bsphack\if@filesw {\let\thepage\relax
   \xdef\@gtempa{\write\@auxout{\string
      \newlabel{#1}{{\@currentlabel}{\thepage}}}}}\@gtempa
   \if@nobreak \ifvmode\nobreak\fi\fi\fi\@esphack}
        \gdef\@eqnlabel{#1}}
\def\@eqnlabel{}
\def\@vacuum{}
\def\draftmarginnote#1{\marginpar{\raggedright\scriptsize\tt#1}}
\def\draftlabel#1{{\@bsphack\if@filesw {\let\thepage\relax
   \xdef\@gtempa{\write\@auxout{\string
      \newlabel{#1}{{\@currentlabel}{\thepage}}}}}\@gtempa
   \if@nobreak \ifvmode\nobreak\fi\fi\fi\@esphack}
        \gdef\@eqnlabel{#1}}
\def\@eqnlabel{}
\def\@vacuum{}
\def\draftmarginnote#1{\marginpar{\raggedright\scriptsize\tt#1}}

\def\draft{\oddsidemargin -.5truein
        \def\@oddfoot{\sl preliminary draft \hfil
        \rm\thepage\hfil\sl\today\quad\militarytime}
        \let\@evenfoot\@oddfoot \overfullrule 3pt
        \let\label=\draftlabel
        \let\marginnote=\draftmarginnote
   \def\@eqnnum{(\theequation)\rlap{\kern\marginparsep\tt\@eqnlabel}%
\global\let\@eqnlabel\@vacuum}  }

\def\numberbysection{\@addtoreset{equation}{section}
        \def\theequation{\thesection.\arabic{equation}}}

\def\underline#1{\relax\ifmmode\@@underline#1\else
        $\@@underline{\hbox{#1}}$\relax\fi}

\def\titlepage{\@restonecolfalse\if@twocolumn\@restonecoltrue\onecolumn
     \else \newpage \fi \thispagestyle{empty}\c@page\z@
        \def\thefootnote{\fnsymbol{footnote}} }

\def\endtitlepage{\if@restonecol\twocolumn \else  \fi
        \def\thefootnote{\arabic{footnote}}
        \setcounter{footnote}{0}}  
\relax

\numberbysection
\hybrid

\newfont{\Bbb}{msbm10 scaled 1\@ptsize00}
\newfont{\Bbbb}{msbm7 scaled 1\@ptsize00}

\newcommand{\CC}{\mathbb{C}}
\newcommand{\CCC}{\mbox{\Bbbb C}}

\newcommand{\DDD}{\raise-1pt\hbox{$\mbox{\Bbbb D}$}}

\newcommand{\RRR}{\mbox{\Bbbb R}}

\newcommand{\UUU}{\raise-1pt\hbox{$\mbox{\Bbbb U}$}}

\newcommand{\ZZ}{\mathbb{Z}}
\newcommand{\z}{\raise-1pt\hbox{$\mbox{\Bbbb Z}$}}
\def\beq{\begin{equation}}
\def\eeq{\end{equation}}
\def\p{\partial}

\def\<{\langle}
\def\>{\rangle}

\def\i2{\frac{i}{2}}
\def\tr{{\rm tr~}}
\def\diag{{\rm diag~}}

\def\psistar{\psi^{*}}

\def\normord{ {\scriptstyle {{\bullet}\atop{\bullet}}} }
\def\normordsmall{ {\scriptscriptstyle {{\bullet}\atop{\bullet}}} }
\def\normordbare{ {\scriptstyle {{ \times}\atop{ \times}}} }

\def\normordboson{ {\scriptstyle {{*}\atop{*}}} }

\def\lbr{\left <}
\def\rbr{\right >}

\def\lvac{\left <0\right |}
\def\rvac{\left |0\right >}

\def\lvacn{\left <n\right |}
\def\rvacn{\left |n\right >}
\def\lvacN{\left <N\right |}
\def\rvacN{\left |N\right >}

\newtheorem{prop}{Proposition}[section]
\newtheorem{theor}[prop]{Theorem}
\newtheorem{cor}[prop]{Corollary}
\usepackage{amssymb}
\usepackage{amsbsy}
\usepackage{amsmath}

\usepackage{graphicx,epsf,tikz,wrapfig}
\usetikzlibrary{shapes.geometric}
\usetikzlibrary{decorations.text}

\begin{document}

\begin{titlepage}

\title{Free fermions and tau-functions}

\author[1,2]{Alexander Alexandrov \thanks{E-mail:  {\tt alexandrovsash at gmail.com}}}
\author[2,3,4]{Anton Zabrodin \thanks{E-mail:  {\tt zabrodin at itep.ru}}}
\affil[1]{\small{Mathematisches Institut, Albert-Ludwigs-Universit\"{a}t,
Freiburg 79104, Germany} }
\affil[2]{\small{ITEP, 25 Bol. Cheremushkinskaya Ulitsa, Moscow 117259, Russia}}
\affil[3]{Institute of Biochemical Physics,
4 Kosygina Ulitsa, Moscow 119991, Russia}
\affil[4]{National Research University Higher School of Economics,
20 Myasnitskaya Ulitsa, Moscow 101000, Russia}

\date{December 2012}
\maketitle

\vspace{-9cm} \centerline{ \hfill ITEP-TH-60/12} \vspace{9cm}

\begin{abstract}

We review the formalism of free fermions 
used for construction of tau-functions of classical 
integrable hierarchies 
and give a detailed derivation of 
group-like properties of the normally ordered exponents, 
transformations between different normal orderings,
the bilinear relations, the generalized Wick
theorem and the bosonization rules. 
We also consider various examples of tau-functions
and give their fermionic realization.

\end{abstract}
\vfill
{\bf Keywords:} Integrable systems; tau-functions; free fermions; Hirota bilinear equations; matrix models.

\end{titlepage}
\newpage

\tableofcontents

\newpage
\section{Introduction}

In this paper we review the formalism of
free fermions introduced by the Kyoto school 
\cite{DJKM83}-\cite{KM81} and
used for construction of tau-functions of classical 
integrable hierarchies of nonlinear PDE's
such as Kadomtsev-Petviashvili (KP), 
modified Kadomtsev-Petviashvili (MKP), 2D Toda lattice (2DTL)
and their multicomponent versions.
The operator approach to the integrable hierarchies appears to be 
an extremely useful tool in dealing with classical integrability
because the bilinear equations of the hierarchies acquire the transparent
meaning as the standard relations for correlation 
functions of free fermions like the
Wick theorem. The $\tau$-function itself is a vacuum expectation value
of special operators from the 
infinite-dimensional Clifford 
algebra generated by free fermions $\psi_n$,
$\psistar_n$.

We give a detailed and self-contained 
derivaton of the key properties and relations: 
group-like properties of the normally ordered exponents, 
transformations between different normal orderings,
the bilinear relations, the generalized Wick
theorem and the boson-fermion correspondence, paying attention 
to particular cases and examples. 
Our motivation comes from the feeling that their
precise formulations and exhaustive proofs 
seem to be missing in
readily available literature. In view of this,
we believe that our review might be of interest for those who are going to 
use the operator methods in their own research work.

Another motivation is to collect 
together various examples of $\tau$-functions that play 
an important role in the theory and/or applications. 
Among them are familiar ones (characters, 
quasi-polynomial and multi-soliton 
$\tau$-functions, partition 
functions of different matrix models) as well as
less known examples which were addressed in the literature only
recently (such as expectation values with non-standard time
evolution and Hurwitz partition functions).
In each case we present explicit fermionic realizations 
of the $\tau$-functions.

In order not to overload the paper, we restrict ourselves by 
one-component charged fermions and 
leave aside integrable hierarchies of the types other 
than $A_{\infty}$. 
Some other related interesting topics, 
such as fermionic representation of 
the generalized Kontsevich model and 
the melting crystal model, remain out of the 
scope of this review because of the space limitations. 
Various aspects of the free fermionic approach to 
integrable systems are also discussed in the books 
\cite[chapter 9]{BBT}, \cite[chapter 14]{Kac} and reviews 
\cite[section 4]{Morozov}, \cite[section 1]{Zinn}.

The organization of the paper is clear from the table of contents.
It should be emphasized that central in our presentation is the
basic bilinear condition (BBC)
\beq\label{basic}
\sum_k \psi_k G\otimes \psistar_k G=\sum_k 
G\psi_k \otimes G\psistar_k
\eeq
which defines an important class of elements $G$ of the Clifford algebra.
Exponents (or normally ordered exponents) of bilinear expressions 
in the fermionic operators $\psi_n$,
$\psistar_n$ are well-known solutions to the BBC.
There are also solutions which can not be represented in this 
form -- for example, any linear combinations of the $\psi_n$'s
(or $\psistar_n$'s). 
Correlation functions of fermions with insertions of
elements $G$ that obey the BBC have some special properties.
In particular, the generalized Wick theorem
holds for them. Taking matrix elements of the operator 
relation (\ref{basic})
between some properly chosen states in the fermionic Fock
space, one can derive the bilinear identity generating
various bilinear equations of the 
Hirota type for the $\tau$-functions.
The general $\tau$-function for one-component fermions
is usually defined as the expectation
value of the form
\beq\label{basic1}
\tau_n ({\bf t_+}, {\bf t}_-)=
\lvacn e^{J_+({\bf t_+})} \, G \, e^{-J_-({\bf t_-})}\rvacn ,
\eeq
where ${\bf t_{\pm}}=\{t_{\pm k}\}_{k=1}^{\infty}$
and $n=t_0 \in \ZZ$ 
are time variables,
$J_{\pm}$ are current operators and $G$ is any solution 
to the BBC with zero charge. An extension of this definition
to solutions of the BBC with non-zero charge is straightforward. 
By freezing some of the time variables at particular values
(say, ${\bf t}_- =0$ or ${\bf t}_- =n=0$) one obtains
$\tau$-functions of the MKP and KP hierarchies respectively.

\section*{Acknowledgements}


We are grateful to V.Kazakov, S.Kharchev, S.Leurent, A.Orlov, T.Takebe,
Z.Tsuboi and P.Wiegmann
for discussions. 
The work of both authors was partly supported 
by Ministry of Education and Science of Russian Federation
under contract 8207 and by grant NSh-3349.2012.2 for support of 
leading scientific schools.
The work of A.A. was supported in part by RFBR grant 11-01-00962 
and by ERC Starting Independent Researcher Grant 
StG No. 204757-TQFT. The work of A.Z. was supported in part by
RFBR grant 11-02-01220 and by joint RFBR grants 12-02-91052-CNRS,
12-02-92108-JSPS.

\newpage
\section{Free fermions}

\subsection{The algebra of fermionic operators}

Let $\psi_n , \psistar_{n}$, $n\in \ZZ$, be free fermionic
operators with usual anticommutation relations
\beq\label{fermop1}
[\psi_n , \psi_m ]_+ = [\psistar_n, \psistar_m]_+=0, \quad
[\psi_n , \psistar_m]_+=\delta_{mn}.
\eeq
They generate
an infinite dimensional Clifford algebra. We also use their
generating series
\beq\label{ferm0}
\psi (z)=\sum_{k\in \z}\psi_k z^k, \quad \quad
\psistar (z)=\sum_{k\in \z}\psistar_k z^{-k}.
\eeq

From the fact that the anticommutator of any linear combinations
of the fermionic operators is a number it follows that the commutator
of any bilinear expressions in $\psi_n$ and $\psistar_{n}$ 
is again bilinear in $\psi_n$ and $\psistar_{n}$. For example,
$$
[\psi_m \psistar _n , \, \psi_{m'} \psistar _{n'}]=
\delta _{nm'} \psi_{m} \psistar _{n'}-
\delta _{mn'} \psi_{m'} \psistar _{n}.
$$
We see that the commutation law for 
the expressions $\psi_m \psistar _n$ is the same as for 
the matrices $E_{mn}$ with matrix elements
$(E_{mn})_{ij}=\delta_{im}\delta_{jn}$
which are generators of the algebra 
$gl (\infty )$ of infinite-size matrices with only 
finite (but arbitrary) number of non-zero elements.
More generally, consider the bilinear expression
\beq\label{XA}
X_A =\sum_{ij}A_{ij}\psi _i \psistar_j
\eeq
with some matrix $A$, 
then $[X_A , \, X_B ]=X_{[A, B]}$ and
$$
[X_A , \, \psi _n ]=\sum_i A_{in}\psi _i \,, \quad \quad
[X_A , \, \psistar _n ]=-\sum_i A_{ni}\psi^*_i .
$$

In order to find the adjoint action of the operator
$X_A$ on fermions, we use the well known formula
\beq\label{ABA}
e^A B e^{-A}=B + [A, B] +\frac{1}{2!}[A, [A,B]] +\ldots
\eeq
valid for any two operators $A$, $B$. We get
\beq\label{ferm}
e^{X_A}\psi_n e^{-X_A}=\sum_i \psi _i \, R_{in}\,, \quad \quad
e^{X_A}\psistar_n e^{-X_A}=\sum_i (R^{-1})_{ni}\psistar _i,
\eeq
where $R^{\pm 1}=e^{\pm A}=1\pm A +\frac{1}{2!}A^2 +\ldots$.
We see that exponents of the expressions 
bilinear in $\psi_n$ and $\psistar_{n}$ possess a rather specific property:
the result of 
their adjoint action on the form linear in fermions is again linear.

\subsection{Dirac vacua and excited states}

Next, we introduce a vacuum state $\left |0\rbr$ which is
a ``Dirac sea'' where all negative mode states are empty
and all positive ones are occupied:
$$
\psi_n \rvac =0, \quad n< 0; \quad \quad \quad
\psistar_n \rvac =0, \quad n\geq 0.
$$
(For brevity, we call indices $n\geq 0$ {\it positive}.)
Similarly, the dual vacuum state has the properties
$$
\lvac \psistar_n  =0, \quad n< 0; \quad \quad \quad
\lvac \psi_n  =0, \quad n\geq 0.
$$
With respect to the vacuum $\rvac$, the operators $\psi_n$ with
$n<0$ and $\psistar_n$ with $n\geq 0$ are annihilation operators
while the operators $\psistar_n$ with $n<0$ and
$\psi_n$ with $n\geq 0$ are creation operators.

We also need ``shifted'' Dirac vacua $\rvacn$ and $\lvacn$
defined as
\begin{align}\label{vacdefr}
\rvacn = \left \{
\begin{array}{l}
\psi_{n-1}\ldots \psi_1 \psi_0 \rvac , \,\,\,\,\, n> 0
\\ \\
\psistar_n \ldots \psistar_{-2}\psistar_{-1}\rvac , \,\,\,\,\, n<0
\end{array} \right.
\end{align}
\begin{align}\label{vacdefl}
\lvacn = \left \{
\begin{array}{l}
\lvac \psistar_{0}\psistar_{1}\ldots \psistar_{n-1} , \,\,\,\,\, n> 0
\\ \\
\lvac \psi_{-1}\psi_{-2}\ldots \psi_{n} , \,\,\,\,\, n<0
\end{array} \right.
\end{align}
In particular, we have 
\begin{align}
 \psi_m \rvacn &=0, \quad m < n; 
\qquad 
\psistar_m \rvacn =0, \quad m \ge n, \\
\lvacn  \psi_{m}&=0 , \quad m \ge n; 
\qquad 
\lvacn  \psistar_{m}=0 , \quad m < n.
\end{align}
It is also clear that
\begin{align}
\psi_n \rvacn &= \left|n+1 \rbr,
\qquad  \psistar_n \left|n+1 \rbr = \left|n \rbr,
\\
 \lbr n+1 \right|\psi_n &= \lbr n \right|
\qquad  
\lbr n \right|\psistar_n = \lbr n+1 \right| .
\end{align}

Excited states (over the vacuum 
$\rvac$) are obtained by filling some empty states
(acting by the operators $\psistar_j$)
and creating some holes (acting by the $\psi _j$'s). 
A particle 
carries the charge $-1$ while a hole 
carries the charge $+1$, so any state with a definite number 
of particles and holes has the definite charge.
Let us introduce a convenient basis
of states with definite charge in the fermionic Fock space ${\cal H}_{F}$.
The basis states $\left |\lambda , n\rbr$ are
parametrized by $n$ and Young diagrams $\lambda$ in the following way.
Given a Young diagram $\lambda =
(\lambda_1 , \ldots , \lambda_{\ell})$ with $\ell =\ell (\lambda )$
nonzero rows, let
$(\vec \alpha |\vec \beta )=(\alpha_1, \ldots , \alpha_{d(\lambda )}|
\beta_1 , \ldots , \beta_{d(\lambda )})$ be the Frobenius notation
for the diagram $\lambda$ (see Appendix A). 
Here $d(\lambda )$ is the number of
boxes in the main diagonal and $\alpha_i =\lambda_i -i$,
$\beta_i =\lambda'_i -i$, where $\lambda'$ is the transposed
(reflected about the main diagonal) diagram $\lambda$. Then
\beq\label{lambda1}
\begin{array}{l}
\left |\lambda , n\rbr :=
\psistar_{n-\beta_1 -1}\ldots \psistar_{n-\beta_{d(\lambda )}\! -1}\,
\psi_{n+\alpha_{d(\lambda )}}\ldots \psi_{n+\alpha_1}\rvacn ,
\\ \\
\lbr \lambda , n \right |:=
\lvacn \psistar_{n+\alpha_1}\ldots \psistar_{n+\alpha_{d(\lambda )}}\,
\psi_{n-\beta_{d(\lambda )}\! -1}\ldots \psi_{n-\beta_1 -1} .
\end{array}
\eeq
The state $\left |\lambda , n\rbr$ has the charge $n$
with respect to the vacuum state $\rvac$. For the empty diagram
$\left < \emptyset , n\right |=\lvacn$, 
$\left | \emptyset , n\right >=\rvacn$.

The states $\left |\lambda , n \right >$ can be constructed from a vacuum
in another, equivalent way which is sometimes more convenient.
\begin{prop}\label{states}
We have:
\beq\label{lambda21}
\begin{array}{l}
\left |\lambda , n \right >=(-1)^{b(\lambda )}
\, \psi_{n+\lambda_1 -1}\psi_{n+\lambda_2 -2}\ldots 
\psi_{n+\lambda_{\ell} -\ell}\left | n-\ell \right >
\\ \\
\left <\lambda , n \right |=(-1)^{b(\lambda )}
\left < n-\ell \right | \psistar_{n+\lambda_{\ell} -\ell}
\ldots \psistar_{n+\lambda_2 -2}\psistar_{n+\lambda_1 -1}
\end{array}
\eeq
and
\beq\label{lambda31}
\begin{array}{l}
\left |\lambda , n \right >=(-1)^{|\lambda |-b(\lambda )}
\, \psistar_{n-\lambda_1'}\psistar_{n-\lambda_2' +1}\ldots 
\psistar_{n-\lambda_{m}' +m-1}\left | n+m \right >
\\ \\
\left <\lambda , n \right |=(-1)^{|\lambda |-b(\lambda )}
\left < n+m \right | \psi_{n-\lambda_{m}' +m-1}
\ldots \psi_{n-\lambda_2' +1}\psi_{n-\lambda_1'}
\end{array}
\eeq
where $\ell =\ell (\lambda )=\lambda_1'$, $m=\lambda_1$ and 
\beq\label{boflambda}
b(\lambda )=\sum_{i=1}^{d(\lambda )}(\beta_i +1).
\eeq
\end{prop}

\noindent
This statement is basically a rephrasing of Proposition (1.7) 
from Macdonald's book \cite{Macdonald}, with an additional input 
of commutaton relations for fermionic operators.
The proof is a simple combinatorial exercise.
We give an idea how to prove (\ref{lambda21}).
Denote the r.h.s. by $\left |\lambda ,n\right >'$, then we can write
$$
\left |\lambda ,n\right >'=\psi_{n+\lambda_1 -1}\ldots 
\psi_{n+\lambda_d -d}\,\, \underbrace{\psi_{n+\lambda_{d+1} -(d+1)}\ldots 
\psi_{n+\lambda_\ell -\ell}}_{\ell -d}\,\,
\underbrace{\psistar_{n-\ell}\ldots \psistar_{n-1}}_{(\ell -d)+d}
\rvacn
$$
where $d=d(\lambda )$ is the size of the maximal square of the diagram.
Since $\lambda_i-i=\alpha_i$ at $i=1, \ldots , d$, we see that the first 
$d$ $\psi$-operators are exactly those standing in (\ref{lambda1}).
The $\ell -d$ $\psi$-operators from the next group have indices 
which are strictly less than $n$, and each $\psi$-operator 
$\psi_{n-i}$ from 
this group has its $\psistar$-partner $\psistar_{n-i}$ among the 
$\psistar$-operators from the third group. These pairs
$\psi_{n-i}\psistar_{n-i}$ cancel when acting to the vacuum $\rvacn$
and we are left with just $d$ $\psistar$-operators which
are exactly those standing in (\ref{lambda1}). 
(It may be helpful to 
note that the Frobenius notation corresponds to
$\lambda_i$'s as follows: $\{\lambda_i -i\}_{i=1}^{d(\lambda )}=
\{\alpha_i \}_{i=1}^{d(\lambda )}$, 
$\{i-\lambda_i \}_{i=d(\lambda )+1}^{\beta_1 +1}=
\{1,2, \ldots , \beta_1 +1 \}\setminus 
\{\beta_i +1\}_{i=1}^{d(\lambda )}$.)
Therefore,
$\left |\lambda ,n\right >'$ differs from
$\left |\lambda ,n\right >$ only by a sign.

\noindent
{\bf Remark.}
Note that if one formally adds some ``zero parts'' with 
$\lambda_i=0$ to the partition $\lambda$, then the state
(\ref{lambda21}) remains unchanged.

From the definition of the vacuum states $\rvacn$ it is 
obvious that for any such state and arbitrary $k\in \mathbb{Z}$ 
either  $\psi_k$ or $\psistar_k$ is an annihilation operator: either $\psi_k\rvacn=0$ or $\psistar_k\rvacn=0$.
From definition (\ref{lambda1}) it is also
obvious that the same is true for any basis state:
\beq\label{fermnt1}
\begin{array}{l}
\psi_k \left|\lambda,n\right> =0, \quad k\in I_{n,\lambda}; \quad \quad \quad
\psistar_k \left|\lambda,n\right> =0, \quad k\notin I_{n,\lambda},
\\ \\
\psi_k \left|\lambda,n\right> \neq0, \quad k\notin I_{n,\lambda}; \quad \quad \quad
\psistar_k \left|\lambda,n\right> \neq0, \quad k\in I_{n,\lambda},
\end{array}
\eeq
and 
\beq\label{fermnt2}
\begin{array}{l}
\left<\lambda,n\right| \psi_k =0, \quad k\notin I_{n,\lambda}; 
\quad \quad \quad
\left<\lambda,n\right| \psistar_k =0, \quad k\in I_{n,\lambda},
\\ \\
\left<\lambda,n\right|  \psi_k  \neq0, \quad k\in I_{n,\lambda}; \quad \quad \quad
\left<\lambda,n\right| \psistar_k  \neq0, \quad k\notin I_{n,\lambda},
\end{array}
\eeq
where $I_{n,\lambda}$ is the set 
$$
I_{n,\lambda}=\left \{ k\in \ZZ \phantom{\Bigl |}\right |
\left. \phantom{\Bigr |}
k<n, \, k\notin \bigl \{ n\! -\! 
\beta_i \! -\! 1\bigr \}_{i=1}^{d(\lambda )}\right \}
\cup \bigl \{ n+\alpha_i \bigr \}_{i=1}^{d(\lambda )}.
$$

\subsection{The vacuum expectation values}

The vacuum expectation value $\lvac \ldots \rvac$ is a
Hermitian linear form on the Clifford algebra fixed
by 
$$
\left. \lvac \! 0 \right > =1.
$$
Then, from the commutation relations (\ref{fermop1}) and definitions of the ``shifted'' Dirac vacua (\ref{vacdefr}), (\ref{vacdefl}) it follows that $\left. \lvacn \! n\right > =1$ for any $n$.
Bilinear combinations of fermions
satisfy the properties 
$\lvacn \psi_i \psi_j\rvacn = \lvacn \psistar_i \psistar_j \rvacn =0$
for all $i,j$ and
$$
\lvacn \psi_i \psistar_j\rvacn =\delta_{ij}\quad
\mbox{for \(j<n\)}, \quad \quad
\lvacn \psi_i \psistar_j\rvacn =0 \quad \mbox{for \(j\geq n\)}.
$$
The expectation value of any operator with non-zero charge 
is zero. The basis vectors (\ref{lambda1})
are orthonormal with respect to the
scalar product induced by the expectation value:
$$
\lbr \lambda , n\right | \left. \mu , m \rbr =
\delta_{mn}\delta_{\lambda \mu}.
$$
This can be directly seen by taking the scalar product
of the vectors of the type (\ref{lambda1}) and moving the 
operators $\psi_{n-\beta_i -1}$ to the right, taking into account
that the sequences $\alpha_1, \alpha_2, \ldots , \alpha_d$ and
$\beta_1, \beta_2, \ldots , \beta_d$ are strictly increasing.

From orthonormality of the basis states it 
follows that if $\lbr \lambda , n\right. \left |  U \rbr = 0$ 
(respectively $\lbr U \right | \left. \lambda , n \rbr =0$) for all 
$\lambda$ and $n$ then $\left| U \rbr=0$ (respectively $\lbr U \right | =0 $). Moreover, if a Clifford algebra element $X$ 
kills all basis states, that is $X | \left. \lambda , n \rbr =0$ (or $\lbr \lambda , n\right|  X = 0$) for all $\lambda$ and $n$, then $X=0$. 

\noindent
\begin{prop}\label{unidec}
The coefficients $\lbr \lambda , n\right. \left |  U \rbr$  
(respectively $\lbr U \right | \left. \lambda , n \rbr$) 
uniquely determine the state $\left |  U \rbr$ 
(respectively $\lbr U \right |  $):
$$
\left |  U \rbr=\sum_{n,\lambda} \left |\lambda , n \rbr
\lbr \lambda , n\right. 
\left |  U \rbr  
$$
Similarly, the 
coefficients $ \lbr \lambda , n\right. | X | \left. \mu , m \rbr$ 
uniquely determine the Clifford algebra element $X$:
$$
X= \sum_{n,\lambda ; m,\mu} \left |\lambda , n \rbr
\left < \lambda , n\right |X\left |\mu , m\right >
\lbr \mu , m \right |.
$$
\end{prop}
The explicit form of the operators 
$\left |\lambda , n \rbr \left <\mu , m\right |$ 
in terms of fermions is given below in section \ref{Proop}.

In general, expectation values of products of
fermionic operators are given by the Wick theorem.
Let $v_i =\sum_j v_{ij}\psi_j$ be linear combinations
of $\psi_j$'s only and
$w_i^*=\sum_j w^{*}_{ij}\psistar_j$
be linear combinations of $\psistar_j$'s only,
then the standard Wick theorem states that
$$
\lvacn v_1 \ldots v_m w_m^* \ldots w_1^* \rvacn =
\det_{i,j =1,\ldots , m}\lvacn v_i w_j^* \rvacn ,
$$
$$
\lvacn  w_1^*  \ldots w_m^* v_m  \ldots  v_1  \rvacn =
\det_{i,j =1,\ldots , m}\lvacn w_i^* v_j  \rvacn .
$$
The Wick
theorem can be proved by induction. We will not give the proof here
because the proof of a more general statement will be presented  below.

For the generating series $\psi (z)$, $\psistar (\zeta )$ we have:
$$
\lvacn \psistar (\zeta ) \psi (z)  \rvacn =
\sum_{j,k}\zeta^{-j}z^k \lvacn \psistar_j \psi_k \rvacn =
\sum_{k\geq n}(z/\zeta )^k =
\frac{z^n\zeta^{1-n}}{\zeta -z}
$$
(assuming that $|\zeta |>|z|$)
and
$$
\lvacn \psi (z) \psistar (\zeta )   \rvacn =
\sum_{j,k}\zeta^{-j}z^k \lvacn \psi_k \psistar_j  \rvacn =
\sum_{k< n}(z/\zeta )^k =\frac{z^n\zeta^{1-n}}{z-\zeta}
$$
(assuming that $|z|>|\zeta |$).
More generally,
\beq\label{ferm5}
\begin{array}{c}
\displaystyle{
\lvacn \psistar (\zeta_1)\ldots \psistar (\zeta_m)
\psi (z _m)\ldots \psi (z _1)\rvacn =
\prod_{l=1}^{m} (z_l / \zeta_l)^n \cdot \, \det_{i,j}
\frac{\zeta _i}{\zeta_i-z_j}}
\\  \\
\displaystyle{
\phantom{aaaaaaaaaaaaaaaaaaaa}
=\,\, \frac{\prod\limits_{i<i'}(z_i-z_{i'})
\prod\limits_{j>j'}(\zeta _j-\zeta _{j'})}{\prod\limits_{i,j}(\zeta_i-z_j)}
\prod_l z_{l}^{n}\zeta_{l}^{1-n}}
\end{array}
\eeq
(assuming that $|\zeta _i|>|z_j|$) and
\beq\label{ferm5a}
\begin{array}{c}
\displaystyle{
\lvacn \psi (z_1)\ldots \psi (z_m)
\psistar (\zeta _m)\ldots \psistar (\zeta _1)\rvacn =
\prod_{l=1}^{m} (z_l / \zeta_l)^n \cdot \, \det_{i,j}
\frac{\zeta _i}{z_i-\zeta_j}}
\\  \\
\displaystyle{
\phantom{aaaaaaaaaaaaaaaaaaaa}
=\,\, \frac{\prod\limits_{i<i'}(z_i-z_{i'})
\prod\limits_{j>j'}(\zeta _j-\zeta _{j'})}{\prod\limits_{i,j}(z_i-\zeta_j)}
\prod_l z_{l}^{n}\zeta_{l}^{1-n}}
\end{array}
\eeq
(where $|z_i|>|\zeta _j|$).
One can also calculate
\beq\label{fermi5c}
\left <n\! +\! 
l\right | \psistar (\zeta _{m-l}) \ldots \psistar (\zeta _{1})
\psi (z_1)\ldots \psi (z_m)
\rvacn =
\frac{\displaystyle{\prod_{i<i'}^{m}(z_i-z_{i'})\!\!
\prod_{j>j'}^{m-l}(\zeta_j -\zeta_{j'})}}{\displaystyle{
\prod_{r=1}^{m-l}\prod_{s=1}^{m}
(\zeta_r -z_s)}}\prod_{k=1}^m z_k^n \!
\prod_{k'=1}^{m-l}\zeta_{k'}^{1-n}
\eeq
(where $|\zeta _i|>|z_j|$) as a limiting case of
(\ref{ferm5}) by tending subsequently $\zeta_m$, 
$\zeta_{m-1}\ldots ,\zeta_{m-l+1}$ to infinity.
In particular,
\beq\label{ferm5e}
\left <n+m\right | 
\psi (z_1)\ldots \psi (z_m) \rvacn =
\prod_{i<j}(z_i-z_j)\, \prod_{k=1}^m z_k^n .
\eeq
In a similar way, we get from (\ref{ferm5a})
\beq\label{ferm5d}
\left <n\! -\! 
l\right | \psi (z_{m-l}) \ldots \psi (z_{1})
\psistar (\zeta_1)\ldots \psistar (\zeta_m)
\rvacn =
\frac{\displaystyle{\prod_{i<i'}^{m}(\zeta_i-\zeta_{i'})\!\!
\prod_{j>j'}^{m-l}(z_j -z_{j'})}}{\displaystyle{
\prod_{r=1}^{m-l}\prod_{s=1}^{m}
(z_r -\zeta_s)}}\prod_{k=1}^m \zeta_k^{1-n} \!
\prod_{k'=1}^{m-l}z_{k'}^{n}
\eeq
(where $|z_i|>|\zeta _j|$) and
\beq\label{ferm5e1}
\left <n\! -\! 
m\right |
\psistar (\zeta _{1}) \ldots \psistar (\zeta _{m})  \rvacn=
\prod_{i<j}(\zeta_i-\zeta_j)\, \prod_{k=1}^m \zeta_k^{1-n}.
\eeq

\subsection{Normal ordering}

Once the vacuum state 
is fixed, a useful notion is the normal ordering of operators.
The normal ordering $\normord (\ldots )\normord $
with respect
to the Dirac vacuum $\rvac$ is defined as 
follows: all annihilation operators
are moved to the right and all creation operators are moved to
the left taking into account that the factor $(-1)$ appears 
each time two neighboring 
fermionic operators exchange their positions.
For example:
$\normord  \psistar_{1}\psi_{1}\normord =
-\psi_{1} \psistar_{1}$, 
$\normord  \psi_{-1}\psi_{0}\normord =
-\psi_{0} \psi_{-1}$, 
$\normord  \psi_{2}\psistar_{1}\psi_1 \psistar_{-2}\normord =
\psi_{2}\psi_1 \psistar_{-2}\psistar_{1}$,
etc. We also note the identity
$$
e^{\alpha \psi_k \psistar_k}=1+(e^{\alpha}-1)\psi_k \psistar_k=
\normord e^{(e^{\alpha}-1)\psi_k \psistar_k}\normord \,,
\quad k\geq 0.
$$
The normally ordered expressions are always well-defined when acting
to the corresponding basis states of the fermionic Fock space 
${\cal H}_F$.

Under the sign of normal ordering, all fermionic 
operators $\psi_j$ and $\psistar_j$ anticommute.
In other words, it is wrong to use the commutation relations
of the Clifford algebra under the sign of normal ordering,
i.e., for example, $\normord \psistar_{1}\psi_{1}\normord 
\neq \normord (1-\psi_{1} \psistar_{1})\normord$.

Using the normal ordering, one can introduce the charge operator
$Q$ as
\beq\label{chargeoper}
Q=\sum_{k\in \z}\normord \psi_k \psistar _k\normord .
\eeq
This operator counts the charge of the state:
$Q\left |\lambda , n\right > =n \left |\lambda , n\right >$
and so $\left <\mu , m\right |Q\left |\lambda , n\right >=
n\delta_{nm}\delta_{\lambda \mu}$ (note that without normal
ordering this matrix element would be 
ill-defined!).
The charge operator
has the commutation relations
$
[Q, \psi_n ]=\psi_n$, 
$[Q, \psistar _n]=-\psistar_n$ which mean that 
$\psi_n$, $\psistar_n$ have charges $\pm 1$.
More generally, we say that a Clifford algebra element $X$
has definite charge $q$ if $[Q, X]=qX$. 

In a similar way, one may define the normal ordering with respect to
any vacuum state. The general rule is that the annihilation operators
are moved to the right and creation operators to the left (with 
the appropriate sign factor).
With respect to the vacuum $\rvacn$, the annihilation 
operators are $\psistar_j$ with $j\geq n$ and $\psi_j$ with
$j<n$ while $\psistar_j$ with $j<n$ and $\psi_j$ with
$j\geq n$ are creation operators. We denote the 
corresponding normal ordering by $\normord (\ldots )\normord _{n}$
(in this notation $\normord (\ldots )\normord =
\normord (\ldots )\normord _{0}$). One can go even further and define the 
normal ordering with respect to any basis state (\ref{lambda1}): the 
operators $\psi_k$ with $k\in I_{n,\lambda}$ and $\psistar_k$ 
with $k\notin I_{n,\lambda}$ are annihilation 
operators for the state $\left|\lambda,n\right>$, 
while all other fermions are creation 
operators, see (\ref{fermnt1}) and (\ref{fermnt2}).

One may also consider the bare vacuum
$\left |\infty \right >$ which is the absolutely empty state.
With respect to the bare vacuum, all $\psi_j$'s are annihilation
operators while all $\psistar_j$'s are creation operators. 
Let us denote the corresponding normal ordering by
$\normordbare (\ldots )\normordbare$ (formally,
$\normordbare (\ldots )\normordbare = 
\normord (\ldots )\normord _{\infty}$).
In this normal ordering, all $\psistar$'s are moved to the left 
and all $\psi$'s to the right, with taking into account the sign 
factor appearing each time one operator is permuted with another.
For example: 
$\normordbare  \psistar_{m}\psi_{n}\normordbare =
\psistar_{m}\psi_{n}$, $\normordbare \psi_{n}\psistar_{m}\normordbare =
-\psistar_{m}\psi_{n}$ 
and
\beq\label{Bex}
\normordbare
\exp \Bigl (\sum_{ik}B_{ik}\psistar_i \psi_k\Bigr )
\normordbare =
1+ \sum_{i,k}B_{ik}\psistar_i \psi_k +\frac{1}{2!}
\sum_{i,i'k,k'}B_{ik}B_{i'k'}\psistar_i \psistar _{i'}
\psi_{k'} \psi_k +\ldots
\eeq

The definition of the normal ordering 
is closely related to the definition of the vacuum expectation value:
$\normord \psistar_k \psi_l\normord _n =
\psistar_k \psi_l -\lvacn \psistar_k \psi_l \rvacn$.
More generally, for any linear combinations $f_0, f_1,
\ldots , f_m$ of the fermion operators $\psi_i, \psistar_j$
we have the recursive formula
\beq\label{normord}
f_0 \normord  f_1 f_2 \ldots
f_m  \normord _n =\normord
f_0 f_1 f_2 \ldots  f_m \normord _n
+\sum_{j=1}^{m}(-1)^{j-1}
\lvacn f_0 f_j \rvacn \normord f_1 f_2 \ldots
\not{\!\!f}_{\!\!j} \ldots f_m\normord _n ,
\eeq
where $\not{\!\!f}_{\!\!j}$ means that this factor should be omitted.
This recursive relation allows one to express the normally ordered
monomials with arbitrary number of fermions as linear combinations
of monomials without normal ordering and vice versa. For example:
$$
\normord f_0 f_1 f_2 \normord _n =
f_0 f_1 f_2 - \left < f_1 f_2 \right >_n f_0 +
\left < f_0 f_2 \right >_n f_1
-\left < f_0 f_1 \right >_n f_2 ,
$$
$$
\begin{array}{ll}
\normord f_0 f_1 f_2 f_3\normord _n \, &=\, f_0 f_1 f_2 f_3
-\left <f_2 f_3 \right >_n f_0 f_1  +
\left <f_1 f_3 \right >_n f_0 f_2 -
\left <f_1 f_2 \right >_n f_0 f_3 
\\& \\
&\,\,\,\, -
\left <f_0 f_1 \right >_n f_2 f_3 +
\left <f_0 f_2 \right >_n f_1 f_3 -
\left <f_0 f_3 \right >_n f_1 f_2
\\ &\\
&\,\,\,\, + \left <f_0 f_1 \right >_n \left <f_2 f_3 \right >_n -
\left <f_0 f_2 \right >_n \left <f_1 f_3 \right >_n
+ \left <f_0 f_3 \right >_n \left <f_1 f_2 \right >_n ,
\end{array}
$$
where $\left <f_i f_j \right >_n := \lvacn f_i f_j \rvacn$.

\subsection{Group-like elements}

\subsubsection{Group elements}

Bilinear combinations
$\sum_{mn} b_{mn}\psistar_m \psi_n$
of the fermions, with certain conditions
on the matrix $b = (b_{mn})$, generate an
infinite-dimensional Lie algebra \cite{JM83}.
Exponentiating these expressions, one obtains
an infinite dimensional group (a version
of $GL(\infty )$).
Elements of this group can be represented
in the form
\begin{equation}\label{gl}
G=\exp \Bigl (\sum_{i, k \in {\z }}b_{ik}\psistar_i \psi_k\Bigr ).
\end{equation}
The inverse element can be written in the same way with the matrix
$(-b_{mn} )$.

As it was already stated, the group elements of the form
(\ref{gl}) obey a rather special property that the adjoint
action of such elements preserves the linear space spanned
by the fermion operators $\psi _n$ and the space
spanned by $\psistar _n$. More precisely, we have:
$$
G\psistar _n G^{-1} = \sum_l \psistar_l R_{ln}\,, \quad
G\psi _n G^{-1} = \sum_l  (R^{-1})_{nl}\psi_l
$$
or
\beq\label{rotation}
G\psistar_n =\sum_{l} R_{ln} \psistar_l  G\,, \quad \quad
\psi_n G =  \sum_l R_{nl} G\psi_l\,,
\eeq
where the matrix
$R = (R_{nl})$ of the induced linear transformation
is given by $R=e^b$. 
The product of two group elements is also a group
element:
\beq\label{comp1}
\exp \Bigl (\sum_{i, k \in {\mathbb Z}}b'_{ik}\psistar_i \psi_k\Bigr )
\exp \Bigl (\sum_{i, k \in {\mathbb Z}}b_{ik}\psistar_i \psi_k\Bigr )=
\exp \Bigl (\sum_{i, k \in {\mathbb Z}}b''_{ik}\psistar_i \psi_k\Bigr )
\eeq
with $e^{b'} e^b =e^{b''}$. Clearly, multiplication of $G$ of the form 
(\ref{gl}) by any constant number preserves the characteristic 
property (\ref{rotation}).

We refer to the 
transformation (\ref{rotation}) as {\it rotation} of the 
fermionic operators with the rotation matrix $R$.
From the fact that the center of the Clifford algebra is
formed by numbers $c$ \cite{SMJ} it follows that two group elements 
$G, G'$ with the same rotation matrix $R$ may differ by a 
scalar factor only: $G'=c G$. In particular, sometimes 
it is more convenient to consider the group elements 
with exponentiated normally ordered 
bilinear fermionic combinations, for example:
$$
\exp \Bigl (\sum_{i, k \in {\z }}b_{ik}\normord
\psistar_i \psi_k\normord\Bigr )=c_b \exp 
\Bigl (\sum_{i, k \in {\z}}b_{ik}\psistar_i \psi_k\Bigr ),
$$
where $c_b=\exp\left(-\sum_{k\geq 0} b_{kk}\right)$.

Note also that the charge 
operator $Q$ (\ref{chargeoper}) commutes with any group
element.

\subsubsection{Normally ordered exponents}

Let us note,
following the works
of the Kyoto school \cite{JM83,SMJ}, that the group elements
can be equivalently represented
as {\it normally ordered} exponents of bilinear forms.
For example, one can directly prove that the element
$G=\normordbare e^{ B_{ik}\psistar_i \psi_k}\normordbare$
(here and below summation over repeated indices is implied)
satisfies the first commutation relation (\ref{rotation})
with $R_{ln}=\delta_{ln}+B_{ln}$:
$$
\begin{array}{c}
\normordbare e^{ B_{ik}\psistar_i\psi_k}\normordbare \psistar_n =
\left ( 1+ B_{a_1 b_1}\psistar_{a_1}\psi _{b_1}+\frac{1}{2!}
B_{a_1 b_1}B_{a_2 b_2}\psistar_{a_1}\psistar_{a_2}\psi _{b_2}\psi _{b_1}+
\ldots  \right ) \psistar_n
\\  \\
=\psistar_n \normordbare e^{ B_{ik}\psistar_i\psi_k}\normordbare
+B_{a_1 n}\psistar_{a_1}\! +\! 
B_{a_1 n}B_{a_2 b_2}\psistar_{a_1}\! \psistar_{a_2}
\psi _{b_2}\! +\! \frac{1}{2!}B_{a_1 n}B_{a_2 b_2}B_{a_3 b_3}
\psistar_{a_1}\! \psistar_{a_2}\! \psistar_{a_3}
\psi _{b_3}\psi _{b_2}+\ldots 
\\ \\
=\psistar_n \normordbare e^{ B_{ik}\psistar_i\psi_k}\normordbare
+B_{an}\psistar_a \normordbare e^{ B_{ik}\psistar_i\psi_k}\normordbare \, =\,
(\delta_{an}+B_{an})\psistar_a 
\normordbare e^{ B_{ik}\psistar_i\psi_k}\normordbare \,.
\end{array}
$$
The second commutation relation (\ref{rotation}) can be 
proved in the same way. Moreover, any solution to (\ref{rotation}) can be represented by a normally ordered exponent
$\normordbare e^{B_{ik}\psistar_i \psi_k}\normordbare$.
It can be also checked, in a similar way, that 
\begin{equation}\label{ferm1a}
\exp \Bigl (b_{ik}\psistar_i \psi_k\Bigr )=
\normordbare \exp \Bigl ((e^b -I)_{ik}\psistar_i \psi_k\Bigr )
\normordbare ,
\eeq
where $I$ is the unity
matrix.
The composition law is given by
\beq\label{comp2}
\normordbare \exp \Bigl (
B'_{ik}\psistar_i \psi_k\Bigr )
\normordbare
\normordbare \exp \Bigl (
B_{ik}\psistar_i \psi_k\Bigr )
\normordbare =
\normordbare \exp \Bigl (
(B\! +\! B'\! +\! B'B)_{ik}\psistar_i \psi_k\Bigr )
\normordbare 
\eeq
which directly follows from the composition law (\ref{comp1})
and the formula $B=e^b -I$.

Let us prove another useful formula which allows one to
represent a group element 
as a normally
ordered exponent with respect to different vacua:

\noindent
\begin{prop}\label{diffno}
Whenever both sides  are well-defined the following is true
\begin{equation}\label{no1}
\normordbare \exp \Bigl (
B_{ik}\psistar_i \psi_k\Bigr )
\normordbare
=\det (I\! +\! P_{+} B)\,
\normord \exp \Bigl (
A_{ik}\psistar_i \psi_k\Bigr )\normord,
\end{equation}
or, equivalently,
\begin{equation}\label{no1a}
\normord \exp \Bigl (
A_{ik}\psistar_i \psi_k\Bigr ) 
\normord =\det (I\! -\! P_{+} A)\,
\normordbare \exp \Bigl (
B_{ik}\psistar_i \psi_k\Bigr )
\normordbare.
\end{equation}
Here $P_{+}$ is the projector on the positive mode space
($(P_{+})_{ik}=\delta_{ik}$ for $i,k\geq 0$ and 0 otherwise)
and the matrices $A,B$ are connected by
\beq\label{no2}
B-A =AP_{+} B\,, \quad \mbox{i.e.,}\quad
B=(I\! - \! AP_+)^{-1}A \quad \mbox{or}\quad 
A=B(I\! + \! P_+B)^{-1}.
\eeq
\end{prop}

\noindent
The idea of the proof is as in \cite{SMJ}.
First we notice that we can write the
normally ordered expression $\normord e^{A_{ik}\psistar_i \psi_k} 
\normord$ as the composition of three operators:
\beq\label{nodecomp}
\normord e^{A_{ik}\psistar_i \psi_k} 
\normord =\underbrace{\phantom{\normord}
e^{A_{\bar a b}\psistar_{\bar a} \psi_b}\phantom{\normord}}_{G_1}\,\cdot \,
\underbrace{ \, \normord
e^{A_{a b}\psistar_{a} \psi_b}\cdot
e^{A_{\bar a \bar b}\psistar_{\bar a} 
\psi_{\bar b}}\normord }_{G_2} \, \cdot \,
\underbrace{\phantom{\normord}
e^{A_{a \bar b}\psistar_{a} \psi_{\bar b}}\phantom{\normord} }_{G_3}
\eeq
where the repeated indices $a,b$ ($\bar a, \bar b$) 
in the r.h.s. imply summation over 
non-negative (respectively, negative) 
integers (summation 
in the repeated $i,k$ in the l.h.s. is over $\ZZ$). 
The operator $G_1$ contains the creation operators only
while $G_3$ contains annihilation operators only. Note also that the
two operators under the normal ordering commute with each other.
It is not difficult to 
find explicitly which rotations are performed by 
the elements $G_1, G_2, G_3$. For $G_{1,3}$ this is 
especially simple:
$$
\psi_n G_1 =G_1 \left \{
\begin{array}{l}\psi_n +A_{nb}\psi_b \,, \quad n<0
\\ \psi_n \,, \quad \quad \quad \quad \quad \! n\geq 0,
\end{array}\right.
\quad \quad 
\psi_n G_3 =G_3 \left \{
\begin{array}{l}
\psi_n \,, \quad \quad \quad \quad \quad \! n< 0
\\
\psi_n +A_{n\bar b}\psi_{\bar b} \,, \quad n\geq 0.
\end{array}\right.
$$
For $G_2$ we write $G_2 = \normord G_2^+ G_2^-\normord$
with $G_2^+=e^{A_{a b}\psistar_{a} \psi_b}$,
$G_2^-=e^{A_{\bar a \bar b}\psistar_{\bar a}\psi_{\bar b}}$.
When we move $\psi_n$ through this element, we can ignore
either $G_2^+$ or $G_2^-$ depending on whether $n$ is 
negative or positive. The rest of the calculation is 
similar to the one with the normal ordering 
$\normordbare (\ldots )\normordbare$ given above.
It gives:
\beq\label{g2rot} 
\left \{ \begin{array}{l}
\psi_n G_2 =G_2 (\psi_n +A_{n\bar b}\psi_{\bar b}),
\quad n<0
\\ \\
G_2\psi_n =(\psi_n -A_{n b}\psi_b )G_2 , \quad n\geq 0.
\end{array} \right.
\eeq
It is instructive to write these rotations in the block matrix
form:
$$
\left ( \begin{array}{c}\psi_{-}G_1\\ \psi_+ G_1\end{array}\right )
=\left (\begin{array}{cc} I&A^-_{+}\\0& I\end{array}\right )
\left ( \begin{array}{c}G_1\psi_{-}\\ G_1\psi_+ \end{array}\right ),
$$
$$
\left ( \begin{array}{c}\psi_{-}G_3\\ \psi_+ G_3\end{array}\right )
=\left (\begin{array}{cc} I&0\\A^+_{-}& I\end{array}\right )
\left ( \begin{array}{c}G_3\psi_{-}\\ G_3\psi_+ \end{array}\right ),
$$
$$
\left ( \begin{array}{c}\psi_{-}G_2\\ \psi_+ G_2\end{array}\right )
=\left (\begin{array}{cc} I\! +\! A^-_{-}&0\\0& (I\! -\! A^+_{+})^{-1}
\end{array}\right )
\left ( \begin{array}{c}G_2\psi_{-}\\ G_2\psi_+ \end{array}\right ),
$$
where the self-explanatory notation is used
(we assume that the matrix $I\! -\! A^+_{+}$ is invertible).
In this notation 
$\displaystyle{
P_+ = \left (\begin{array}{cc}0&0\\ 0&I\end{array}
\right )}
$. The full rotation matrix is then obtained as the product 
of these three:
$$
\begin{array}{lll}
R&=&\displaystyle{\left (\begin{array}{cc} I&A^-_{+}\\0& I
\end{array}\right )
\left (\begin{array}{cc} I\! +\! A^-_{-}&0\\0& (I\! -\! A^+_{+})^{-1}
\end{array}\right )
\left (\begin{array}{cc} I&0\\A^+_{-}& I\end{array}\right )}
\\ &&\\
&=& \left (\begin{array}{cc}I&0\\ 0&I\end{array}
\right )+
\left (\begin{array}{cc} A_{-}^{-}\! +\!  A_{+}^{-}(I\! -\! A^+_{+})^{-1}
 A_{-}^{+}
&A^-_{+} (I\! -\! A^+_{+})^{-1}\\(I\! -\! A^+_{+})^{-1} A^+_-& 
(I\! -\! A^+_{+})^{-1}A^+_+
\end{array}\right ).
\end{array}
$$
One can check that the second matrix in the last line 
(which is $R-I=B$) is exactly $(I\! -\! AP_+)^{-1}A$, in agreement
with (\ref{no2}). It remains to calculate the scalar factor 
in front of (\ref{no1a}). Let us take the expectation value 
of the both sides
with respect to the bare vacuum. Then we should prove that 
$$
\left <\infty \right |\normord \exp \Bigl (
A_{ik}\psistar_i \psi_k\Bigr ) 
\normord \left |\infty \right > =\det (I\! -\! P_{+} A).
$$
Using the decomposition of the operator in the l.h.s., 
we write:
$$
\begin{array}{c}
\left <\infty \right |\normord e^{
A_{ik}\psistar_i \psi_k}
\normord \left |\infty \right >=
\left <\infty \right |\normord e^{-A_{ab}\psi_b \psistar_a}
\normord \left |\infty \right >
\\ \\
=\, \displaystyle{
\sum_{k\geq 0}\frac{(-1)^k}{k!}A_{a_1 b_1}\ldots 
A_{a_k b_k}\left <\infty \right | \psi_{b_1}\ldots \psi_{b_k}
\psistar_{a_k}\ldots \psistar_{a_1}\left |\infty \right >}
\\ \\
=\, \displaystyle{
\sum_{k\geq 0}\frac{(-1)^k}{k!}
\sum_{a_1, \ldots , a_k \geq 0}
\left |\begin{array}{cccc}A_{a_1 a_1}&A_{a_1 a_2} & \ldots &A_{a_1 a_k}
\\  
A_{a_2 a_1}&A_{a_2 a_2} & \ldots &A_{a_2 a_k}
\\  
\ldots &  \ldots & \ldots & \ldots 
\\  
A_{a_k a_1}&A_{a_k a_2} & \ldots &A_{a_k a_k}
\end{array}\right |}
\\ \\
=\, \det (I-A_+^+)=\det (I-P_+A).
\end{array}
$$

Proposition \ref{diffno} can be 
easily generalized to the normal ordering 
$\normord (\ldots )\normord _n$.
For example, for $G$ given in (\ref{gl}) it can be written as
\begin{equation}\label{ferm1}
G=
\normordbare \exp \Bigl (
B_{ik}\psistar_i \psi_k\Bigr )
\normordbare
=\det (I\! +\! P_{\geq n} B)\,
\normord \exp \Bigl (
A_{ik}\psistar_i \psi_k\Bigr )
\normord _{n}
\end{equation}
with the matrices $A$, $B$ determined by
the matrix $b$ in (\ref{gl}) according to the formulas \cite{SMJ}
\beq\label{matrices}
B=e^b -I\,, \quad  \quad B-A =AP_{\geq n} B\,.
\eeq
Here $P_{\geq n}$ is the projector on the $\geq n$ mode space
($(P_{\geq n})_{ik}=\delta_{ik}$ for $i,k\geq n$ and 0 otherwise).

\subsubsection{Non-invertible (group-like) elements}

The normal ordering allows one
to represent in the form (\ref{ferm1}) not only group elements but
also some non-invertible elements of the Clifford algebra that satisfy
the commutation relations
\beq\label{rotation-a}
G\psistar_n =\sum_{l} R_{ln} \psistar_l  G\,, \quad \quad
\psi_n G =  \sum_l R_{nl} G\psi_l\,,
\eeq
or
\beq\label{rotation-b}
\psistar_n G=\sum_{l} R'_{ln} G \psistar_l  \quad \quad
G \psi_n  =  \sum_l R'_{nl} \psi_l \, G
\eeq
with some (not necessarily invertible)
matrices $R, R'$ (if they are invertible, then both pairs 
of these relations
hold with $R'R=I$ but
for non-invertible elements only one pair of these relations
holds). Any normally ordered 
exponent $G=\normord e^{B_{ik}\psistar_i \psi_k}
\normord _{-\infty}$ satisfies (\ref{rotation-b}) 
and, vice versa, if $G$ is a solution of (\ref{rotation-b}) then it can be represented by a normally ordered 
exponent of this form. 

We call elements $G$ of the Clifford algebra such that
the commutation relations (\ref{rotation-a}) or 
(\ref{rotation-b})
hold the 
{\it group-like elements} (in the next section this definition
will be further extended).
If the matrix
$R$ fails to be invertible, so does $G$, as an element
of the Clifford algebra. In this case it can not be represented
in the exponential form (\ref{gl}). Some of such elements
still admit a representation as a {\it normally ordered} exponent of
bilinear forms in the fermionic operators. However, there are 
normally ordered exponents such that neither $R$ nor $R'$ matrices
exist for them.

\noindent
{\bf Examples.} 
\begin{itemize}
\item[a)] Let $\Psi$, $\Phi^*$ be arbitrary
linear combinations of the fermion operators $\psi_n$,
$\psistar_n$, respectively, and consider the element
$$
G=e^{\beta \Phi^* \Psi}=
\normordbare e^{\alpha \Phi^* \Psi}\normordbare
= 1+\alpha \Phi^* \Psi = 1+\alpha \gamma -\alpha \Psi \Phi^* ,
$$
where
$\gamma := \lbr \infty \right | \Psi \Phi^* \left |\infty \rbr $
and $\alpha $, $\beta $ are related as
$e^{\gamma \beta} =1+\gamma \alpha $.
For general values of $\alpha $ this element is invertible and the
two representations are equivalent. However, for $\gamma \neq 0$ at $ \alpha =-1/\gamma $
the invertibility breaks down and the element $G$ becomes
$$
G= \frac{\Psi \, \Phi^*}{\lbr \infty
\right | \! \Psi \Phi^* \! \left |\infty \rbr}.
$$
which can not be written in the exponential form (\ref{gl}) but
can be represented as the normally ordered exponent (so that the matrix $R$ exists, but it is not invertible). Group-like elements of this type are used in section \ref{Characqp} below.
\item[b)] For any element $G$ of the form $G=\normord \exp \Bigl (
A_{ik}\psistar_i \psi_k \Bigr ) \normord$ one can find
$$R=I+(I-AP_+)^{-1}A \quad \mbox{and} \quad  
R'= I-(I+AP_-)^{-1}A,$$ where $P_-=I-P_+$.
Then for 
$
G=\normord \exp \Bigl (
\psistar_1 \psi_1-\psistar_{-1} \psi_{-1}\Bigr )
\normord =\psistar_1 \psi_1 \psi_{-1}\psistar_{-1}
$
both matrices $I\! -\! AP_+$ and $I\! +\! AP_-$ 
are degenerate, so neither $R$ nor $R'$ exist.
Nevertheless, this element $G$ obeys the BBC (\ref{commute}) 
below. Various projection operators, described in section \ref{Proop}, 
are also elements of this type. 
\end{itemize}

\subsection{The basic bilinear condition}

It is easy to see that any element satisfying (\ref{rotation-a}) or (\ref{rotation-b}) obeys
the commutation relation
\begin{equation}
\sum_{k \in {\mathbb Z}} \psi_{k} G \otimes  \psi_{k}^{*} G =
\sum_{k \in {\mathbb Z}}G\psi_{k} \otimes G \psi_{k}^{*}
\label{commute}
\end{equation}
which we call the basic bilinear condition (BBC).
It means that $G\otimes G$ commutes with 
$\sum_{k}\psi_k \otimes \psistar_k$.
In terms of matrix elements it reads
\begin{equation}
\sum_{k \in {\mathbb Z}} \lbr U \right|  \psi_{k} G \left|V \rbr
 \lbr U^{\prime} \right|  \psistar_{k} G \left|V^{\prime} \rbr =
\sum_{k \in {\mathbb Z}} \lbr U \right| G \psi_{k} \left|V \rbr
 \lbr U^{\prime} \right| G \psistar_{k} \left|V^{\prime} \rbr 
 \label{bilinear-fermi}
\end{equation}
for any states $\left|V \rbr , \left|V^{\prime} \rbr$,
$\lbr U \right|,  \lbr U^{\prime} \right|$
from the space
${\cal H}_{F}$ and its dual. Indeed, substituting (\ref{rotation-a}) 
or (\ref{rotation-b}) 
instead of $\psi_k G$ and $G\psistar_k$ 
or $\psistar_k G$ and $G\psi_k$ in the left and right hand sides
of (\ref{commute}) respectively, we get the identity.

\noindent
\begin{prop}
All normally ordered exponents of bilinear forms solve the BBC.
\end{prop}

\noindent
It is enough to prove the statement for the normal 
ordering with respect to one particular vacuum, say $\rvac$. 
Let us consider the decomposition (\ref{nodecomp}) of general 
normal ordered exponent. Then, $G_1$ and $G_3$ are group elements, so they satisfy BBC and all we have to prove is that $G_2$ also does. 
This is obvious from (\ref{g2rot}) and 
corresponding rotation for $\psistar_k$'s:
\beq
\left \{ \begin{array}{l}
\psi_n G_2 =G_2 (\psi_n +A_{n\bar b}\psi_{\bar b}),
\quad n<0
\\ \\
G_2\psi_n =(\psi_n -A_{n b}\psi_b )G_2 , \quad n\geq 0,
\end{array} \right.
\quad \,\,
\left \{ \begin{array}{l}
G_2 \psistar_n = (\psistar_n +A_{n\bar b}\psistar_{\bar b}) G_2,
\quad n<0
\\ \\
\psistar_n G_2=G_2(\psistar_n -A_{n b}\psistar_b ) , \quad n\geq 0.
\end{array} \right.
\eeq
Indeed, one can easily check that (\ref{commute}) is 
valid separately for the sums with $k<0$ and $k\geq 0$.

It turns out that in addition to the
group elements and normally ordered exponents 
there are other solutions 
to the BBC. For example, it is easy to check 
that $G=\psi_n$ as well as
$G=\psistar_n$ solve (\ref{commute}). 
More generally, so does any linear combination of $\psi$'s 
(as well as $\psistar$'s). Indeed, set $v=\sum_k v_k \psi_k$,
then 
$$
\sum_k v\psi_k \otimes v\psistar_k =-\sum_k \psi_k v\otimes
(-\psistar_k v +v_k)=\sum_k \psi_k v \otimes \psistar_k v -
\sum_k v_k \psi_k v \otimes 1.
$$
The last term vanishes because $\sum_k v_k \psi_k v =v^2=0$.
At the same time, the element $G=\psi_n$ (for example)
does not generate any linear transformation in the linear
space spanned by $\psistar_k$, neither of the 
form (\ref{rotation-a}) nor (\ref{rotation-b}).

It appears that the only important property of an element $G$
for what follows is not the induced rotation like 
(\ref{rotation-a}) or (\ref{rotation-b}) but the BBC 
(\ref{commute}). Therefore, {\it from now on we extend the definition
of the group-like elements including in this class all 
solutions to the BBC}.

We have the following general properties of the group-like elements.

\noindent
\begin{prop}  The elements that satisfy the 
BBC (\ref{commute}) form a semigroup: if $G$ and $G'$ satisfy it
then so does $GG'$.
\end{prop}

\noindent
This is obvious from (\ref{commute}).

\noindent
\begin{prop}\label{dcharge}  
All solutions of the BBC 
(\ref{commute}) have definite charge, i.e., 
$[Q, G]=qG$ with some integer $q$.
\end{prop}

\noindent
The proof is given in Appendix B.

\subsection{The generalized Wick theorem}

Let $v_i =\sum_j v_{ij}\psi_j$ be
linear combinations of $\psi_j$'s only and
$w_i^*=\sum_j w^{*}_{ij}\psistar_j$
be linear combinations of $\psistar_j$'s only,
as before.

\noindent
\begin{theor} Let $G, G'$ be any two group-like elements
with zero charge. Then
for any $v_j, w^*_i$ and any $n$ such that $\lvacn G'G\rvacn \neq 0$ it holds
\begin{equation}\label{Wick1}
\frac{\lvacn G' v_1 \ldots v_m w^{*}_m \ldots w^{*}_1 G\rvacn }{\lvacn
G'G\rvacn }=\det_{i,j =1,\ldots , m}
\frac{\lvacn G'v_j w^{*}_iG\rvacn }{\lvacn
G'G\rvacn }\,.
\end{equation}
\end{theor}

\noindent
This is the generalized Wick theorem. A similar statement
holds when $w^*_j$'s stand to the left of $v_j$'s.
Writing
$\lbr n-m \right | = \lvacn \psi_{n-1}\ldots \psi_{n-m}$
or $\lbr n+m \right | = \lvacn \psistar_{n}\ldots \psistar_{n+m-1}$
with $m>0$,
we get from (\ref{Wick1}):

\noindent
\begin{cor} Let $G$ be any group-like element. Then
\begin{equation}\label{Wick2}
\frac{\lbr n-m \right | w^{*}_m \ldots w^{*}_1 G\rvacn }{\lvacn
G\rvacn }=\det_{i,j =1,\ldots , m}
\frac{\lvacn \psi_{n-j} w^{*}_i G\rvacn }{\lvacn
G\rvacn }\,,
\end{equation}
\begin{equation}\label{Wick2a}
\frac{\lbr n+m \right | v_m \ldots v_1 G\rvacn }{\lvacn
G\rvacn }=\det_{i,j =1,\ldots , m}
\frac{\lvacn \psistar_{n+j-1} v_i G\rvacn }{\lvacn
G\rvacn }\,.
\end{equation}
\end{cor}

The theorem can be proved by induction using the BBC in the form
(\ref{bilinear-fermi}). Suppose (\ref{Wick1}) is valid
for some $m \geq 1$ (it is trivially valid at $ m=1$).
Set
$$
\lbr U\right |=\lvacn G' w_{1}^{*}, \quad
\lbr U'\right |=\lvacn G' v_1 v_2
\ldots v_{m+1} w_{m+1}^{*}w_{m}^{*}\ldots
w_{2}^{*},
\quad \left |V\rbr = \left |V'\rbr =\rvacn .
$$
Plugging this in the BBC, we see that its
r.h.s. vanishes because either $\psi_k \rvacn =0$ or 
$\psistar_k \rvacn =0$ while the l.h.s. gives
$$
\sum_k \lvacn G' w_{1}^* \psi_k G\rvacn \lvacn G' v_1 \ldots 
v_{m+1} w^*_{m+1}\ldots w^*_2 \psistar_k G\rvacn =0.
$$
Substituting $w_{1}^* \psi_k = w^{*}_{1k}-\psi_kw_{1}^*$ 
in the first factor and moving $\psistar_k$ 
in the second factor through the chain of the 
$w_j^*$'s, we get in the l.h.s.
$$
\begin{array}{c}
\lvacn G'G\rvacn \lvacn G'v_1 \ldots 
v_{m+1} w^*_{m+1}\ldots w^*_1 G\rvacn
\\ \\\displaystyle{
-\, (-1)^m \sum_k \lvacn G' \psi_k w_{1}^*  G\rvacn
\lvacn G' v_1 \ldots 
v_{m+1} \psistar_k w^*_{m+1}\ldots w^*_2  G\rvacn .
}
\end{array}
$$
We proceed by moving $\psistar_k$ to the left 
through the chain of the $v_j$'s using $v_j\psistar_k=
v_{jk}-\psistar_k v_j$ at each step. The result is
$$
\begin{array}{c}
\lvacn G'G\rvacn \lvacn G'v_1 \ldots 
v_{m+1} w^*_{m+1}\ldots w^*_1 G\rvacn
\\ \\ \displaystyle{
+\, \sum_{j=1}^{m+1}(-1)^j \lvacn G' v_j w_{1}^*  G\rvacn
\lvacn G' v_1 \ldots \not \! v_j \ldots 
v_{m+1} w^*_{m+1}\ldots w^*_2  G\rvacn }
\\ \\
\displaystyle{+\, \sum_k \lvacn G' \psi_k w_{1}^*  G\rvacn
\lvacn G' \psistar_k v_1 \ldots 
v_{m+1} w^*_{m+1}\ldots w^*_2  G\rvacn }.
\end{array}
$$
In the last line we can again use the
BBC to write it as
$$
\sum_k \lvacn \psi_k G'  w_{1}^*  G\rvacn
\lvacn \psistar_k G'  v_1 \ldots 
v_{m+1} w^*_{m+1}\ldots w^*_2  G\rvacn
$$
which is 0 because either $\lvacn \psi_k =0$ or
$\lvacn \psistar_k =0$. Therefore, we conclude that
$$
\begin{array}{c}
\displaystyle{\lvacn G'G\rvacn \lvacn G'v_1 \ldots 
v_{m+1} w^*_{m+1}\ldots w^*_1 G\rvacn}
\\ \\ \displaystyle{
+\, \sum_{j=1}^{m+1}(-1)^{j} \lvacn G' v_j w_{1}^*  
G\rvacn
\lvacn G' v_1 \ldots \not \! v_j \ldots 
v_{m+1} w^*_{m+1}\ldots w^*_2  G\rvacn =0}
\end{array}
$$
or
$$
\begin{array}{c}
\displaystyle{\frac{\lvacn G'v_1 \ldots 
v_{m+1} w^*_{m+1}\ldots w^*_1 G\rvacn}{\lvacn G'G\rvacn}}
\\ \\ \displaystyle{
=\, \sum_{j=1}^{m+1}(-1)^{j-1} \frac{\lvacn G' v_j w_{1}^*  
G\rvacn}{\lvacn G'G\rvacn}
\, \frac{\lvacn G' v_1 \ldots \not \! v_j \ldots 
v_{m+1} w^*_{m+1}\ldots w^*_2  G\rvacn}{\lvacn G'G\rvacn}. }
\end{array}
$$
By the assumption, the second ratio in the r.h.s. is 
the $m$ by $m$ determinant. We see that the r.h.s. is the 
expansion of the $m+1$ by $m+1$ determinant in the first 
column, so the theorem is proved.

In a similar way 
one can prove that for arbitrary group-like elements $G, G', G''$ 
it holds
\begin{equation}\label{WickG}
\frac{\lvacn G' v_1 \ldots v_m G'' w^{*}_m \ldots w^{*}_1 G\rvacn }{\lvacn
G'G''G\rvacn }=\det_{i,j =1,\ldots , m}
\frac{\lvacn G'v_j G''w^{*}_iG\rvacn }{\lvacn
G'G''G\rvacn }\,.
\end{equation}
Moreover, since for the derivation only the 
BBC (\ref{commute})
is used, this version of the Wick theorem can be 
immediately extended to solutions with non-zero charge. 
Namely, for any three solutions of (\ref{commute}) $G_q$, $G'_{q'}$ and $G''_{q''}$ with charges $q$, $q'$ and $q''$ respectively, one has:
$$
\frac{\lvacn G'_{q'} v_1 \ldots v_m G''_{q''} w^{*}_m 
\ldots w^{*}_1 G_q\left |\tilde{n}\right > }{\lvacn
G'_{q'}G''_{q''}G_{q}\left |\tilde{n}\right > }=\det_{i,j =1,\ldots , m}
\frac{\lvacn G'_{q'}v_j G''_{q''}w^{*}_iG_q\left |\tilde{n}\right > }{\lvacn
G'_{q'}G''_{q''}G_{q}\left |\tilde{n}\right > }\,,
$$
where $\tilde{n}=n-q-q'-q''$. For a general version of the 
Wick theorem see also 
\cite{Perk}.

There exists an alternative determinant
representation of the expectation value in the l.h.s. of
(\ref{Wick2}), which is another form of the generalized
Wick theorem.

\noindent
\begin{cor} Let $G$ be any group-like element. Then
\begin{equation}\label{Wick3}
\frac{\lbr n-m \right | w^{*}_m \ldots w^{*}_1 G\rvacn }{\lvacn
G\rvacn }=\det_{i,j =1,\ldots , m}
\frac{\lbr n\! -\! j\right | w^{*}_i G\left | n\! -\! j\! +\! 1
\rbr }{\lbr n\! -\! j\! +\! 1\right |
G \left | n\! -\! j\! +\! 1\rbr }\,,
\end{equation}
\begin{equation}\label{Wick3a}
\frac{\lbr n+m \right | v_m \ldots v_1 G\rvacn }{\lvacn
G\rvacn }=\det_{i,j =1,\ldots , m}
\frac{\lbr n\! +\! j\right | v_i G\left | n\! +\! j\! -\! 1
\rbr }{\lbr n\! +\! j\! -\! 1\right |
G \left | n\! +\! j\! -\! 1\rbr }\,,
\end{equation}
\begin{equation}\label{Wick3aa}
\frac{\left < n \right |  G v_1 \ldots v_m 
\left |n-m\right > }{\lvacn G\rvacn }=\det_{i,j =1,\ldots , m}
\frac{\lbr n\! -\! j\! +\! 1\right | Gv_i \left | n\! -\! j
\rbr }{\lbr n\! -\! j\! +\! 1\right |
G \left | n\! -\! j\! +\! 1\rbr }\,,
\end{equation}
\begin{equation}\label{Wick3aaa}
\frac{\left < n \right | Gw_1^* \ldots w_m^*  
\left |n+m\right > }{\lvacn G\rvacn }=\det_{i,j =1,\ldots , m}
\frac{\lbr n\! +\! j\! -\! 1\right | Gw_i^* \left | n\! +\! j
\rbr }{\lbr n\! +\! j\! -\! 1\right |
G \left | n\! +\! j\! -\! 1\rbr }\,.
\end{equation}
\end{cor}

Let us outline the proof of (\ref{Wick3}). The idea 
of the proof is as follows\footnote{This proof was suggested 
by Z.Tsuboi.}.
The first columns of the matrices in the r.h.s.
of (\ref{Wick2}) and (\ref{Wick3}) coincide and one can show that
the $j$-th column of the matrix in (\ref{Wick2}) is equal to
the $j$-th column of
the matrix in (\ref{Wick3}) plus a linear combination of the
first $j-1$ columns.
To show this, we use the BBC. Let us take
\beq
\lbr U \right| =\lbr n+1 \right| \psi_{l}w_{i}^{*} ,
\quad
\lbr U^{\prime} \right| = \lbr n \right| ,
\quad
\left|V \rbr = \left| n \rbr ,
\quad
\left|V^{\prime} \rbr = \left| n+1 \rbr
\eeq
in \eqref{bilinear-fermi}.
It is easy to see
that only one term in the r.h.s. remains:
$$
\sum_k \lbr n+1\right | \psi_l w_{i}^{*}\psi_k G \rvacn
\lvacn \psistar_k G \left |n+1 \rbr =
\lbr n+1\right | \psi_l w_{i}^{*}G\left |n+1 \rbr
\lvacn G\rvacn .
$$
In the l.h.s. we plug $w_i^*=\sum_j w^{*}_{ij}\psistar_j$
and move $\psi_k$ in the first factor to the left position.
Using the anti-commutaton relation for the fermion operators,
and transforming the sum $\sum_k w^{*}_{ik}\psistar_k$
back to $w_i^*$ in the second factor, we arrive at the
3-term identity
$$ 
\begin{array}{l}
\lbr n \right| G \left| n \rbr
\lbr n+1 \right| \psi_{l} w_{i}^{*} G \left| n+1 \rbr
=
\lbr n+1 \right| G \left| n+1 \rbr
\lbr n\right| \psi_{l} w_{i}^{*} G \left| n \rbr
\\ \\
\hspace{4cm}+\, 
\lbr n+1 \right| \psi_{l} G \left| n \rbr
\lbr n \right| w_{i}^{*} G \left| n+1 \rbr .
\end{array}
$$
Divide it by
$\lbr n \right | G \left | n \rbr
\lbr n+1 \right | G \left | n+1 \rbr $, take a sum over the values
of $n$ equal to
$n-j,n-j+1,\dots, n-1$, and then put $l = n-j$.
We obtain the relation
{\small $$
\frac{\lbr n \right | \psi_{n-j} w^{*}_i G\left | n \rbr }{\lbr n \right |
G \left | n \rbr }
=
\frac{\lbr n\! -\! j\right | w^{*}_i G\left | n \! -\! j\! +\! 1
\rbr }{\lbr n\! -\! j+\!1 \right |
G \left | n\! -\! j +\! 1\rbr }
+
\sum_{k=1}^{j-1}
\frac{\lbr n\! -\! k + \!1 \right | \psi_{n-j} G\left | n \! -\! k
\rbr }{\lbr n\! -\! k \right |
G \left | n\! -\! k \rbr }
\frac{\lbr n\! -\! k \right | w^{*}_i G\left | n \! -\! k \! +\! 1
\rbr }{\lbr n\! -\! k \! +\! 1\right |
G \left | n\! -\! k \! +\! 1\rbr }
$$}
which shows that the columns of the matrix in
\eqref{Wick2} are the linear combinations of the
columns of the matrix in \eqref{Wick3} and their determinants
are thus equal. The proof of (\ref{Wick3a}) is similar.

Again, formulas (\ref{Wick3})--(\ref{Wick3aaa}) 
can be easily generalized to the case of group-like elements 
with non-zero charge. 

\subsection{Projection operators}\label{Proop}

In some applications (in particular, to
models of random matrices), an important 
role is played by the non-invertible group-like 
elements
\beq\label{PF1}
\begin{array}{l}
\displaystyle{
{\sf P}^+
= \, \normord \exp \left (\sum_{i<0}\psi_i \psistar_i\right )\normord }
=\, \prod_{i<0}(1-\psistar_i \psi_i )=\, \prod_{i<0}\psi_i \psistar_i ,
\\ \\
\displaystyle{
{\sf P}^-= \normord \exp \left (-\! 
\sum_{i\geq 0}\psi_i \psistar_i\right )\normord }
=\prod_{i\geq 0}(1-\psi_i \psistar_i )=\prod_{i\geq 0}\psistar_i \psi_i .
\end{array}
\eeq
In a sense, these operators are projectors to the spaces of states with 
positive (i.e., $\geq 0$) and
negative (i.e., $\leq 0$) charge respectively. Their properties (extensively
used in what follows)
can be easily seen from the definition. Both operators
obey the projector property:
$({\sf P}^{\pm})^2 ={\sf P}^{\pm}$. The operator 
${\sf P}^+$ kills negative
creation modes standing to the right and negative annihilation modes
standing to the left and commutes with all positive modes:
\beq\label{ferm9}
\begin{array}{l}
\phantom{a}{\sf P}^+ \psistar_k =\psi_k  {\sf P}^+=0, 
\quad \quad \quad k<0,
\\ \\ \phantom{a} [{\sf P}^+ , \psistar_k ]
=[{\sf P}^+ ,\psi_k ]=0, \quad k \geq 0.
\end{array}
\eeq
The operator ${\sf P}^-$ kills positive
creation modes standing to the right and positive annihilation modes
standing to the left and commutes with all negative modes:
\beq\label{ferm9a}
\begin{array}{l}
\phantom{a}{\sf P}^- \psi_k =\psistar_k {\sf P}^- 
=0, \quad \quad \quad k\geq 0,
\\ \\ \phantom{a} [{\sf P}^- , \psistar_k ]=[{\sf P}^- , 
\psi_k ]=0, \quad k < 0.
\end{array}
\eeq
From this it is obvious that ${\sf P}^+ \! \rvacn =0$ at $n<0$ and
${\sf P}^+ \! \rvacn =\rvacn$ at $n\geq 0$. Similarly,
${\sf P}^- \! \rvacn =0$ at $n> 0$ and
${\sf P}^- \! \rvacn =\rvacn$ at $n\leq0$, so that ${\sf P}^+ \! 
\rvac ={\sf P}^- \! \rvac =\rvac$. 
Somewhat less obvious properties, also used in what follows, are 
\beq
\begin{array}{l}\displaystyle{
{\sf P}^+ e^{J_-({\bf t})}\rvacn=
\sum_{\ell (\lambda)\leq n} (-1)^{b(\lambda )}
s_{\lambda}({\bf t})\left |\lambda , n\rbr, \!\! \quad n\geq 0,}
\\ \\\displaystyle{
\lvacn e^{J_+({\bf t})}{\sf P}^+=
\sum_{\ell (\lambda)\leq n} (-1)^{b(\lambda )}
s_{\lambda}({\bf t})\lbr \lambda , n \right |, \,\, \quad n\geq 0,}
\end{array}
\eeq
(see section \ref{currentss} for definitions) in particular,
${\sf P}^+ e^{J_-}\rvac =\rvac$,
$\lvac e^{J_+}{\sf P}^+ =\lvac$.

Extending definition (\ref{PF1}) to other vacuum states,
we can also introduce the projectors
\beq\label{PF1a}
{\sf P}^+_n = \prod_{i<n}\psi_i \psistar_i \,,
\quad \quad
{\sf P}^-_n = \prod_{i\geq n}\psistar_i \psi_i
\eeq
with similar properties. Moreover, 
similar operators can be introduced for any basis state:
\beq\label{ferm10}
\begin{array}{l}
\displaystyle{
{\sf P}^+_{n,\lambda}=\prod_{k=1}^\infty \psi_{n+\lambda_k-k} \psistar_{n+\lambda_k-k}=
\prod_{k\in I_{n,\lambda}} \psi_k \psistar_k,
}
\\ \\
\displaystyle{
{\sf P}^-_{n,\lambda}=\prod_{k=1}^\infty \psistar_{n-\lambda'_k +k -1 } \psi_{n-\lambda'_k +k -1 }=
\prod_{k\notin I_{n,\lambda}} \psistar_k \psi_k
}
\end{array}
\eeq
(in particular, 
${\sf P}^+_n={\sf P}^+_{n,\emptyset}$). They
also obey the projector property
$\left({\sf P}^\pm_{n,\lambda}\right)^2={\sf P}^\pm_{n,\lambda}$ 
and commute with each other, 
$[{\sf P}^-_{n,\lambda},{\sf P}^+_{n,\lambda}]=0$. 
Properties  (\ref{ferm9}) and (\ref{ferm9a}) are naturally 
generalized for the corresponding creation and annihilation operators. 
Their product ${\sf P}^+_{n,\lambda}{\sf P}^-_{n,\lambda}$
is the projector to the state $\left| n,\lambda\right>$:
\beq\label{ferm11}
{\sf P}^+_{n,\lambda}{\sf P}^-_{n,\lambda}\left|m,\mu\right> =
\delta_{n,m}\delta_{\lambda,\mu}\left| n,\lambda\right>.
\eeq
Thus for the basis states we can write
$
\left |\lambda , n \rbr \left <\lambda , n\right |= 
{\sf P}^+_{n,\lambda}{\sf P}^-_{n,\lambda}
$ and, in particular,
$\left| n\right> \left< n \right| = {\sf P}^+_n {\sf P}^-_n$ 
and $\left| 0\right> \left< 0 \right| = {\sf P}^+ {\sf P}^-$.
Note that
$$
\sum_{n,\lambda}\left |\lambda , n \rbr \left <\lambda , n\right |=\sum_{n,\lambda}{\sf P}^-_{n,\lambda}{\sf P}^+_{n,\lambda}=\prod_{k=-\infty}^\infty
\left(\psi_k\psistar_k+\psistar_k\psi_k\right)=1,
$$
as it should be. 

From (\ref{fermnt1}) and (\ref{fermnt2}) it follows that 
the operator $\left| n,\lambda\right>\left<m,\mu\right|$ 
is represented by an infinite product 
\beq\label{genpro}
\left| n,\lambda\right>\left<m,\mu\right|=\prod_{k=-\infty}^\infty R_k^{(n,\lambda;m,\mu)},
\eeq
where $R_k^{(n,\lambda;\, m,\mu)}$ is given by
\begin{align}
R_k^{(n,\lambda;\, m,\mu)} = \left \{
\begin{array}{l}
\psi_k\psistar_k, \,\,\,\,\,\,\, k \in I_{n,\lambda},\,\, k \in I_{m,\mu}
\\ \\
\psistar_k\psi_k , \,\,\,\,\,\,\, k \notin I_{n,\lambda},\,\, 
k \notin I_{m,\mu}
\\ \\ 
\psi_k, \,\,\,\,\,\,\, k \in I_{n,\lambda},\,\, k \notin I_{m,\mu}
\\ \\ 
\psistar_k ,  \,\,\,\,\,\,\, k \notin I_{n,\lambda},\,\, k \in I_{m,\mu}\,.
\end{array} \right.
\end{align}
Any operator $\left| n,\lambda\right>\left<m,\mu\right|$  is group-like.

\subsection{Currents $J_{\pm}({\bf t})$}\label{currentss}

Among bilinear combinations of fermions of particular
importance are the operators
\beq\label{Jk}
J_k =\sum_{j\in \z}\normord \psi_j \psistar_{j+k}\normord
=\mbox{res}_z \Bigl ( \normord \psi (z)
z^{k-1}\psistar (z)\normord \Bigr )
\eeq
which are Fourier modes of the ``current operator''
$J(z)=\normord \psi (z)\psistar (z)\normord $. Note that 
the normal ordering in (\ref{Jk}) is essential at $k=0$ only,
and in this case $J_0$ coincides with the charge operator 
$Q$ (\ref{chargeoper}). At $k\neq 0$ the
normal ordering can be ignored:
\beq\label{Jka}
J_k =\sum_{j\in \z}\psi_j \psistar_{j+k}\,, \quad \quad
k\neq 0.
\eeq
These operators are of the form (\ref{XA})
with the matrix $A_{ij}=\delta_{i, j-k}$. This matrix 
is not of the class discussed in section 2.1 because 
it has an infinite
non-zero diagonal, and thus one should take some care when working with
formal expressions containing infinite sums (see the example 
below in this section). It is easy to 
see that $[J_k , \psi _m]=\psi_{m-k}$, $[J_k , 
\psistar _m]=-\psistar_{m+k}$.

Let $t_k$, $k\in \ZZ$, be arbitrary parameters (called times).
It is convenient to denote the collection of times
with positive (negative) indices
by ${\bf t}_+ = \{t_1, t_2 , \ldots \}$ and 
${\bf t}_- = \{t_{-1}, t_{-2} , \ldots \}$
respectively. Set
\beq\label{ferm2}
J_+ =J_+({\bf t}_+)= \sum_{k\geq 1}t_k J_k, \quad \quad
J_- =J_-({\bf t}_-)= \sum_{k\geq 1}t_{-k} J_{-k}
\eeq
and introduce the generating function
$$
\xi ({\bf t}, z)=\sum_{k\geq 1}t_{k}z^k.
$$
Here and below we write simply ${\bf t}$
for any half-infinite set of times (either ${\bf t}_+$ or ${\bf t}_-$)
when it does not lead to a 
misunderstanding.

It is easy to check that the series $\psi (z)$, $\psistar (z)$
transform diagonally under the adjoint action
of the elements
$e^{J_+}$, $e^{J_-}$:
\beq\label{ferm3}
\begin{array}{l}
e^{J_+({\bf t})}\psi (z)e^{-J_+({\bf t})}=e^{\xi ({\bf t} ,\, z)}\psi (z),
\\ \\
e^{J_{+}({\bf t})}\psistar (z)e^{-J_{+}({\bf t})}=e^{-\xi ({\bf t} , \,
z)}\psistar (z)
\end{array}
\eeq
and
\beq\label{ferm3a}
\begin{array}{l}
e^{J_{-}({\bf t})}\psi (z)e^{-J_{-}({\bf t})}=
e^{\xi ({\bf t} ,\, z^{-1})}\psi (z),
\\ \\
e^{J_{-}({\bf t})}\psistar (z)e^{-J_{-}({\bf t})}=e^{-\xi ({\bf t} , \,
z^{-1})}\psistar (z).
\end{array}
\eeq
In terms of the polynomials $h_k({\bf t})$ defined by
\beq\label{schur1}
e^{\xi ({\bf t} ,\, z)}=\sum_{k\geq 0}h_k ({\bf t})z^k
\eeq
the adjoint action of $e^{J_{\pm}}$ to $\psi_n , \psistar_{n}$
is given by the formulas
\beq\label{ferm4}
\begin{array}{l}
\displaystyle{
e^{J_{+}({\bf t})}\psi _n e^{-J_{+}({\bf t})}=
\sum_{k\geq 0}\psi_{n- k}h_k({\bf t}),\quad
e^{J_{-}({\bf t})}\psi _n e^{-J_{-}({\bf t})}=
\sum_{k\geq 0}\psi_{n+ k}h_k({\bf t})},
\\ \\
\displaystyle{
e^{J_{+}({\bf t})}\psistar _n e^{-J_{+}({\bf t})}=
\sum_{k\geq 0}\psistar_{n+k}h_k(-{\bf t}), \quad
e^{J_{-}({\bf t})}\psistar _n e^{-J_{-}({\bf t})}=
\sum_{k\geq 0}\psistar_{n- k}h_k(-{\bf t})}.
\end{array}
\eeq
Their easy consequence is
\beq\label{hj}
\left < 1\right | e^{J_{+}({\bf t})}\psi_j \rvac =h_j({\bf t})\,,
\quad j\geq 0,
\eeq
and, more generally, 
\beq\label{hji}
\left < 0\right | \psistar_i 
e^{J_{+}({\bf t})}\psi_j \rvac =h_{j-i}({\bf t})\,,
\quad i,j\geq 0.
\eeq

The commutator $[J_k, J_l]$ can be calculated as
$$
[J_k, J_l]=\sum_j [J_k, \psi_j \psistar_{j+l}]=
\sum_j \left ( [J_k, \psi_j]\psistar_{j+l}+
\psi_j [J_k, \psistar_{j+l}]\right ).
$$
Using the formulas obtained above, we get
\beq\label{JkJl1}
[J_k, J_l]=\sum_j \left (\psi_{j-k}\psistar_{j+l}-\psi_j
\psistar_{j+l+k}\right ).
\eeq
Naively, one can shift the summation index $j\to j+k$ in the 
first sum to get 0. In fact this can be done only when $k+l\neq 0$.
At $k=-l$ the r.h.s. appears to be ill-defined because of the 
infinite summation. 
Let us
fix the r.h.s. of equation (\ref{JkJl1}). When $k+l\neq 0$, 
the r.h.s. is well-defined (and equal to zero) as it stands because
the result of its action to any basis state contains only a finite 
number of terms. When $k+l=0$, one should first rewrite the r.h.s. 
in terms of the normally ordered expressions:
$$
[J_k, J_{-k}]=\sum_j \normord 
\left (\psi_{j-k}\psistar_{j-k}-\psi_j
\psistar_{j}\right ) \normord +\sum_j \Bigl (\theta (j<k)-\theta (j<0)
\Bigr ),
$$
where $\theta (j<k)=1$ for $j<k$ and $0$ otherwise. The normally 
ordered expressions are well-defined and the summation index can be shifted
thus yielding $0$. The rest gives the commutation law
\beq\label{JkJl2}
[J_k, J_l]=k\delta_{k+l, 0}\,.
\eeq
We see that the Fourier modes of the fermionic currents commute
as bosonic operators. 
We thus have
$$
[J_+({\bf t}_+), \, J_-({\bf t}_-)]=\sum_{k\geq 1}kt_kt_{-k}
$$
and
\beq\label{JkJl3}
e^{J_+({\bf t}_+)}e^{J_-({\bf t}_-)}=
\exp \Bigl (\sum_{k\geq 1}kt_kt_{-k}\Bigr )
e^{J_-({\bf t}_-)}
e^{J_+({\bf t}_+)}.
\eeq

\subsection{Expansion of the states 
$e^{J_{\pm}({\bf t})}\left |\lambda , n\right >$ and Schur
functions}

The definition of the vacua
implies that $\lvacn J_-=J_+ \rvacn =0$ and thus
$\lvacn e^{J_-}=\lvacn$, $e^{J_+}\rvacn =\rvacn$.
The coherent
states $e^{J_-}\rvacn$ and $\lvacn e^{J_+}$
(and, more generally, the states 
$e^{J_{\pm}}\left |\lambda , n\right >$)
can be expanded \cite{JM83,HO03}
as linear combinations of the basis states. 
The coefficients are the famous Schur polynomials.
This expansion is important since it provides a link 
between hierarchies of integrable equations and the theory
of symmetric functions.

\subsubsection{Expansion of the states 
$e^{J_{\pm}({\bf t})} \rvacn$}

Given a Young diagram $\lambda = (\vec \alpha |\vec \beta )$,
one can introduce the Schur polynomials (or Schur functions) 
via the
Jacobi-Trudi formulas \cite{Macdonald}:
\beq\label{schur2}
s_{\lambda}({\bf t})=\det_{i,j=1, \ldots , \ell (\lambda )}
h_{\lambda_i -i +j}({\bf t}),
\eeq
where $h_k({\bf t})$ are defined in (\ref{schur1}) (they are
Schur polynomials for one-row diagrams). 
For details see \cite{Macdonald} and Appendix A.
We recall that $b(\lambda )=\sum_{i=1}^{d(\lambda )}(\beta_i +1)$
(see (\ref{boflambda})).
For the empty diagram
$s_{\emptyset}({\bf t})=1$, $b(\emptyset )=0$.

\noindent
\begin{prop}\label{chardec} 
It holds
\beq\label{lambda2}
\displaystyle{
e^{J_-({\bf t})}\rvacn = \sum_{\lambda} (-1)^{b(\lambda )}
s_{\lambda}({\bf t})\left |\lambda , n\rbr},
\eeq
\beq\label{lambda3}
\displaystyle{
\lvacn e^{J_+({\bf t})}=\sum_{\lambda} (-1)^{b(\lambda )}
s_{\lambda}({\bf t})\lbr \lambda , n \right |},
\eeq
where 
the sums run over all Young diagrams $\lambda$ including the
empty one.
\end{prop}

\noindent
From orthogonality of the basis states it follows that (\ref{lambda2}) and (\ref{lambda3}) are equivalent to
\beq\label{schur3}
\left < \lambda , n\right | e^{J_-({\bf t})}\rvacn =
(-1)^{b(\lambda )} s_{\lambda }({\bf t}),
\eeq
\beq\label{schur4}
\lvacn  e^{J_+({\bf t})}\left |\lambda , n \right >=
(-1)^{b(\lambda )} s_{\lambda }({\bf t}).
\eeq

Let us sketch a proof of the first formula of the proposition.
First of all we recall that the Schur polynomials 
for an arbitrary Young diagram $\lambda = (\vec \alpha |\vec \beta )$
are expressed
through the Schur polynomials $s_{(\alpha|\beta)}({\bf t})$
corresponding to the hook diagrams 
$\lambda =(\alpha+1,1^{\beta})$
with the help of the Giambelli formula
\beq\label{Gamb}
s_{\lambda}({\bf t})=\det_{i,j=1,\ldots,d(\lambda)} 
s_{(\alpha_i|\beta_j)}({\bf t}).
\eeq
Therefore, the r.h.s. of (\ref{lambda2}) can be transformed
as follows:
$$
\begin{array}{ll}
&\displaystyle{ \sum_{\lambda} (-1)^{b(\lambda )}
s_{\lambda}({\bf t})\left |\lambda , n\rbr}
\\ &\\
=& \displaystyle{ \sum_{d\geq 0}\sum_{{\alpha_1 >\alpha_2>
\ldots >\alpha_d\geq 0\atop \beta_1 >\beta_2>
\ldots >\beta_d\geq 0}}(-1)^{b(\lambda )}\det_{1\leq i,j\leq d}
s_{(\alpha_i|\beta_j)}({\bf t})\,
\psistar_{n-\beta_1 -1}\ldots \psistar_{n-\beta_d -1}
\psi_{n+\alpha_d}\ldots \psi_{n+\alpha_1}\! \rvacn}
\\ &\\
=& \displaystyle{ \sum_{d\geq 0}\frac{1}{d!}\!\!
\sum_{{\alpha_1 , \alpha_2 ,
\ldots ,\alpha_d\geq 0\atop \beta_1 ,\beta_2,
\ldots ,\beta_d\geq 0}}\!\! (-1)^{b(\lambda )}
s_{(\alpha_1|\beta_1)}\! ({\bf t})\ldots 
s_{(\alpha_d|\beta_d)}\! ({\bf t})
\psistar_{n\! -\! \beta_1 \! -\! 1}\ldots 
\psistar_{n\! -\! \beta_d \! -\! 1}
\psi_{n+\alpha_d}\ldots \psi_{n+\alpha_1}\! \rvacn}
\\ &\\
=& \displaystyle{\exp \Bigl (
\sum_{\alpha , \beta \geq 0}
(-1)^{\beta +1}s_{(\alpha |\beta )}({\bf t}) 
\psistar_{n-\beta -1}\psi_{n+\alpha}\Bigr ) \rvacn}.
\end{array}
$$
Next we use the relation
$$
s_{(\alpha |\beta )}({\bf t}) =(-1)^{\beta}
\sum_{m=0}^{\beta}h_{\beta -m}(-{\bf t})h_{\alpha +m+1}({\bf t})
$$
which is a consequence of the Jacobi-Trudi formulas
(see Appendix A). With the help of equations (\ref{ferm4})
we can write, continuing the chain of equalities above:
$$
\sum_{\lambda} (-1)^{b(\lambda )}
s_{\lambda}({\bf t})\left |\lambda , n\rbr =
\exp \Bigl ( -\sum_{m<n}\psistar _m (J_-({\bf t}))\left (
\psi_m (J_-({\bf t}))\right )_{\geq n}\Bigr )\rvacn ,
$$
where 
\beq\label{short}
\begin{array}{l}
\displaystyle{
\psi _m (J_-({\bf t})):=\sum_{k\geq 0}h_{k}({\bf t})\psi_{m+k}
=e^{J_-({\bf t})}\psi_m e^{-J_-({\bf t})}}
\\ \\
\displaystyle{
\psistar _m (J_-({\bf t})):=\sum_{k\geq 0}h_{k}({-\bf t})\psistar_{m-k}
=e^{J_-({\bf t})}\psistar_m e^{-J_-({\bf t})}}
\end{array}
\eeq
and 
$\left (
\psi_m (J_-({\bf t}))\right )_{\geq n}$ means that in the series
$\psi_m (J_-({\bf t}))$ only $\psi$-operators $\psi_k$ with 
$k\geq n$ are kept. The next step is to notice that the 
operator in the r.h.s. has the same effect when acting to the
right vacuum as the normally ordered operator 
$$\displaystyle{\normordbare
\exp \Bigl ( -\sum_{m<n}\psistar _m (J_-({\bf t}))
\psi_m (J_-({\bf t}))\Bigr )\normordbare}$$ (which was 
considered in \cite{KM81}), so
$$
\begin{array}{c}\displaystyle{
\sum_{\lambda} (-1)^{b(\lambda )}
s_{\lambda}({\bf t})\left |\lambda , n\rbr =
\normordbare
\exp \Bigl ( -\sum_{m<n}\psistar _m (J_-({\bf t}))
\psi_m (J_-({\bf t}))\Bigr )\normordbare \rvacn}
\\ \\
\displaystyle{=\, e^{J_-({\bf t})} \normordbare
\exp \Bigl ( -\sum_{m<n}\psistar _m 
\psi_m \Bigr )\normordbare \, e^{-J_-({\bf t})} \rvacn}.
\end{array}
$$
(A comment on the last step is in order. 
The substitution 
$\psi _m (J_-({\bf t}))
=e^{J_-({\bf t})}\psi_m e^{-J_-({\bf t})}$,
$\psistar _m (J_-({\bf t}))
=e^{J_-({\bf t})}\psistar_m e^{-J_-({\bf t})}$ can not be done
under the sign of normal ordering. It can be done
only after expanding the normally ordered exponent
in a series of the form (\ref{Bex}).)
Finally, it is easy to see that the operator 
\beq\label{Pn}
\begin{array}{c}\displaystyle{
\normordbare
\exp \Bigl ( \! -\! \! \sum_{m<n}\psistar _m 
\psi_m \Bigr )\normordbare =
\normordbare \prod_{m<n}e^{-\psistar_m \psi_m}\normordbare
=\normordbare \prod_{m<n}(1\! -\! \psistar_m \psi_m )\normordbare}
\\ \\
\displaystyle{
=\,  \prod_{m<n}(1\! -\! \psistar_m \psi_m ) =
\prod_{m<n}\psi_m \psistar_m \, =\, {\sf P}^{+}_{n}}
\end{array}
\eeq
(see (\ref{PF1a}))
has the property ${\sf P}^{+}_{n}e^{-J_-({\bf t})}\rvacn =
\rvacn$, which proves the desired formula.

Using the Cauchy-Littlewood identity (\ref{id}), we can write
$$
e^{J_-({\bf t})}\rvacn =\exp \Bigl ( \sum_{k\geq 1}J_{-k}t_k\Bigr )
\rvacn =\sum_{\lambda} s_{\lambda}({\bf t})
s_{\lambda}(\tilde {\bf J}_-)
\rvacn ,
$$
hence
\beq\label{lambda2a}
\left |\lambda , n\right > =(-1)^{b(\lambda )}
s_{\lambda}(\tilde {\bf J}_-)\rvacn
\eeq
and, similarly,
\beq\label{lambda2b}
\left <\lambda , n\right | =(-1)^{b(\lambda )}
\lvacn s_{\lambda}(\tilde {\bf J}_+),
\eeq
where $\tilde {\bf J}_{\pm}=(J_{\pm 1}, J_{\pm 2}/2,  J_{\pm 3}/3,
\ldots )$. These formulas for the basis vectors were obtained in 
\cite{Orlov02}.

Because of importance of Proposition \ref{chardec}
we find it instructive to give another proof. 
Instead of proving (\ref{lambda2}) we will prove the equivalent 
formula (\ref{schur3}): 
\beq
\begin{array}{ll}
\displaystyle{\left < \lambda , n\right | e^{J_-({\bf t})}\rvacn} 
&=\displaystyle{\lvacn \psistar_{n+\alpha_1}\ldots 
\psistar_{n+\alpha_{d(\lambda )}}\,
\psi_{n-\beta_{d(\lambda )}\! -1}\ldots 
\psi_{n-\beta_1 -1}e^{J_-({\bf t})}\rvacn}
\\ &\\
&\displaystyle{={\det_{i,j=1,\ldots,d(\lambda)}} \lvacn \psistar_{n+\alpha_i}\psi_{n-\beta_j-1} e^{J_-({\bf t})}\rvacn}
\\ &\\
&\displaystyle{={\det_{i,j=1,\ldots,d(\lambda)}} \lvacn \psistar_{n+\alpha_i}(J_-(-{\bf t}))\psi_{n-\beta_j-1} (J_-(-{\bf t}))\rvacn}
\\ &\\
&\displaystyle{={\det_{i,j=1,\ldots,d(\lambda)}} \sum_{k=0}^{\alpha_i} h_{\alpha_i-k}({\bf t}) h_{\beta_j+1+k}(-{\bf t})}
\\ &\\
&\displaystyle{={\det_{i,j=1,\ldots,d(\lambda)}} 
(-1)^{\beta_j+1} s_{(\alpha_i |\beta_j )}({\bf t}) }
\\ &\\
&\displaystyle{= (-1)^{b(\lambda )} s_{\lambda }({\bf t}) },
\end{array}
\eeq
where we have used the Wick theorem (\ref{Wick1}), 
the expression for the Schur polynomial corresponding to a hook diagram (\ref{hook1}) and the Giambelli formula (\ref{Gamb}).

\subsubsection{Expansion of the states 
$e^{J_{\pm}({\bf t})} \left | \lambda , n\right >$ in the skew
Schur functions}

The previous proposition can be generalized as follows:

\noindent
\begin{prop} It holds
\beq\label{lambda33}
\begin{array}{l}
\displaystyle{
e^{J_-({\bf t})}\left |\lambda , n\right > = 
\sum_{\mu} (-1)^{b(\mu \setminus \lambda )}
s_{\mu \setminus \lambda}({\bf t})\left |\mu , n\rbr},
\\ \\
\displaystyle{
e^{J_+({\bf t})}\left |\lambda , n\right > = 
\sum_{\mu} (-1)^{b(\lambda \setminus \mu )}
s_{\lambda \setminus \mu}({\bf t})\left |\mu , n\rbr},
\\ \\
\displaystyle{
\left <\lambda , n \right |
e^{J_+({\bf t})}=\sum_{\mu} (-1)^{b(\mu \setminus \lambda )}
s_{\mu \setminus \lambda}({\bf t})\lbr \mu , n \right |},
\\ \\
\displaystyle{
\left <\lambda , n \right |
e^{J_-({\bf t})}=\sum_{\mu} (-1)^{b(\lambda \setminus \mu )}
s_{\lambda \setminus \mu}({\bf t})\lbr \mu , n \right |},
\end{array}
\eeq
where $b(\lambda \setminus \mu )=b(\lambda )-b(\mu )$ and
\beq\label{lambda34}
s_{\lambda \setminus \mu}({\bf t})=\det_{1\leq i,j\leq \ell (\lambda )}
h_{\lambda_i -\mu_j -i+j}({\bf t})
\eeq
are skew Schur functions
(see \cite{Macdonald} and Appendix A).
\end{prop}

\noindent
Note that the sums in the second and the fourth lines are finite
because $s_{\lambda \setminus \mu}({\bf t})$ is non-zero only if
$\mu \subseteq \lambda$. These formulas imply
\beq\label{lambda4}
\left < \mu , n\right | e^{J_+({\bf t})}\left |\lambda , n\right >
=\left < \lambda , n\right | e^{J_-({\bf t})}\left |\mu , n\right >
=(-1)^{b(\lambda \setminus \mu )}s_{\lambda \setminus \mu}({\bf t}).
\eeq
Let us sketch the proof of (\ref{lambda33}). The first formula
is proved by the following chain of equalities:
$$
\begin{array}{lll}
e^{J_-({\bf t})}\left |\lambda , n\right >&=&
(-1)^{b(\lambda )}e^{\sum_{k\geq 1}t_k J_{-k}}
s_{\lambda}(\tilde {\bf J}_-)\rvacn
\\ && \\
&=&\displaystyle{(-1)^{b(\lambda )}\sum_{\nu}s_{\nu }({\bf t}) 
s_{\nu}(\tilde {\bf J}_-)
s_{\lambda}(\tilde {\bf J}_-)\rvacn}
\\ && \\
&=&\displaystyle{(-1)^{b(\lambda )}\sum_{\mu , \nu}
c_{\nu \lambda}^{\mu}s_{\nu }({\bf t}) s_{\mu}(\tilde {\bf J}_-)\rvacn}
\\ && \\
&=&\displaystyle{(-1)^{b(\lambda )}\sum_{\mu}
s_{\mu \setminus \lambda}({\bf t}) s_{\mu}(\tilde {\bf J}_-)\rvacn}
\\ && \\
&=&\displaystyle{\sum_{\mu}(-1)^{b(\mu )-b(\lambda )}
s_{\mu \setminus \lambda}({\bf t})\left |\mu , n\right >},
\end{array}
$$
where in the second line the Cauchy-Littlewood identity is used.
In the third line we use the definition
$$
s_{\lambda \setminus \mu} ({\bf t})=
\sum_{\nu}c_{\mu \nu}^{\lambda}s_{\nu}({\bf t}),
$$
where the Littlewood-Richardson coefficients $c_{\mu \nu}^{\lambda}$ are determined by
$$
s_{\mu}({\bf t})s_{\nu}({\bf t})=\sum_{\lambda}
c_{\mu \nu}^{\lambda}s_{\lambda}({\bf t}).
$$
The other formulas are proved in a similar way.

The determinant formula (\ref{lambda34}) follows from the Wick theorem.
Put $\ell =\ell (\lambda )$ and add
$\ell (\lambda )-\ell (\mu )$ 
zero lines to the bottom of the diagram $\mu$ if 
$\ell (\mu )<\ell (\lambda )$. Using (\ref{lambda21}), we write:
$$
\begin{array}{ll}
s_{\lambda \setminus \mu} ({\bf t})&=\left <-\ell \right |
\psistar_{\mu_{\ell}-\ell}  \ldots  \psistar_{\mu_1-1}
e^{J_+({\bf t})} \psi_{\lambda_1 -1} \ldots 
\psi_{\lambda_{\ell}-\ell}\left |-\ell \right >
\\ &\\
&=\displaystyle{
\, \det_{1\leq i,j\leq \ell} \lvac \psistar_{\ell +\mu_j-j}\,
e^{J_+({\bf t})} \psi_{\ell +\lambda_i-i}\rvac}
\\ &\\
&=\displaystyle{
\det_{1\leq i,j\leq \ell}h_{\lambda_i -\mu_j -i+j}({\bf t})}.
\end{array}
$$
The last equality follows from (\ref{hji}).

\subsection{The boson-fermion correspondence}\label{Boson}

\subsubsection{Bosonization rules}

Consider the current operator
$$
J(z)=\normord \psi (z)\psistar (z)\normord =
\sum_{k\in \z} J_k z^{-k}.
$$
As we have seen, its Fourier modes $J_k$ have the same 
commutation relations as bosonic operator modes: 
$[J_k, J_l]=k\delta_{k+l,0}$, so they generate the Weyl algebra 
and $J_k$ and $J_{-k}/k$ are 
canonically conjugate. 

The operator $J_0=Q$ is special.
Its canonically conjugate partner is an operator $P$ such that
$e^P$ is the shift operator acting as 
$$
e^P \psi_n e^{-P}=\psi_{n+1}\,, \quad \quad
e^P \psistar_n e^{-P}=\psistar_{n+1}
$$
on the fermionic operators 
or $e^{\pm P}\rvacn = \left |n\pm 1\right >$,
$\lvacn e^{\pm P} = \left <n\mp 1\right |$
on the vacuum states
(this definition implies the commutation relation
$[Q,P]=1$). In accordance with Proposition \ref{unidec},
$$
e^{\pm P}=\sum_{n,\lambda} \left |\lambda , n \rbr
\lbr \lambda , n\mp 1 \right |  ,
$$
where the states in r.h.s. are given by (\ref{genpro}).
Note that $e^{\pm P}$ satisfy the BBC.

All this prompts to introduce
the chiral bosonic field
\beq\label{bos1}
\begin{array}{lll}
\phi (z)&=& \displaystyle{\sum_{k>0}\frac{J_{-k}}{k}\, z^k +P +
Q \log z -\sum_{k>0}\frac{J_{k}}{k}\, z^{-k}}
\\ &&\\
&=& \displaystyle{J_{-}([z])+P +J_0 \log z -J_{+}([z^{-1}])}
\end{array}
\eeq
such that $z\p_z \phi (z)=J(z)$ or
$$
\phi (z_2)-\phi(z_1)=\int_{z_1}^{z_2}\!\! J(z)\frac{dz}{z}\,.
$$
In the second line of (\ref{bos1}) we use the notation
$J_{\pm}([z])=J_{\pm 1}z+\frac{1}{2}J_{\pm 2}z^2 +
\frac{1}{3}J_{\pm 3}z^3 +\ldots$.

The operators $J_{+k}$ with $k>0$ kill the right vacuum, so they are
bosonic annihilation operators while $J_{-k}$ with $k>0$ kill
the left vacuum, so they are creation operators.
Let us introduce the bosonic normal ordering 
$\normordboson (\ldots )\normordboson$ defined by the usual rule 
that all annihilation operators are moved to the right and 
all creation ones to the left and consider the normally ordered
exponents of the free bosonic field:
\beq\label{bos2}
\begin{array}{l}
\normordboson e^{\phi (z)}\normordboson =
e^{J_{-}([z])} \, e^P z^Q \, e^{-J_{+}([z^{-1}])},
\\ \\
\normordboson e^{-\phi (z)}\normordboson =
e^{-J_{-}([z])} \, z^{-Q}  e^{-P} \, e^{J_{+}([z^{-1}])}
\end{array}
\eeq
(note that the normal ordering acts only to the current modes 
$J_{k}$ with nonzero $k$). From (\ref{JkJl3}) it follows that
$$
\begin{array}{l}
e^{J_{\pm}({\bf t})}\, \normordboson e^{\phi (z)}\normordboson \, 
e^{-J_{\pm}({\bf t})}
=e^{\xi({\bf t},z^{\pm 1})}\normordboson e^{\phi (z)}\normordboson \,,
\\ \\
e^{J_{\pm}({\bf t})}\, \normordboson e^{-\phi (z)}\normordboson \, 
e^{-J_{\pm}({\bf t})}
=e^{-\xi({\bf t},z^{\pm 1})}\normordboson e^{-\phi (z)}\normordboson \,.
\end{array}
$$
These formulas tell us that $\normordboson e^{\pm \phi (z)}\normordboson$
behave like the fermionic fields $\psi (z), \psistar (z)$.
Moreover, one can show that all matrix elements of 
$\normordboson e^{\pm \phi (z)}\normordboson$
with respect to the basis states coincide with those of
$\psi (z), \psistar (z)$. Therefore, we can identify
\beq\label{bos3}
\psi (z) =\normordboson e^{\phi (z)}\normordboson ,
\quad \quad
\psistar (z) =\normordboson e^{-\phi (z)}\normordboson .
\eeq
Let us outline the proof of the first formula (the second one
is proved in a similar way). We need to compare matrix elements 
of the operators in both sides. It is clear that the matrix 
elements of $\psi (z)$ vanish unless the charge 
difference between the left and right states is 1. 
Consider the generating function
of non-zero matrix elements, which is the matrix element 
between the coherent states
$\lvacn e^{J_{+}({\bf t}_+)}$ and $e^{J_{-}({\bf t}_-)}\rvacn$,
then we should prove that
$$
\lvacn e^{J_{+}({\bf t}_+)}\normordboson e^{\phi (z)}\normordboson
e^{J_{-}({\bf t}_-)}
\left |n-1\right > =
\lvacn e^{J_{+}({\bf t}_+)} \psi (z)e^{J_{-}({\bf t}_-)}
\left |n-1\right >
$$
for all ${\bf t}_{\pm}$ and $n$. It is a matter 
of straightforward calculation,
which uses the explicit form of $\normordboson e^{\phi (z)}\normordboson$
(\ref{bos2}), the commutation relation (\ref{JkJl3}) 
and relations (\ref{ferm3}), (\ref{ferm3a}), to see that the
both sides are equal to
$$
z^{n-1}\exp \Bigl (\xi ({\bf t}_+, z)-\xi ({\bf t}_-, z^{-1})
+\sum_{k\geq 1}kt_k t_{-k}\Bigr ).
$$

Let us check that the bosonization formulas (\ref{bos3}) imply
the relation 
\beq\label{bos4}
\normord \psi (z) \psistar (z)\normord = z\p_{z}\phi (z)
\eeq
and in this sense are consistent with (\ref{bos1}).
We start with
$
\psi (z_2) \psistar (z_1)=\normordboson e^{\phi (z_2)}\normordboson
\normordboson e^{-\phi (z_1)}\normordboson
$
and rewrite the both sides through normally ordered expressions.
An easy calculation shows that
$$
\psi (z_2) \psistar (z_1)=\normord \psi (z_2) \psistar (z_1)\normord
+\frac{z_1}{z_2 -z_1}\,,
$$
$$
\normordboson e^{\phi (z_2)}\normordboson
\normordboson e^{-\phi (z_1)}\normordboson =
\frac{z_1}{z_2 -z_1}\, \normordboson 
e^{\phi (z_2)-\phi (z_1)}\normordboson .
$$
Setting $z_1 =z$, $z_2=z_1+\varepsilon$, we thus get:
$$
\frac{z}{\varepsilon}\Bigl (\normordboson 
e^{\phi (z+\varepsilon )-\phi (z)}\normordboson \, - 1\Bigr )
=\normord \psi (z+\varepsilon ) \psistar (z)\normord
$$
which coincides with (\ref{bos4}) in the limit $\varepsilon \to 0$. 
Next terms of expansion are given by:
\beq\label{bosdecomp}
\begin{array}{l}
\displaystyle{\normord \psi (z+\varepsilon ) 
\psistar (z)\normord=z \left(\p \phi +
\frac{\epsilon}{2!}\left(\normordboson \left(\p \phi\right)^2\normordboson +\p^2\phi \right)\right.
}
\\ \\
\displaystyle{\left. \,\, +\,\, \frac{\epsilon^2}{3!}\left(
\normordboson \left(\p \phi \right)^3\normordboson+
3\normordboson \p \phi \, \p^2 \phi\normordboson +
\p^3\phi \right)+O(\epsilon^3) 
\right).
}
\end{array}
\eeq
where $\p^m \phi:=\p^m_z\phi (z)$.
For more details see \cite{Fukuma}.

Being applied to vacuum states, the operators (\ref{bos2})
simplify because either $e^{\pm J_-([z])}$ or $e^{\pm J_+([z^{-1}])}$
disappears. Acting by both sides 
of (\ref{bos3}) to the left vacuum, we get the bosonization rules
\beq\label{bf3}
\begin{array}{l}
\lvacn \psi (z)= z^{n-1}\left <n\! -\! 1\right | 
e^{-J_+([z^{-1}])},
\\ \\
\lvacn \psi^{*} (z)= z^{-n}\left <n\! +\! 1\right | 
e^{J_+([z^{-1}])}.
\end{array}
\eeq
Their simple consequence is
\beq\label{bf3b}
\lvacn \psistar (\zeta )\psi (z) =\frac{z^n \zeta^{1-n}}{\zeta -z}
\lvacn e^{J_+([\zeta ^{-1}]-[z^{-1}])}.
\eeq
The repeated use of equations (\ref{bf3}) gives the following 
formulas for arbitrary number of fermionic operators:
\beq\label{bf4}
\begin{array}{l}\displaystyle{
\lvacn \psi (z_1)\ldots \psi (z_m)=(z_1 \ldots z_m)^{n-m}\,
\prod_{i<j}
(z_i-z_j)\,
\left <n\! -\! m \right | e^{-J_+([z_1^{-1}])-\ldots -J_+([z_m^{-1}])}},
\\ \\
\displaystyle{
\lvacn \psistar (z_1)\ldots \psistar (z_m)=(z_1 \ldots z_m)^{-n-m+1}\!
\prod_{i<j}
(z_i\! -\! z_j)
\left <n\! +\! m \right | e^{J_+([z_1^{-1}])+\ldots +J_+([z_m^{-1}])}}.
\end{array}
\eeq
Merging the points in these formulas, one gets
\beq\label{bf5}
\begin{array}{l}
\lvacn \p^{m-1}\psi (z)\ldots \p \psi (z)\psi (z)=
a_m z^{m(n-m)} \left < n-m\right | e^{-mJ_+([z^{-1}])},
\\ \\
\lvacn \p^{m-1}\psistar (z)\ldots \p \psistar (z)\psistar (z)=
a_m z^{-m(n+m-1)} \left < n\! +\! m\right | e^{mJ_+([z^{-1}])},
\end{array}
\eeq
where $a_m = 1! \, 2! \ldots (m-1)!$

Similarly, the bosonization formulas for the 
right vacuum are:
\beq\label{bf3a}
\begin{array}{l}
\psi (z)\rvacn = z^{n}\,
e^{J_-([z])}\, \left |n\! +\! 1\right >,
\\ \\
\psi^{*} (z)\rvacn = z^{-n+1}
e^{-J_-([z])}\left |n\! -\! 1\right >
\end{array}
\eeq
from which it follows that
\beq\label{bf4b}
\psistar (\zeta )\psi (z)\rvacn =\frac{z^n \zeta^{1-n}}{\zeta -z}\,
e^{J_-([z]-[\zeta ])}\rvacn
\eeq
and
\beq\label{bf4a}
\begin{array}{l}\displaystyle{
\psi (z_1)\ldots \psi (z_m)\rvacn =(z_1 \ldots z_m)^{n}\,
\prod_{i<j}
(z_i-z_j)\,
e^{J_-([z_1])+\ldots +J_-([z_m])}
\left |n\! +\! m \right > },
\\ \\
\displaystyle{
\psistar (z_1)\ldots \psistar (z_m)\rvacn =(z_1 \ldots z_m)^{-n+1}\,
\prod_{i<j}
(z_i-z_j)\,
e^{-J_-([z_1])-\ldots -J_-([z_m])}
\left |n\! -\! m \right > },
\end{array}
\eeq
\beq\label{bf5a}
\begin{array}{l}
\psi (z)\p \psi (z) \ldots \p^{m-1}\psi (z) \rvacn=
a_m z^{m(n+m-1)} e^{mJ_-([z])} \left | n+m\right >,
\\ \\
\psistar (z) \p \psistar (z) \ldots \p^{m-1}\psistar (z) \rvacn =
a_m z^{m(m-n)} e^{-mJ_-([z])} \left | n\! -\! m\right >.
\end{array}
\eeq

\subsubsection{Vertex operators}

Let us present the bosonization formulas in a different
but equivalent form which is based on an explicit realization
of the bosonic Fock space ${\cal H}_B$ by polynomials in
infinite number of variables
$t_1, t_2, t_3, \ldots$.
More precisely, following 
\cite{MDD}, consider the space
$$
{\cal H}_B = \CC [w,w^{-1}, \, t_1, t_2, t_3, \ldots ]
=\bigoplus_{l\in \z}w^l \,
\CC [t_1, t_2, t_3, \ldots ],
$$
where an extra variable $w$ is added to take into account
the fermionic states with different charges, and the map
$\Phi : {\cal H}_F \rightarrow {\cal H}_B$ defined for an arbitrary
state $\left |U\right > \in {\cal H}_F$
as follows:
\beq\label{bf1}
\Phi (\left |U\right >)=\sum_{l\in \z} w^l  \left <l\right |
e^{J_{+}({\bf t})}\left |U\right >.
\eeq
If the state $\left |U\right >$ has a definite charge $q$, then
the sum in the r.h.s. contains the only non-zero term with $l=q$.
It can be shown that this correspondence means that
the fermionic Fock space ${\cal H}_F$
is isomorphic to the bosonic Fock space ${\cal H}_B$ and
the Fourier components of the current
as well as fermionic creation and annihilation 
operators become some operators acting in the space of functions of
$w$ and $t_i$.

\begin{prop}\label{currents}
For the components of the current we have:
\beq\label{bf102}
\Phi (J_k \left |U\right >)=\left \{\begin{array}{l}
\,\, \p_{t_k}\Phi (\left |U\right >)\,, \quad \,\, \, \, k>0,
\\ \\
\,\, w\p_w \Phi (\left |U\right >)\,, \!\!\!\! \quad \quad k=0,
\\ \\
-kt_{-k}\Phi (\left |U\right >)\,, \quad k<0.
\end{array}\right.
\eeq
\end{prop}
The first two formulas are obvious from the definition (\ref{bf1}).
The third one follows from the commutation relation 
$e^{J_+({\bf t})}J_{-k}=(J_{-k}+kt_k)e^{J_+({\bf t})}$ (where $k>0$)
obtained with the help of formula (\ref{ABA}).

Let us introduce {\it vertex operators} by the formulas
\beq\label{bf2}
\begin{array}{l}
\displaystyle{
X(z)=\exp \left (\sum_{j\geq 1}t_j z^j\right )
\exp \left (-\sum_{j\geq 1}\frac{1}{jz^j}\, \p_{t_j}\right )
e^P z^Q},
\\ \\
\displaystyle{
X^{*}(z)=\exp \left (-\sum_{j\geq 1}t_j z^j\right )
\exp \left (\sum_{j\geq 1}\frac{1}{jz^j}\, \p_{t_j}\right )
z^{-Q} e^{-P}},
\end{array}
\eeq
where the operators $P,Q$ are defined by their action to the functions of
$w$: 
$$
e^P f(w)=wf(w)\,, \quad \quad
z^Q f(w)=f(zw)
$$
(it is easy to check that $[Q,P]=1$).
Using the previously introduced short-hand notation, we can write:
\beq\label{bf2a}
\begin{array}{l}
\displaystyle{
X(z)=e^{\xi ({\bf t},z)} e^{-\xi (\tilde \p , z^{-1})}
e^P z^Q},
\\ \\
\displaystyle{
X^{*}(z)=e^{-\xi ({\bf t},z)} e^{\xi (\tilde \p , z^{-1})}
z^{-Q} e^{-P} },
\end{array}
\eeq
where $\tilde \p =\{\p_{t_1}, \frac{1}{2}\p_{t_2}, 
\frac{1}{3}\p_{t_3}, \ldots \}$.

\begin{prop}\label{vertex}
Under the boson-fermion correspondence,
the fermionic operators $\psi (z)$, $\psistar (z)$ are represented 
by the vertex operators $X(z), X^{*}(z)$:
\beq\label{bf101}
\begin{array}{l}
\Phi (\psi (z)\left |U\right >)=X(z)\Phi (\left |U\right >),
\\ \\
\Phi (\psistar (z)\left |U\right >)=X^{*}(z)
\Phi (\left |U\right >).
\end{array}
\eeq
\end{prop}
In order to prove the first equality, we write down the 
both sides separately,
$$
\begin{array}{l}
\displaystyle{
\Phi (\psi (z)\left |U\right >)=
\sum_n w^n \left <n\right | e^{J_+({\bf t})}\psi (z)\left |U\right > 
=e^{\xi ({\bf t},z)}
\sum_n w^n 
\left <n\right | \psi (z)
e^{J_+({\bf t})}\left |U\right >},
\\ \\
\displaystyle{
X(z)\Phi (\left |U\right >)=e^{\xi ({\bf t},z)}
\sum_n z^{n-1}w^n \left <n\! -\! 1\right | 
e^{-J_+([z^{-1}])} e^{J_+({\bf t})}\left |U\right >},
\end{array}
$$
and apply the
bosonization rule (\ref{bf3}). The second equality 
is proven in a similar way.

\newpage
\section{Tau-functions}

As it has been established in the works of the Kyoto school,
the expectation values of 
group-like elements\footnote{In this section we assume that the group-like elements $G$ have zero charge, if the contrary is not stated
explicitly.} are $\tau$-functions
of integrable hierarchies of nonlinear differential equations.
This means that they obey an
infinite set of Hirota bilinear equations \cite{Hirota81,Miwa82}.

\subsection{The main types of $\tau$-functions}

There are three main types of the $\tau$-functions (which generalize each 
other):
\begin{itemize}
\item The $\tau$-function of the KP hierarchy
depending on
the times ${\bf t}=\{t_1, t_2, \ldots \}$:
\begin{equation}\label{ferm6}
\tau ({\bf t})=\lvac e^{J_+ ({\bf t})}G\rvac .
\end{equation}
\item
The $\tau$-function of the modified KP (MKP)
hierarchy 
\cite{Kupershmidt,Dickey,Tak01,TakTeo06}\footnote{In \cite{TakTeo06}
a slightly more general version called there
a coupled modified KP (cMKP) hierarchy, was considered.}.
\begin{equation}\label{ferm6a}
\tau_n ({\bf t})=\lvacn e^{J_+ ({\bf t})}G\rvacn .
\end{equation}
Equations of the MKP hierarchy are differential-difference equations
which include shifts of the variable $n=t_0$ (the ``zero time'').
At each fixed $n$, the $\tau$-function (\ref{ferm6a})
is a $\tau$-function of the KP hierarchy, so the $n$-evolution
represents an infinite chain of B\"acklund transformations.
\item
The $\tau$-function of the 2D Toda lattice (2DTL) hierarchy 
\cite{UT84,T90}:
\begin{equation}\label{ferm6b}
\tau_n ({\bf t}_+ , {\bf t}_-)=
\lvacn e^{J_+ ({\bf t}_+)}Ge^{-J_- ({\bf t}_-)}\rvacn .
\eeq
It is the most general $\tau$-function associated with the 
one-component fermions.
At fixed $n$, ${\bf t}_-$ it is a $\tau$-function of the KP 
hierarchy (as a function of ${\bf t}_+$) and 
at fixed $n$, ${\bf t}_+$ it is a $\tau$-function of the KP 
hierarchy (as a function of ${\bf t}_-$).
\end{itemize}

The meaning of the variable $n$ depends on the class
of solutions. For some classes of solutions $n$ can be regarded as a 
continuous, or even complex, variable.
Using the operator $P$ introduced in section \ref{Boson},
we can equivalently represent the $n$-dynamics as
\beq\label{ferm6aaa}
\tau_n ({\bf t}_+ , {\bf t}_-)=
\lvac e^{J_+ ({\bf t}_+)}e^{-nP}Ge^{nP}e^{-J_- ({\bf t}_-)}\rvac .
\eeq

\noindent
{\bf Remark.}
More general $\tau$-functions can be obtained as matrix elements 
of group-like elements with non-zero charge between
different Dirac vacua. Any group-like element $G_q$ 
with charge $q$ gives rise to the $\tau$-function
\begin{equation}\label{ferm6d}
\tau_n ({\bf t}_+ , {\bf t}_-)=
\lvacn e^{J_+ ({\bf t}_+)}G_q e^{-J_- ({\bf t}_-)}\left| n-q \right>
\eeq
which, along with 
(\ref{ferm6b}), is a solution to the 
bilinear identity, see (\ref{2D1}) below.
In section \ref{Characqp} we show that a family 
of (quasi)polynomial and soliton
KP and MKP $\tau$-functions corresponding to 
group-like elements with non-zero charge can be understood in terms of 
singular limits of group-like elements 
with zero charge.

\subsection{Bilinear equations for the $\tau$-function}

\subsubsection{The bilinear identity for the MKP and KP hierarchies}

Consider first the MKP $\tau$-function.
It obeys a bilinear identity which is
a direct consequence of the BBC 
in the form (\ref{bilinear-fermi}).
Setting $\left|V \rbr =\rvacn$,
$\left|V^{\prime} \rbr =\left |n'\right >$ with $n\geq n'$,
where $\rvacn$ and $\left |n'\right >$ are two shifted
Dirac vacua, we have
\begin{equation}\label{B1}
\sum_{k \in {\mathbb Z}} \lbr U \right|  \psi_{k} G \left|n \rbr
\lbr U^{\prime} \right|  \psistar_{k} G \left|n' \rbr =0
\end{equation}
because either $\psi_k$ or $\psistar_k$ kills the state
in each term in the r.h.s. of (\ref{bilinear-fermi}).
Now, setting $\lbr U \right| = \lbr n+1 \right | e^{J_{+}({\bf t})}$,
$\lbr U' \right| = \lbr n'-1 \right | e^{J_{+}({\bf t'})}$,
we can write
$$
\begin{array}{lll}
0&=&\displaystyle{\sum_k
\lbr n+1 \right | e^{J_{+}({\bf t})} \psi_k G \rvacn
\lbr n'-1 \right | e^{J_{+}({\bf t'})}\psistar_k G\left |n'\right >}
\\ &&\\
&=& \displaystyle{
\mbox{res}_{z}\left [
z^{-1} \lbr n+1 \right | e^{J_{+}({\bf t})} \psi (z) G \rvacn
\lbr n'-1 \right | e^{J_{+}({\bf t'})}\psistar (z)
G\left |n'\right >\right ]}
\\ &&\\
&=& \displaystyle{
\mbox{res}_{z}\left [
e^{\xi ({\bf t}-{\bf t'},z)}z^{n-n'}
\lbr n \right | e^{J_+({\bf t}-[z^{-1}])}G\rvacn
\lbr n' \right | e^{J_+({\bf t'}+[z^{-1}])}G\left |n' \rbr \right ],}
\end{array}
$$
where 
we have used the commutation relations of the fermion
operators with $e^{J_+({\bf t})}$ and the
bosonization rules (\ref{bf3}). 
Here $\mbox{res}_{z}$
means picking the coefficient in front of $z^{-1}$ of the 
Laurent series.
In this way we arrive at the
bilinear identity
\begin{equation}\label{bi1}
\oint_{{\mathcal C}_{\infty}} e^{\xi ({\bf t}-{\bf t'},z)}z^{n-n'}
\tau_n ({\bf t}-[z^{-1}])\tau_{n'}({\bf t'}+[z^{-1}])dz =0
\end{equation}
for the $\tau$-function valid for all ${\bf t}, {\bf t'}$
and $n\geq n'$. Here we use
the standard
short-hand notation
$$
{\bf t}\pm [z]\equiv \bigl \{ t_1\pm 
z, t_2\pm \frac{1}{2}z^2, 
t_3 \pm \frac{1}{3}z^3, \ldots \bigr \}.
$$
At $n=n'$ (\ref{bi1}) becomes the bilinear identity for the
$\tau$-function of the KP hierarchy,
\begin{equation}\label{bi1KP}
\oint_{{\mathcal C}_{\infty}} e^{\xi ({\bf t}-{\bf t'},z)}
\tau ({\bf t}-[z^{-1}])\tau ({\bf t'}+[z^{-1}])dz =0.
\end{equation}
Note that it implies that if $\tau ({\bf t})$ is a KP 
$\tau$-function, then so is $\tau (-{\bf t})$.
At $n=n'+1$ (\ref{bi1}) gives the bilinear identity for the 
MKP hierarchy:
\begin{equation}\label{bi1MKP}
\oint_{{\mathcal C}_{\infty}} ze^{\xi ({\bf t}-{\bf t'},z)}
\tau_{n+1} ({\bf t}-[z^{-1}])\tau _n ({\bf t'}+[z^{-1}])dz =0.
\end{equation}
The same bilinear identity can be obtained for the $\tau$-functions
constracted as matrix elements of group-like elements with 
non-zero charge.

The choice of the integration contour ${\mathcal C}_{\infty}$
depends on the type of the time evolution. Formally,
the evolution factor $e^{\xi ({\bf t}-{\bf t'},z)}z^{n-n'}$
has an essential singularity at $\infty$, and the contour
should 
then be a small circle around $\infty$
(a big circle in the complex plane).   
This standard prescription does work if 
only a finite number of times $t_k$ are nonzero.
However, in general, when the values of the times are
such that the factor $e^{\xi ({\bf t}-{\bf t'},z)}z^{n-n'}$ has
singularities at finite points of the 
complex plane, then the prescription is as
follows:
the contour ${\mathcal C}_{\infty}$
must encircle all singularities of the
function $\tau_n ({\bf t}-[z^{-1}])\tau_{n'}({\bf t'}+[z^{-1}])$,
leaving all the singularities of 
$e^{\xi ({\bf t}-{\bf t'},z)}z^{n-n'}$ 
outside the contour.

\noindent
{\bf Remark.} There is a freedom to
multiply the $\tau$-function of the MKP hierarchy
by an exponent of any linear form of times
with constant coefficients and by an arbitrary function
of $n$:
\begin{equation}\label{linear}
\tau_n ({\bf t})\rightarrow C(n)\exp \Bigl (
\sum_{k\geq 1}C_k t_k\Bigr )\tau_n ({\bf t}).
\end{equation}
For the KP hierarchy
\begin{equation}\label{linear2}
\tau ({\bf t})\rightarrow C\exp \Bigl (
\sum_{k\geq 1}C_k t_k\Bigr )\tau ({\bf t}).
\end{equation}
Clearly, these transformations preserve the
form of the bilinear identities.
By noting that
$$
C(0)\normord
\exp \left (\sum_{j\in \z} d_j \psistar_j \psi_j \right )
\normord \rvacn =C(n)  \rvacn
$$
with
$d_n =\frac{C(n)}{C(n+1)} -1$ at $n<0$ and 
$d_n = 1-\frac{C(n+1)}{C(n)}$  at $n\geq 0$
we see that the transformation (\ref{linear}) 
means
$$
G \rightarrow C(0)\exp \Bigl (\sum_{k\geq 1} \frac{C_k}{k} J_{-k}
\Bigr ) \, G \, \normord \exp \Bigl (
\sum_{j} d_j \psistar_j \psi_j\Bigr )
\normord
$$ for
the group-like element.

Equation (\ref{bi1KP}) encodes all the PDE's of the KP hierarchy.
They are obtained by expanding the l.h.s. 
in the Taylor series in ${\bf t}'-{\bf t}$ and equating the 
coefficients to zero.
Technically it is convenient to set
$t_i \to t_i -a_i$ è $t_i'\to t_i +a_i$
and expand in $a_i \to 0$:
$$
\begin{array}{ll}
&\mbox{res}_{z} \left [ \tau ({\bf t}-
{\bf a}-[z^{-1}])\, \tau ({\bf t}+{\bf a} +[z^{-1}])\,
e^{-2\xi ({\bf a},z)}\right ]
\\ & \\
=& \mbox{res}_{z} 
\left [ e^{\xi (\tilde \p _a, z^{-1})} (\tau ({\bf t}-{\bf a})\, 
\tau ({\bf t}+{\bf a}))\,
e^{-2\xi ({\bf a},z)}\right ]
\\ & \\
=& \displaystyle{
\mbox{res}_{z} \left [\sum_{j\geq 0} z^{-j}h_j(\tilde \p _a)
(\tau ({\bf t}-{\bf a})\, \tau ({\bf t}+{\bf a}))\,\sum_{l\geq 0} z^l 
h_l(-2{\bf a})\right ]}
\\ & \\
=& \displaystyle{\sum_{j\geq 0}h_j(-2{\bf a})h_{j+1}(\tilde \p _a)\,
\tau ({\bf t}-{\bf a})\, \tau ({\bf t}+{\bf a})\, =\, 0}.
\end{array}
$$
In the second line, the shift by $[z^{-1}]$ is represented 
by action of the exponentiated differential operator
$$
\xi (\tilde \p _a, z^{-1})= \sum_{j\geq 1}
\frac{z^{-j}}{j}\, \p_{a_j}.
$$
The last equality can be written as
$$
\left. \sum_{j\geq 0}h_j(-2{\bf a})h_{j+1}(\tilde \p _X)\,
e^{\sum_{l\geq 1}a_l \p_{X_l}}
\tau ({\bf t}-{\bf X})\, \tau ({\bf t}+{\bf X})\right |_{X_m=0} =\, 0.
$$
Using the symbols $D_i$ for the ``Hirota derivatives'' 
defined by
$$
\left. \phantom{\int}
P(D) f({\bf t})\cdot g({\bf t}) :=P(\p_X)(f({\bf t}+{\bf X})
g({\bf t}-{\bf X}))\right |_{{\bf X}=0},
$$
where $P(D)$ -- is any polynomial in $D_i$, we can rewrite it in the form
\beq\label{hir2}
\sum_{j\geq 0}h_j(-2{\bf a})h_{j+1}(\tilde D)\,
e^{\sum_{l\geq 1}a_l D_{l}}
\tau ({\bf t}) \cdot \tau ({\bf t}) =0
\eeq
The first non-trivial equation contained here is
\beq\label{hir3}
(D_{1}^4 +3D_{2}^2 -4D_1 D_3 )\tau \cdot \tau =0
\eeq
(the KP equation in the bilinear form) or
\beq
\tau\tau_{1111}-4\tau_1\tau_{111}+3\left(\tau_{11}\right)^2+
3\tau\tau_{22}-3\left(\tau_2\right)^2
-4\tau\tau_{13}+4\tau_1\tau_3=0,
\eeq
where $\tau_{i}:=\p_{t_i}\tau$.
The second derivative of this expression gives the 
KP equation in its standard form
$$
3u_{22}=\left(4u_3-12 u u_1- u_{111}\right)_1,
$$
where $u=\p_{t_1}^2 \log (\tau)$.

Equation (\ref{bi1MKP}) encodes the PDE's of the hierarchy
which is sometimes called 1-modified KP hierarchy.
The first non-trivial equation is
\beq\label{MKPfirst}
(D_1^2 -D_2)\tau_{n+1}\cdot \tau_n =0
\eeq
or
\beq\label{MKPfirst1}
\p_{t_2}\log \frac{\tau_{n+1}}{\tau_n}=
\p_{t_1}^2 \log (\tau_{n+1}\tau_n ) +
\Bigl ( \p_{t_1}\log \frac{\tau_{n+1}}{\tau_n}\Bigr )^2.
\eeq

\subsubsection{The bilinear identity for the 2DTL hierarchy}

For the 2DTL case the bilinear identity is written in the form
$$
\begin{array}{c}
\displaystyle{
\sum_k \left <n+1\right | e^{J_+({\bf t}_+)}\psi_k G 
e^{-J_-({\bf t}_-)}\rvacn
\left <n'-1\right |e^{J_+({\bf t}'_+)}\psistar_k G 
e^{-J_-({\bf t}'_-)}\left |n'\right >}
\\ \\
\displaystyle{
=\, \sum_k \left <n+1\right | e^{J_+({\bf t}_+)}G \psi_k  
e^{-J_-({\bf t}_-)}\rvacn
\left <n'-1\right |e^{J_+({\bf t}'_+)}G \psistar_k  
e^{-J_-({\bf t}'_-)}\left |n'\right >}
\end{array}
$$
or
$$
\begin{array}{c}
\displaystyle{
\mbox{res}_{z}\left [z^{-1}
\left <n+1\right | e^{J_+({\bf t}_+)}\psi (z) G 
e^{-J_-({\bf t}_-)}\rvacn
\left <n'-1\right |e^{J_+({\bf t}'_+)}\psistar (z) G 
e^{-J_-({\bf t}'_-)}\left |n'\right >\right ]}
\\ \\
\displaystyle{
=\, \mbox{res}_{z}\left [z^{-1}
\left <n+1\right | e^{J_+({\bf t}_+)}G \psi (z)  
e^{-J_-({\bf t}_-)}\rvacn
\left <n'-1\right |e^{J_+({\bf t}'_+)}G \psistar (z)  
e^{-J_-({\bf t}'_-)}\left |n'\right >\right ]}.
\end{array}
$$
Using formulas (\ref{ferm3}) and (\ref{ferm3a}), we transform this
to
{\small
$$
\begin{array}{c}
\displaystyle{
\mbox{res}_{z}\left [z^{-1}e^{\xi ({\bf t}_+ -{\bf t}'_+,z)}
\left <n+1\right |\psi (z) e^{J_+({\bf t}_+)} G 
e^{-J_-({\bf t}_-)}\rvacn
\left <n'-1\right |\psistar (z) e^{J_+({\bf t}'_+)} G 
e^{-J_-({\bf t}'_-)}\left |n'\right >\right ]}
\\ \\
\displaystyle{
=\, \mbox{res}_{z}\left [z^{-1}e^{\xi ({\bf t}_- -{\bf t}'_-,z^{-1})}
\left <n\! +\! 1\right | e^{J_+({\bf t}_+)}G   
e^{-J_-({\bf t}_-)}\psi (z)\rvacn
\left <n'\! -\! 1\right |e^{J_+({\bf t}'_+)}G   
e^{-J_-({\bf t}'_-)}\psistar (z)\left |n'\right >\right ]}.
\end{array}
$$}

\noindent
Next, bosonization formulas (\ref{bf3}), (\ref{bf3a}) give:
{\small
$$
\begin{array}{c}
\displaystyle{
\mbox{res}_{z}\left [z^{n-n'}e^{\xi ({\bf t}_+ -{\bf t}'_+,z)}
\left <n\right | e^{J_+({\bf t}_+-[z^{-1}])} G 
e^{-J_-({\bf t}_-)}\rvacn
\left <n'\right | e^{J_+({\bf t}'_+ +[z^{-1}] )} G 
e^{-J_-({\bf t}'_-)}\left |n'\right >\right ]}
\\ \\
\displaystyle{
=\, \mbox{res}_{z}\left [z^{n-n'}e^{\xi ({\bf t}_- -{\bf t}'_-,z^{-1})}
\left <n\! +\! 1\right | e^{J_+({\bf t}_+)}G   
e^{-J_-({\bf t}_- -[z])}\left |n\! +\! 1\right >\right.
}
\\ \\
\hspace{7cm}\displaystyle{
\left. \cdot \left <n'\! -\! 1\right |e^{J_+({\bf t}'_+)}G   
e^{-J_-({\bf t}'_- +[z])}\left |n'\! -\! 1\right >\right ]}.
\end{array}
$$}
We can write this as the bilinear identity
\beq\label{2D1}
\begin{array}{c}
\displaystyle{
\oint_{{\cal C}_{\infty}}
z^{n-n'}e^{\xi ({\bf t}_+ -{\bf t}'_+,z)}
\tau_n ({\bf t}_+ \!-\! [z^{-1}], {\bf t}_-)\,
\tau_{n'} ({\bf t}'_+ \!+\! [z^{-1}], {\bf t}'_-)\, dz}
\\ \\
\displaystyle{
=\, \oint_{{\cal C}_{0}}
z^{n-n'}e^{\xi ({\bf t}_- -{\bf t}'_-,z^{-1})}
\tau_{n+1} ({\bf t}_+, {\bf t}_- \!-\! [z])\,
\tau_{n'-1} ({\bf t}'_+ , {\bf t}'_- \!+\! [z])\, dz}
\end{array}
\eeq
valid for any $n,n'$, ${\bf t}_{\pm}$, ${\bf t}'_{\pm}$.
The same bilinear identity can be obtained for the $\tau$-functions
constracted as matrix elements (\ref{ferm6d}) 
of group-like elements with 
non-zero charge.
The contour ${\cal C}_{0}$ encircles all singularities of the 
function $e^{\xi ({\bf t}_- -{\bf t}'_-,z^{-1})}z^{n-n'}$. 
In particular, if only a finite number of times are non-zero, then
it is a small contour around $0$.
Note that at ${\bf t}_- ={\bf t}'_-$ and $n\geq n'$ the r.h.s. 
vanishes and we get the bilinear identity for the MKP hierarchy
for the function $\tau_n ({\bf t})=\tau_n ({\bf t}, {\bf t}_-)$,
as it should be expected. In a similar way, setting 
${\bf t}_+ ={\bf t}'_+$ and $n\leq n'-2$, we see that the l.h.s.
vanishes while the r.h.s. gives, after the change $z\to 1/z$,
the bilinear identity for the MKP hierarchy
for the function $\tau_n ({\bf t}_-)=\tau_n ({\bf t}_+, -{\bf t}_-)$.

The first non-trivial equation contained in bilinear identity
(\ref{2D1}) is
\beq\label{nt1}
\frac{1}{2}\, D_1D_{-1}\tau_n \cdot \tau_n +\tau_{n+1}\tau_{n-1}=0
\eeq
or
\beq\label{nt2}
\p_{t_1}\p_{t_{-1}}\log \tau_n = -\, 
\frac{\tau_{n+1}\tau_{n-1}}{\tau_n^2}\,.
\eeq
Subtracting these equations at $n+1$ and $n$ one gets the 2D 
Toda equation 
\beq\label{nt3}
\p_{t_1}\p_{t_{-1}} \varphi_n =e^{\varphi_{n}-\varphi_{n-1}}-
e^{\varphi_{n+1}-\varphi_{n}}
\eeq
for $\varphi_n = \log (\tau_{n+1}/\tau_n)$.

\noindent
{\bf Remark.} There is a freedom to
multiply the $\tau$-function of the 2DTL hierarchy
by an exponent of any linear form of times
with constant coefficients:
\begin{equation}\label{linear1}
\tau_n ({\bf t}_+ , {\bf t}_-)\rightarrow C\exp \Bigl (C_0 n+
\sum_{k\in \ZZ, \neq 0}C_k t_k\Bigr )\tau_n ({\bf t}_+ , {\bf t}_-)
\end{equation}
corresponding to the following transform of the group-like element:
$$
G\rightarrow C \exp\left(\sum_{k>0} \frac{C_k J_{-k}}{k}\right)
G\exp\left(C_0 Q+ \sum_{k>0} \frac{C_{-k} J_{k}}{k}\right).
$$
Clearly, this transformation preserves the
form of the bilinear identity.

\subsubsection{3-term bilinear equations}

Setting $n'=n$ and
$t_k'=t_k-\frac{1}{k}(z_{1}^{-k}+
z_{2}^{-k}+z_{3}^{-k})$ in (\ref{bi1}), we see that the essential
singularity at $\infty$ splits into 3 simple poles
at $z_1, z_2, z_3$:
$$
e^{\xi ({\bf t}-{\bf t'},z)}=
\frac{1}{
(1-\frac{z}{z_1})(1-\frac{z}{z_2})(1-\frac{z}{z_3})}.
$$
Taking the residues,
we arrive at the 3-term relation
\begin{equation}\label{bi2}
(z_2-z_3)\tau _n\left ({\bf t}-[z_{1}^{-1}]\right )\tau _n
\left ({\bf t}-[z_{2}^{-1}]-[z_{3}^{-1}]\right )
+(231)+(312)=0,
\end{equation}
where the last two terms are obtained from the first one by
the cyclic permutations of the indices.
Another possible choice is $n'=n$ and
$t_k'=t_k+\frac{1}{k} z_{0}^{-k} -\frac{1}{k}(z_{1}^{-k}+
z_{2}^{-k}+z_{3}^{-k})$, then
$$
e^{\xi ({\bf t}-{\bf t'},z)}=
\frac{1-\frac{z}{z_0}}{
(1-\frac{z}{z_1})(1-\frac{z}{z_2})(1-\frac{z}{z_3})}.
$$
This leads to a slightly different (but equivalent) equation
\begin{equation}\label{bi201}
(z_0 -z_1)(z_2-z_3)\tau _n\left ({\bf t}-
[z_{0}^{-1}]-[z_{1}^{-1}]\right )\tau _n
\left ({\bf t}-[z_{2}^{-1}]-[z_{3}^{-1}]\right )
+(231)+(312)=0.
\end{equation}

Setting $n'=n-1$,
$t_k'=t_k-\frac{1}{k}(z_{1}^{-k}+
z_{2}^{-k})$, we see that the essential
singularity at $\infty$ splits into simple poles
at $z_1, z_2$:
$$
ze^{\xi ({\bf t}-{\bf t'},z)}=
\frac{z}{
(1-\frac{z}{z_1})(1-\frac{z}{z_2})}.
$$
Besides the residues at these points, there is also
a contribution from the residue at $\infty$, so
we obtain the 3-term relation
\begin{equation}\label{bi3}
\begin {array}{c}
z_2\tau_{n+1}\left ({\bf t}-[z_{2}^{-1}]\right )
\tau_{n}\left ({\bf t}-[z_{1}^{-1}]\right )-
z_1\tau_{n+1}\left ({\bf t}-[z_{1}^{-1}]\right )
\tau_{n}\left ({\bf t}-[z_{2}^{-1}]\right )
\\ \\
+\, (z_1-z_2)\tau_{n+1}({\bf t})\tau_{n}
\left ({\bf t}-[z_{1}^{-1}]-[z_{2}^{-1}]\right )\, =0.
\end{array}
\end{equation}
It can be formally regarded as a particular case of
(\ref{bi201}) in the limit $z_3\to \infty$, $z_0\to 0$.
The limit $z_3\to \infty$ is smooth while the other one
requires to assign a meaning to
$\lim\limits_{z_0 \to 0} \tau_n ({\bf t}- [z_{0}^{-1}])$.
The form of the evolution multiplier, $ z^n e^{\xi ({\bf t}, z)}$,
suggests to set formally
\begin{equation}\label{adddef}
\lim_{z_0 \to 0} \tau_n ({\bf t}\pm [z_{0}^{-1}])=
\tau_{n\mp 1}({\bf t})
\end{equation}
which actually holds if the function
$\tau_n ({\bf t}-[z^{-1}])$ for arbitrary $n$ can be analytically
continued from a neighborhood of $\infty $ to a
neighborhood of the point $z=0$ and is regular there.
If it is singular, as in the case of solutions relevant to
quantum spin chains, (\ref{adddef}) should be substituted by a more general
prescription which, however, also allows
one to regard (\ref{bi3}) as a particular case of (\ref{bi201}).

Setting $n'=n$, $t'_k=t_k-\frac{1}{k}\, a^{-k}$, 
$t'_{-k}=t_{-k}-\frac{1}{k}\, b^k$ ($k\geq 1$) in 
(\ref{2D1}) and taking the residues, we get the 
3-term bilinear equation for the 2DTL hierarchy:
\beq\label{bi4}
\begin{array}{c}
\tau_n ({\bf t}_+ \! -\! [a^{-1}], {\bf t}_-)
\tau_n ({\bf t}_+, {\bf t}_- \! -\! [b])-
\tau_n ({\bf t}_+, {\bf t}_-)
\tau_n ({\bf t}_+ \! -\! [a^{-1}], {\bf t}_- \! -\! [b])
\\ \\
\displaystyle{
\phantom{aaaaaaa}=\,\, \frac{b}{a}\,
\tau_{n+1} ({\bf t}_+, {\bf t}_- \! -\! [b])
\tau_{n-1} ({\bf t}_+ \! -\! [a^{-1}], {\bf t}_-)}.
\end{array}
\eeq

\subsection{Schur function expansions and Pl\"ucker coordinates}

\subsubsection{Schur function expansions}

One may expand the MKP
$\tau$-function in the Schur polynomials:
\begin{equation}\label{tau1}
\tau_n ({\bf t})=\sum_{\lambda}c_{\lambda}(n) s_{\lambda}({\bf t}).
\end{equation}
The sum is over all Young diagrams including the empty one
(for which $c_{\emptyset}(n)=\tau_n (0)$).
The coefficients $c_{\lambda}(n)$
(``Pl\"ucker coordinates'') can be determined from the
formula (\ref{ferm6}) by inserting the complete set of states
in between the operators $e^{J_+}$ and $G$ and using (\ref{lambda2}):
$$
\tau_n ({\bf t})=\sum_{\lambda}\lvacn e^{J_{+}({\bf t})}\left |
\lambda , n\rbr
\lbr \lambda , n \right | G\rvacn =
\sum_{\lambda}(-1)^{b(\lambda )} s_{\lambda}({\bf t})
\lbr \lambda , n \right | G\rvacn ,
$$
so
\begin{equation}\label{tau2}
\begin{array}{lll}
c_{\lambda}(n)&=&(-1)^{b(\lambda )}
\lbr \lambda , n \right | G\rvacn
\\ && \\
&=&(-1)^{b(\lambda )}\lvacn \psistar_{n+\alpha_1} \ldots
\psistar_{n+\alpha_{d(\lambda )}}
\psi_{n-\beta_{d(\lambda )}-1}  \ldots \psi_{n-\beta_1 -1}G\rvacn ,
\end{array}
\end{equation}
where $b(\lambda )$ is defined in (\ref{boflambda}).
In other words, $c_{\lambda}(n)$ are expansion coefficients 
of the state $G\rvacn$:
$$
G\rvacn =\sum_{\lambda} (-1)^{b(\lambda)}c_{\lambda}(n) \left |\lambda , n\right >.
$$
For the KP $\tau$-function the Schur function expansion 
is
\beq\label{tau1a}
\tau({\bf t}) =\sum_{\lambda}c_{\lambda}s_{\lambda}({\bf t}),
\eeq
where $c_{\lambda}=c_{\lambda}(0)=(-1)^{b(\lambda )}
\left < \lambda , 0\right | G\rvac$.
For the 2DTL $\tau$-function the general form of the expansion is
the double sum
\beq\label{tau1b}
\tau_n({\bf t}_+, {\bf t}_-) =\sum_{\lambda ,\mu}
c_{\lambda \mu}(n)s_{\lambda}({\bf t}_+)
s_{\mu}(-{\bf t}_-)
\eeq
where
$c_{\lambda \mu}(n)=(-1)^{b(\lambda )+b(\mu )}
\left <\lambda , n\right |G\left |\mu , n\right >$. 
The Schur function expansions of $\tau$-function are also
discussed in \cite{OShi05,EH10,Tak84}.

\subsubsection{Restricted Schur function expansions}\label{restsum}

If $\tau ({\bf t})=\sum_{\lambda}c_{\lambda} 
s_{\lambda}({\bf t})$ is a KP $\tau$-function, then the expansion
\beq\label{tau6}
\tau ^{(N)} ({\bf t})=\sum_{\ell (\lambda )\leq N}c_{\lambda} 
s_{\lambda}({\bf t})
\eeq
restricted to Young diagrams with not more than $N$ non-zero lines
is also a KP $\tau$-function for all $N\geq 0$.
This follows from the fermionic representation
$$
\tau ^{(N)} ({\bf t})=\lvac e^{J_+({\bf t})}
{\sf P}_{-N}^{+}\, G\rvac ,
$$
where ${\sf P}_{-N}^{+}$ is the operator introduced in
(\ref{Pn}).

More generally,
if $\tau_n ({\bf t}_+,{\bf t}_-)=\sum_{\lambda,\mu}c_{\lambda\mu} (n)
s_{\lambda}({\bf t}_+)s_{\lambda}(-{\bf t}_-)$ 
is a 2DTL $\tau$-function, then 
\beq\label{tau7}
\tau ^{(N,M)}_n ({\bf t}_+,{\bf t}_-)=
\sum_{\ell (\lambda )\leq N+n}\sum_{\ell (\mu)\leq M+n}c_{\lambda,\mu}(n) 
s_{\lambda}({\bf t}_+)s_{\mu}(-{\bf t}_-)
\eeq
is also a 2DTL $\tau$-function for all $N$ and $M$ 
(in particular, one of them can be equal to $+\infty$, 
so that the corresponding sum is unrestricted).
This follows from the fermionic representation
$$
\tau ^{(N,M)}_n ({\bf t})=\lvacn e^{J_+({\bf t}_+)}
{\sf P}_{-N}^{+}\, G\, {\sf P}_{-M}^{+}\,  e^{-J_-({\bf t}_-)}\rvacn ,
$$
where ${\sf P}_{-N}^{+}$ and ${\sf P}_{-M}^{+}$ are 
the operators introduced in (\ref{Pn}). (Note that 
the operators  ${\sf P}_{-n}^{-}$ 
allow one to apply a similar
restriction on the number of columns of the diagrams in 
the sums, while other projectors introduced in section \ref{Proop} 
make it possible to impose more complicated restrictions.)
In particular, if $\tau_n ({\bf t})=\sum_{\lambda}c_{\lambda} (n)
s_{\lambda}({\bf t})$ is a MKP $\tau$-function, then 
\beq
\tau ^{(N)}_n ({\bf t})=\sum_{\ell (\lambda )\leq N+n}c_{\lambda}(n) 
s_{\lambda}({\bf t})
\eeq
is also a MKP $\tau$-function for all $N$.
This follows from the fermionic representation
$$
\tau ^{(N)}_n ({\bf t})=\lvacn e^{J_+({\bf t})}
{\sf P}_{-N}^{+}\, G\rvacn .
$$

\subsubsection{Determinant formulas 
for $c_{\lambda}(n)$ of the Giambelli type}

\noindent
\begin{prop}
The following 
determinant formulas for 
the coefficients $c_{\lambda}(n)$ of the $\tau$-function
hold:
\begin{equation}\label{tau3a}
c_{\lambda}(n)=(c_{\emptyset}(n))^{-d(\lambda )+1}\,
\det_{i,j =1, \ldots , d(\lambda )}
c_{(\alpha _i |\beta _j)}(n)
\eeq
(assuming that $c_{\emptyset}(n)\neq 0$).
\end{prop}

\noindent
These are the Giambelli-like formulas
where $c_{(\alpha _i |\beta _j)}(n)$ are the
Pl\"ucker coordinates corresponding to
the hook diagrams $(\alpha _i |\beta _j)=(\alpha_i+1, 1^{\beta_j})$.
In the last expression, we have taken into account that
$
\lvacn G \rvacn =\tau_n (0)= c_{\emptyset}(n).
$
The proof consists in
applying 
Wick's theorem in the form (\ref{Wick1}) to (\ref{tau2}):
\begin{equation}\label{tau3}
\begin{array}{lll}
c_{\lambda}(n)&=&\displaystyle{
(-1)^{b(\lambda )}
(\lvacn G \rvacn )^{-d(\lambda )+1}\,
\det_{i,j =1, \ldots , d(\lambda )}
\lvacn \psistar_{n+\alpha_i}
\psi_{n-\beta_j -1}G\rvacn }
\\ && \\
&=&\displaystyle{(c_{\emptyset}(n))^{-d(\lambda )+1}\,
\det_{i,j =1, \ldots , d(\lambda )}
c_{(\alpha _i |\beta _j)}(n)}.
\end{array}
\end{equation}

Formulas of the Giambelli type for the
expansion coefficients, or Pl\"ucker coordinates of the $\tau$-function
were given in \cite{EH10}. Using the Jacobi identity for
determinants, it is easy to see that they are equivalent to
the 3-term bilinear relations for the coefficients $c_{\lambda}(n)$
given in \cite{JM83} (the Pl\"ucker relations):
\beq\label{tau5}
c_{(\vec \alpha |\vec \beta )}(n)
c_{(\vec \alpha_{\not r\not s} |\vec \beta_{\not r\not s} )}(n)=
c_{(\vec \alpha_{\not r} |\vec \beta_{\not r} )}(n)
c_{(\vec \alpha_{\not s} |\vec \beta_{\not s} )}(n)-
c_{(\vec \alpha_{\not r} |\vec \beta_{\not s} )}(n)
c_{(\vec \alpha_{\not s} |\vec \beta_{\not r} )}(n).
\eeq
Here 
$$
\begin{array}{l}
(\vec \alpha_{\not  \, i}|\vec \beta_{ \not  \, k})=
(\alpha_1, \ldots , \not \! \alpha_i , \ldots , \alpha_d|
\beta_1 , \ldots , \not \! \beta_k , \ldots , \beta_d),
\\ \\
(\vec \alpha_{\not \,  r \not \, s} |\vec \beta_{\not \, r \not \, s})
= (\alpha_1, \ldots , \not \! \alpha_r , \ldots , 
\not \! \alpha_s, \ldots , \alpha_d|
\beta_1 , \ldots , \not \! \beta_r , \ldots , \not \! \beta_s, \ldots ,
\beta_d)\,.
\end{array}
$$
Another way to obtain this relation is to apply the Wick theorem
to $$\left < \lambda , n\right | \psi_{n+\alpha_r}\psi_{n+\alpha_s}
\psistar_{n-\beta_s -1}\psistar_{n-\beta_r -1}G\rvacn .$$

Equation (\ref{tau3}) allows one to obtain a useful
representation for the state $G\rvacn$, where $G$ is any group-like
element with zero charge.
Using (\ref{tau3}), we can write:
\beq\label{tau401}
\begin{array}{c}
\hspace{-4cm}G\rvacn =\displaystyle{
\sum_{\lambda}(-1)^{b(\lambda)}c_{\lambda}(n)\left |\lambda , n\right >}
\\ \\
=\displaystyle{c_{\emptyset}(n)+\! \sum_{d\geq 1}(c_{\emptyset}(n))^{1-d}
\!\!\!\!\!\!\!\!\!\sum_{{\alpha_1 >\alpha_2 >\ldots >\alpha_d\geq 0\atop
\beta_1 >\beta_2 >\ldots >\beta_d\geq 0}} (-1)^{\sum_{i=1}^d(\beta_i+1)}
\det_{1\leq i,k\leq d}c_{(\alpha_i|\beta_k )}(n)
\psistar_{n\! -\! \beta_1 \! -\! 1} \ldots
\psistar_{n\! -\! \beta_{d}\! -\! 1}
\psi_{n\! +\! \alpha_{d}}  \ldots \psi_{n\! +\! \alpha_1}\rvacn }
\\ \\
=\displaystyle{c_{\emptyset}(n)\Bigl [
1+\! \sum_{d\geq 1}\frac{(c_{\emptyset}(n))^{-d}}{d!}
\!\!\!\!\!\! \sum_{{\alpha_1 ,\alpha_2 ,\ldots ,\alpha_d\geq 0\atop
\beta_1 ,\beta_2 ,\ldots ,\beta_d\geq 0}}
\prod_{j=1}^d \left((-1)^{\beta_j+1}c_{(\alpha_j |\beta_j )}(n)\right)
\psistar_{n\! -\! \beta_1 \! -\! 1} \ldots
\psistar_{n\! -\! \beta_{d}\! -\! 1}
\psi_{n\! +\! \alpha_{d}}  \ldots \psi_{n\! +\! \alpha_1}\rvacn \Bigr ]}
\\ \\\hspace{-2cm}
=\displaystyle{c_{\emptyset}(n)
\exp \Bigl ( \sum_{\alpha , \beta \geq 0}
(-1)^{\beta+1}\frac{c_{(\alpha |\beta )}(n)}{c_{\emptyset}(n)}\, \psistar_{n-\beta -1}
\psi_{n+\alpha}\Bigr ) \rvacn }
\\ \\
=\, \displaystyle{\lvacn G \rvacn 
\exp \Bigl ( \sum_{\alpha , \beta \geq 0}
\frac{ \lvacn \psistar _{n+\alpha}\psi_{n-\beta -1}G
\rvacn}{\lvacn G \rvacn } \, \psistar_{n-\beta -1}
\psi_{n+\alpha}\Bigr ) \rvacn }
\end{array}
\eeq
(it is assumed that $\lvacn G \rvacn \neq 0$).

\subsubsection{Determinant formulas 
for $c_{\lambda}(n)$ of the Jacobi-Trudi type}

There are also determinant formulas of another type which connect 
the coefficients $c_{\lambda}(n)$ with different $n$.
They were obtained in \cite{AKLTZ}.

\noindent
\begin{prop}
The coefficients $c_{\lambda}(n)$ of the $\tau$-function
obey the relations
\begin{equation}\label{tau4a}
c_{\lambda}(n)
=\displaystyle{
\left ( \prod_{k=1}^{\ell (\lambda )-1}c_{\emptyset}(n-k)\right )^{-1}
\det_{i,j =1,\ldots , \ell (\lambda )}
c_{\lambda_i -i+j}(n-j+1)},
\end{equation}
where
$c_s (n):= c_{(s-1|0)}(n)=\left < n\! -\! 1\right |
\psistar_{n+s-1}G\rvacn $ are
the expansion coefficients for one-row diagrams
and it is assumed that $c_{\emptyset}(n)\neq 0$.
An equivalent set of relations is 
\begin{equation}\label{tau4b}
c_{\lambda}(n)
=\displaystyle{
\left ( \prod_{k=1}^{\lambda _1 -1}c_{\emptyset}(n+k)\right )^{-1}
\det_{i,j =1,\ldots , \lambda _1}
c^{\lambda^{\prime}_i -i+j}(n+j-1)},
\end{equation}
where $c^a(n):=c_{(0|a-1)}(n)=(-1)^a \left <n\! +\! 1 \right | \psi_{n-a}
G\rvacn$
are expansion coefficients for one-column diagrams.
\end{prop}

\noindent
These relations are
sometimes called quantum Jacobi-Trudi formulas.
Note that the transformation $c_{\lambda}(n) \rightarrow
C(n) c_{\lambda}(n)$ with arbitrary $C(n)$ preserves
the determinant formulas (\ref{tau4a}), (\ref{tau4b}). Clearly, this freedom
corresponds to the possibility of multiplying the $\tau$-function
by any $C(n)$, see (\ref{linear}).

Let us outline the proof of (\ref{tau4a}).
The main step is to recall Proposition \ref{states} and apply
formula (\ref{lambda21}) for the state $\left <\lambda , n\right |$
in (\ref{tau2}). Using this we get
$$
c_{\lambda}(n)=\left < n\! -\! \beta_1 \! -\! 1\right |
\psistar_{n+\lambda_{\beta_1+1}\! -\! (\beta_1 \! +\! 1)}\, \ldots \,
\psistar_{n+\lambda_2 -2}\psistar_{n+\lambda_1 -1}
G\rvacn
$$
(recall that $\beta_1 +1 =\ell (\lambda )$).
Applying the Wick theorem in the form (\ref{Wick3}), we
arrive at the desired result. The proof of (\ref{tau4b}) is similar.

\noindent
{\bf Remark.} Let $\lambda = (s^a)$ be the rectangular Young 
diagram of height $a$ and length $s$, and let $c_s^a(n)$ be 
the corresponding coefficients $c_{\lambda}(n)$, then application
of the Jacobi identitity for determinants to (\ref{tau4a}) yields
the following 3-term bilinear relation:
\beq\label{3-term}
c_{s}^{a}(n)c_{s}^{a}(n+1)-c_{s+1}^{a}(n)c_{s-1}^{a}(n+1)=
c_{s}^{a-1}(n)c_{s}^{a+1}(n+1).
\eeq

\subsection{Examples of $\tau$-functions}

In this section we consider the following familiar examples
of $\tau$-functions:
\begin{itemize}
\item
Characters (Schur functions)
\item
(Quasi)polynomial $\tau$-functions
\item
Multi-soliton $\tau$-functions
\item
Partition functions of matrix models
\end{itemize}
and give their fermionic realization.

\subsubsection{Characters and (quasi)polynomial $\tau$-functions}
\label{Characqp} 

The Schur polynomials themselves are $\tau$-functions
of the KP hierarchy. This follows from equation (\ref{schur3}),
\beq\label{ch1}
\tau ({\bf t})=\lvac e^{J_+({\bf t})}\left |\lambda , 0\right >=
\lvac e^{J_+({\bf t})}\psistar_{-\beta_1 \! -\! 1}\ldots
\psistar_{-\beta_d \! -\! 1}\psi_{\alpha_d}\ldots \psi_{\alpha_1}
\rvac =(-1)^{b(\lambda )}s_{\lambda}({\bf t})
\eeq
and the fact that the product of $\psi$- and $\psistar$-operators
in the correlator
satisfies the BBC (\ref{commute}).
Note that one can represent the state
$\left |\lambda , 0\right >$ as action of a normally 
ordered exponent to the vacuum \cite{KMMM93}:
$$
\left |\lambda , 0\right >=
\prod _{i=1}^{d(\lambda )}
\normordbare e^{(\psi_{-\beta_i -1}-\psi_{\alpha_i})(
\psistar_{-\beta_i -1}-\psistar_{\alpha_i})}\normordbare
\rvac .
$$
We call such $\tau$-functions characters because $s_{\lambda}({\bf t})$ 
with $t_k =\frac{1}{k}\, \mbox{tr}\, g^k$ is the character of a 
group element $g\in GL(N)$ in the representation with the highest 
weight $\lambda$. If $t_k =\frac{1}{k}\, \mbox{str}\, g^k$
for $g\in GL(N|M)$, then $s_{\lambda}({\bf t})$ are super-characters.
There are different possibilities 
to embed these KP $\tau$-functions into the MKP and the Toda hierarchies. 
One of them is to use the projectors (\ref{genpro}), for example
$
\tau_n({\bf t})=\lvacn e^{J_+({\bf t})} \left |\lambda , 
0\right > \left< 0 | n \right> =\delta_{n,0} 
(-1)^{b(\lambda )}s_{\lambda}({\bf t})
$.

In fact the Schur polynomials are very special cases of 
more general (quasi)polynomial $\tau$-functions. 
In the case of MKP hierarchy these $\tau$-functions can 
be obtained as either mean values of group-like elements with
non-zero charge between different vacua or
as some
special limiting cases of (\ref{ferm6a}).

Fix two finite sets of distinct points $p_i \in \CC$, $q_j\in \CC$,
and construct (finite) linear combinations of the operators
$\psi (z)$ and their $z$-derivatives at the points $p_i$
and of the operators
$\psistar (z)$ and their $z$-derivatives at the points $q_j$:
\begin{equation}\label{I1}
\Psi_{i}(p_i):=
\sum _{m\geq 0} a_{im}\, \p_z^m \psi (z)\Bigr |_{z=p_i}\,, \quad \quad
\Phi^*_{j}(q_j):=
\sum _{m\geq 0} b_{jm}\, \p_z^m \psistar (z)\Bigr |_{z=q_j}.
\end{equation}
The product of $N$ $\Psi$'s and $M$ $\Phi^*$'s 
is a group-like element with charge $N-M$. Therefore,
\beq\label{pol100}
\tau_n ({\bf t})=\lvacn e^{J_{+}({\bf t})}
\Psi_1(p_1) \ldots \Psi_N(p_N)\Phi_1^*(q_1)\ldots 
\Phi_M^*(q_M)\left |n\! -\! N\! +\! M\right >
\eeq
is a $\tau$-function of the MKP hierarchy.
It is not difficult to see that it is a polynomial 
in all the variables $n$, $t_k$ multiplied 
by an exponential function of a linear combination
of these variables (a quasi-polynomial). 
These solutions are sometimes
called {\it rational} because ratios of such $\tau$-functions are 
rational functions.
One can see that the Schur function expansion of the $\tau$-function
(\ref{pol100}) is restricted to Young diagrams that do not
contain the rectangular $N\times M$ diagram. In particular,
at $M=0$ the summation is restricted to diagrams $\lambda$ such that
$\ell (\lambda )\leq N$.

Let us show how to obtain $\tau$-functions of the form
(\ref{pol100}) by taking a
singular limit of (\ref{ferm6a}). 
We illustrate the point by a very simple example.
Consider
the operator $\normord e^{\beta \psi '(p) \psistar (r)}\normord$
(with $|r| < |p|$), where $\psi '(p)\equiv\p_p \psi (p)$. 
It is a group-like element with zero charge for any $\beta $
but at $\beta =\beta_0 =(\lvac  \psi '(p) 
\psistar (r)\rvac )^{-1}$ it is not
invertible and equals
$$
\normord e^{\beta _0 \psi '(p) \psistar (r)}\normord =
1+ \beta _0\psi '(p) \psistar (r) -\beta_0 \lvac \! \psi '(p) 
\psistar (r)\rvac
=\beta_0 \psi '(p) \psistar (r).
$$
Therefore,
$
\tau_n ({\bf t}) = \lvacn e^{J_+({\bf t})}\psi '(p) \psistar (r)\rvacn
$
is a $\tau $-function of the MKP hierarchy.
Then one can consider the limit $r\to 0$.
The limit is singular and requires some care.
From
$$
r^{n-1}
\psistar (r)\! \rvacn = \sum_l r^{n-l-1}\psistar_{l}\! \rvacn
=\sum_{l\leq n-1} r^{n-l-1}\psistar_{l}\! \rvacn
=\left | n-1\right > +O(r)
$$
we see that in order to get a well-defined limit,
one should multiply the $\tau$-function by $r^{n-1}$
before tending $r\to 0$ (this is just a transformation
of the form (\ref{linear})). Then we obtain the $\tau$-function
of the form (\ref{pol100}):
$$
\tau_n ({\bf t})=\lvacn e^{J_+({\bf t})}\psi '(p)\left |n-1\right >
=\Bigl ( n-1+\sum_{k\geq 1}kt_k p^k\Bigr ) p^{n-1}e^{\xi ({\bf t}, p)}.
$$

This example can be further generalized.
Take $N$
operators $\Psi_{i}(p_i)$ 
of the form (\ref{I1}).
As in the previous example, we can construct the group-like
elements $\normord e^{\beta_j \Psi_{j}(p_j)\psistar (r_j)}\normord$  
with zero charge and 
consider the $\tau$-function 
$
\lvacn e^{J_+({\bf t})}
\normord e^{\beta_N \Psi_N(p_N) \psistar (r_N)} \normord 
 \ldots
\normord e^{\beta_1 \Psi_{1}(p_1) \psistar (r_1)} \normord 
 \rvacn
$.
Again, we choose
$\beta_j =(\lvac \Psi_{j}(p_j) \psistar (r_j)\rvac )^{-1}$,
then each operator becomes the product
$\beta_j \Psi_{j}(p_j)\psistar (r_j)$
(non-invertible). 
Thus the above $\tau$-function reduces to 
a $\tau$-function of the form 
$$
\beta_{1} \ldots \beta_{N}
 \lvacn e^{J_+({\bf t})}
\Psi_{1}(p_1) \ldots \Psi_{N}(p_N)
\psistar (r_N) \ldots \psistar (r_1)
 \rvacn .
$$
In order to implement the limit
$r_j \to 0$, we redefine $r_j \to \varepsilon r_j$ with
$\varepsilon \to 0$, then it is easy to verify that
\begin{equation}\label{I2}
\varepsilon^{(n-1)N-\frac{1}{2}\, N(N-1)}
\psistar (\varepsilon r_N)\, \ldots \, \psistar (\varepsilon r_1)
\rvacn = (r_1 \ldots r_N)^{-n+1}
\Delta_{N}(r_i) \left |n-N\right >  + 
O(\varepsilon )
\end{equation}
where
$\displaystyle{
\Delta_N (r_i)=\det\limits_{i,j=1, \ldots , N}
r_{i}^{j-1}=\prod_{i>j}(r_i -r_j)}
$
is the Vandermonde determinant. This formula
says that
in order to get a well-defined limit,
one should multiply the $\tau$-function by
$\varepsilon^{(n-1)N-\frac{1}{2}\, N(N-1)}$
before tending $\varepsilon \to 0$.
The $r_i $-dependent factors $(r_1 \ldots r_N)^{-n+1}\Delta_N (r_i)$
in the r.h.s. of (\ref{I2}) are irrelevant because they can be eliminated
by a transformation of the form (\ref{linear}).
We thus obtain the $\tau$-function
\begin{equation}\label{I3}
\tau_n ({\bf t})=\lvacn e^{J_+({\bf t})}\Psi_{1}(p_1)\ldots
\Psi_{N}(p_N) \left |n-N\right >
\end{equation}
of the form (\ref{pol100}).

{\bf Remark.} 
In this case the prescription (\ref{adddef}) is not directly
applicable.
The behavior of the function $\tau ({\bf t}\pm [z_{0}^{-1}])$
as $z_0 \to 0$ can be found
by using the bosonization formulas (\ref{bf3})
and moving the fermion operators
$\psi (z_0)$ or $\psistar (z_0)$ to the very right position.
In this way we obtain a more general
version of the prescription (\ref{adddef}):
\begin{equation}\label{adddef1}
\lim_{z_0\to 0}\left [
(-z_0)^{\pm N} \tau_n ({\bf t}\mp [z_{0}^{-1}])\right ]
=\tau_{n \pm 1} ({\bf t})
\end{equation}
which is equally enough to deduce the bilinear equation
(\ref{bi3}) from (\ref{bi201}) as a limiting case.

\subsubsection{Multi-soliton $\tau$-functions}

We begin with the most general soliton-like MKP $\tau$-function
$\tau_n ({\bf t})=\lvacn e^{J_+({\bf t})}G \rvacn$, where 
$G$ is given by 
\beq\label{sol1}
G=\normordbare \exp \Bigl ( \sum_{i,k=1}^{N} A_{ik}\psistar (q_i)
\psi (p_k)\Bigr )\normordbare
\eeq
with a nondegenerate matrix $A$ and $q_i, p_k \in \CC$ such that
$|q_i|>|p_j|$. We have:
$$
\tau_n ({\bf t})=1+\! \sum_{i,k}\! A_{ik}\! \lvacn e^{J_+}
\psistar (q_i)\psi (p_k)\rvacn +\frac{1}{2!}\!\!
\sum_{i,i', k,k'} \!\! A_{ik}A_{i'k'}\!
\lvacn e^{J_+}\psistar (q_i)\psistar (q_{i'})
\psi(p_{k'})\psi(p_k) \rvacn +\ldots
$$
Reorganizing the summation and using (\ref{ferm5}), we get:
\beq\label{sol2}
\tau_n ({\bf t})=1+\! \sum_{d=1}^{N}\! \sum_{{1\leq i_1<i_2<
\ldots <i_d \leq N\atop 1\leq k_1<k_2<
\ldots <k_d \leq N}}
\det_{1\leq r,s\leq d}\! A_{i_rk_s}\det_{1\leq r,s\leq d}
\Bigl ( \frac{q_{i_r}}{q_{i_r}\! -\! p_{k_s}}\Bigr ) 
\prod_{m=1}^d \Bigl (\frac{p_{i_m}}{q_{k_m}}\Bigr )^n 
\! e^{\xi ({\bf t}, p_{i_m})-
\xi ({\bf t}, q_{k_m})}.
\eeq
In this expression one can recognize determinant of the sum
of two matrices:
\beq\label{sol3}
\tau_n ({\bf t})=\det_{N\times N}\left (I+AQ^{n,{\bf t}}\right ),
\eeq
where the matrix $Q^{n,{\bf t}}$ is
\beq\label{sol4}
(Q^{n,{\bf t}})_{ik}=\frac{q_k}{q_k-p_i}\, 
\Bigl (\frac{p_i}{q_k}\Bigr )^n
e^{\xi ({\bf t}, p_{i})-
\xi ({\bf t}, q_{k})}.
\eeq

The standard $N$-soliton solutions of the 
MKP hierarchy are obtained in the case
when the matrix $A$ is diagonal:
\beq\label{sol5}
G=\normordbare \exp \Bigl ( \sum_{k=1}^{N} a_{k}\psistar (q_k)
\psi (p_k)\Bigr )\normordbare .
\eeq
In this case
\beq\label{sol6}
\tau_n ({\bf t})=\det_{N\times N}\left (\delta_{ik}+
\frac{a_k q_k}{q_k-p_i}\, (p_i/q_k)^ne^{\xi ({\bf t}, p_{i})-
\xi ({\bf t}, q_{k})}
\right ).
\eeq
More explicitly, this function has the form
\beq\label{sol7}
\tau_n ({\bf t})=1+\sum_{i}e^{\eta_i}+
\sum_{i<j}c_{ij}e^{\eta_i +\eta_j}+
\sum_{i<j<k}c_{ij}c_{ik}c_{jk}e^{\eta_i +\eta_j+\eta_k}+\ldots \,,
\eeq
where
$$
e^{\eta_i}=\frac{a_i q_i }{q_i \! -\! p_i}
\Bigl ( \frac{p_i}{q_i}\Bigr )^n
e^{\xi ({\bf t}, p_i)-\xi ({\bf t}, q_i)},\quad \quad
c_{ij}=\frac{(p_i-p_j)(q_i-q_j)}{(p_i-q_j)(q_i-p_j)}\,.
$$
The $\tau$-function of the $N$-soliton solutions to the 
2DTL hierarchy has a similar structure:
\beq\label{sol8}
\tau_n ({\bf t}_+, {\bf t}_-)=
e^{\sum_{k\geq 1}k t_k t_{-k}}
(1+\sum_{i}e^{\eta_i}+
\sum_{i<j}c_{ij}e^{\eta_i +\eta_j}+\ldots ),
\eeq
where now
$$
e^{\eta_i}=\frac{a_i q_i }{q_i \! -\! p_i}
\Bigl ( \frac{p_i}{q_i}\Bigr )^n
e^{\xi ({\bf t}_+, p_i)-\xi ({\bf t}_+, q_i)+
\xi ({\bf t}_-, p_i^{-1})-\xi ({\bf t}_-, q_i^{-1})}.
$$

There is another fermionic realization of the $N$-soliton solutions
to the MKP hierarchy.
Let us introduce the following fermionic
operators:
\begin{equation}\label{fermi1}
\Psi _i (p_i, q_i):= \psi (q_i)+b_i \psi (p_i).
\end{equation}
They are group-like elements with charge $+1$.
Therefore,
\begin{equation}\label{fermi2}
\tau_n({\bf t})=\lvacn e^{J_+({\bf t})} \Psi _1 (p_1, q_1)\ldots
\Psi _N (p_N, q_N) \! \left |\, n\! -\! N\right >
\end{equation}
is a $\tau$-function which
can be calculated 
using the Wick theorem in the form (\ref{Wick3aa}):
\begin{equation}
\begin{array}{lll}
\tau_n({\bf t})&=&\displaystyle{\det_{i,j=1, \ldots , N}
\left < n-j+1\right |
e^{J_+({\bf t})}\Psi_i(p_i, q_i)\left | n-j\right >}
\\ && \\
&=&\displaystyle{\det_{i,j=1, \ldots , N}
\left < n\! -\! j\! +\! 1\right |e^{J_+({\bf t})}\left (
\psi(q_i)+b_i \psi (p_i)\right )
\psistar_{n-j}\left | n\! -\! j\! +\! 1\right >}
\\ && \\
&=&\displaystyle{\det_{i,j=1, \ldots , N}
\left < n\! -\! j\! +\! 1\right | \left (
e^{\xi ({\bf t}, q_i)}\psi(q_i)+b_i e^{\xi ({\bf t}, p_i)}\psi (p_i)\right )
\psistar_{n-j}\left | n\! -\! j\! +\! 1\right >  }
\\ && \\
&=&\displaystyle{\det_{i,j=1, \ldots , N}
\left (e^{\xi ({\bf t}, q_i)}q_i^{n-j}+b_i e^{\xi ({\bf t}, p_i)}
p_i^{n-j} \right )}.
\end{array} 
\label{tau-fermi0}
\end{equation}
One can show that this determinant representation of the 
$N$-soliton solution is equivalent to (\ref{sol6}).

It should be noted that the multi-soliton solutions are 
degenerate cases of much more general algebro-geometric 
(or finite-gap) solutions associated with smooth 
complex algebraic curves (Riemann surfaces) according to the 
Krichever's construction. Their $\tau$-functions are 
essentially Riemann's theta-functions.
The fermionic representation of the algebro-geometric 
$\tau$-functions was addressed in \cite{Z89}.

\subsubsection{Partition functions of matrix models}\label{Pfmm}

\paragraph{The unitary matrix model.}
As the first example, let us calculate the $\tau$-function corresponding
to the group-like element ${\sf P}^+$ (see (\ref{PF1})):
$$
\tau_N ({\bf t}_+,  {\bf t}_-)
=\lvacN e^{J_+({\bf t}_{+})} {\sf P}^+ e^{-J_-({\bf t}_{-})}\rvacN .
$$
Here we follow \cite{KMMOZ91,Z10}.
First of all, it is
not difficult to see that
$\tau_N =0$ at $N<0$ and $\tau_0 =1$. For $N>1$ we have:
$$
\tau_N = \lvac \psistar_{N-1}\ldots \psistar_0 \, e^{J_+}{\sf P}^+
e^{-J_-}\psi_0 \ldots \psi_{N-1} \rvac .
$$
Below we use the short hand notation introduced in 
(\ref{short}). Using this notation, we can write
$$
\tau_N = \lvac e^{J_+} \psistar_{N-1}(-J_+)\ldots \psistar_0 (-J_+) 
{\sf P}^+
{\sf P}^+ \psi_0 (-J_-)\ldots \psi_{N-1}(-J_-) e^{-J_-} \rvac .
$$
Equations (\ref{ferm4}) imply that
the operators $\psi _n (-J_-)$, $\psistar _n (-J_+)$ in the
formula for $\tau_N$ contain only positive modes and, therefore,
commute with ${\sf P}^+$. Moving one ${\sf P}^+$ to the right and another
one to the left, and using the properties mentioned above,
we obtain
$$
\begin{array}{lll}
\tau_N &=&\displaystyle{
\lvac  \psistar_{N-1}(-J_+)\ldots \psistar_0 (-J_+)\,
\psi_0 (-J_-)\ldots \psi_{N-1}(-J_-)  \rvac }
\\ && \\
&=&\displaystyle{
\det_{1\leq j,k \leq N}\lvac \psistar_{j-1}(-J_+)\psi_{k-1} (-J_-)\rvac }
\end{array}
$$
by the Wick theorem. The expectation value under the determinant
can be represented as a contour integral as follows:
$$
\begin{array}{c}
\displaystyle{
\lvac \psistar_{j}(-J_+)\psi_{k} (-J_-)\rvac =
\sum_{a,b\geq 0}h_a ({\bf t}_+)h_b (-{\bf t}_-)\lvac \psistar_{j+a}\psi_{k+b}\rvac}
\\ \\
\displaystyle{ =\,
\sum_{a,b\geq 0}h_a ({\bf t}_+)h_b (-{\bf t}_-)\delta_{j+a, k+b}=
\oint_{|z|=1} z^{j-k} e^{\xi ({\bf t}_+, z)-\xi ({\bf t}_- , 1/z)}
\frac{dz}{2\pi i z}}\, .
\end{array}
$$
The whole determinant can then be written as an $N$-fold
contour integral:
\beq\label{unitary}
\tau_N = \frac{1}{N!}\oint \ldots \oint
\prod_{j<k}(z_j -z_k)(z_{j}^{-1}-z_{k}^{-1})
\prod_{l=1}^{N}e^{\xi ({\bf t}_+, z_l)-
\xi ({\bf t}_- , 1/z_l)}\frac{dz_l}{2\pi i z_l}\, .
\eeq
In accordance with subsection \ref{restsum} the Schur function expansion of this $\tau$-function
coincides with a restricted version of the expansion of the trivial $\tau$-function (\ref{id}):
$$
\tau_N ({\bf t}_+,  {\bf t}_-)=\sum_{l(\lambda)\leq N} s_\lambda({\bf t}_+)s_\lambda({\bf t}_-).
$$

When $t_{-k}=-\bar t_k$ (the bar means complex conjugation),
the expression $\xi ({\bf t}_+, z)-\xi ({\bf t}_- , 1/z)$ is purely real
for $z$ on the unit circle and
$\tau_N$ coincides with the partition function of the unitary
random matrix model written in terms of the eigenvalues.
In this form, it can be treated also as the partition function
of a canonical ensemble of $N$ 2D Coulomb particles confined
on a circle.

\paragraph{Partition functions of normal and Hermitian 
random matrices.}
Let us fix an arbitrary measure $d\mu (z)$ in the complex plane
and consider the following group-like element:
\beq\label{can1}
G_0 = \normord \exp \left (\int_{\CCC}\psi_+(z)\psistar _{+}(1/\bar z)
d\mu (z) -\sum_{j\geq 0} \psi_j \psistar_j \right )\normord
\eeq
Here $\displaystyle{\psi_+(z)= \sum_{n\geq 0}\psi_n z^n}$,
$\displaystyle{\psistar_+(z)= \sum_{n\geq 0}\psistar_n z^{-n}}$ are
truncated Fourier series containing only positive modes. Obviously,
$G_0$ commutes with ${\sf P}^+$. Expending the exponent into a series,
one can represent $G_0$ in a more explicit form:
\beq\label{can2}
G_0= \sum_{m=0}^{\infty} \frac{1}{m!}
\int_{\CCC ^m} \psi_+ (z_1) \ldots \psi_+(z_m){\sf P}^-
\psistar_+ (1/\bar z_m)\ldots \psistar_+ (1/\bar z_1)
d\mu_1 \ldots d\mu_m\,,
\eeq
where $d\mu_j \equiv d\mu (z_j)$ and the first term in the sum is ${\sf P}^-$.

Let us consider the expectation value
\beq\label{can3}
\tau_N ({\bf t}_+,{\bf t}_-)=\lvacN e^{J_+({\bf t}_+)}G_0 {\sf P}^+ 
e^{-J_-({\bf t}_-)}\rvacN
\eeq
and apply to it a chain of transformations similar to the ones
made in the simpler case $G_0=1$ previously considered. 
Again, $\tau_N =0$ at $N<0$ and $\tau_0 =1$.
For $N>1$ we have:
$$
\begin{array}{lll}
\tau_N &=& \lvac \psistar_{N-1}\ldots \psistar_0 \, e^{J_+}  G_0 {\sf P}^+
e^{-J_-}\psi_0 \ldots \psi_{N-1} \rvac
\\ && \\
&=&\lvac e^{J_+} \psistar_{N-1}(-J_+) \ldots \psistar_0 (-J_+) {\sf P}^+ 
G_0 {\sf P}^+
\psi_0 (-J_-)\ldots \psi_{N-1}(-J_-) e^{-J_-} \rvac
\\ && \\
&=&\lvac  \psistar_{N-1}(-J_+) \ldots \psistar_0 (-J_+)   G_0
\psi_0 (-J_-)\ldots \psi_{N-1}(-J_-)  \rvac .
\end{array}
$$
Substituting the explicit form of $G_0$, we get:
$$
\begin{array}{c}
\displaystyle{\tau_N =
\sum_{m\geq 0} \frac{1}{m!}
\int_{\CCC ^m} d\mu_1 \ldots d\mu_m
\lvac \psistar_{N-1}(-J_+) \ldots \psistar_0 (-J_+)
\psi_+ (z_1) \ldots \psi_+(z_m)}
\\ \\
\displaystyle{\times \,\, {\sf P}^-
\psistar_+ (1/\bar z_m)\ldots \psistar_+ (1/\bar z_1)
\psi_0 (-J_-)\ldots \psi_{N-1}(-J_-)  \rvac}.
\end{array}
$$
The next step is to notice that only the term with
$m=N$ contributes to the sum and all other terms vanish.
Indeed, at $m>N$ the state
$$
\psistar_+ (1/\bar z_m)\ldots
\psistar_+ (1/\bar z_1)
\psi_0 (-J_-)\ldots \psi_{N-1}(-J_-)  \rvac
$$
is in fact the null state because the number of annihilation
operators exceeds the number of creation operators while at
$m<N$ the operator
$$
{\sf P}^- \psistar_+ (1/\bar z_m)\ldots \psistar_+ (1/\bar z_1)
\psi_0 (-J_-)\ldots \psi_{N-1}(-J_-)
$$
is in fact the null operator because ${\sf P}^-$ multiplied by the
uncompensated positive $\psi$-modes from the right gives $0$
(see (\ref{ferm9a})). Therefore, the expression simplifies to
$$
\begin{array}{c}
\displaystyle{\tau_N =
 \frac{1}{N!}
\int_{\CCC ^N} d\mu_1 \ldots d\mu_N
\lvac \psistar_{N-1}(-J_+) \ldots \psistar_0 (-J_+)
\psi_+ (z_1) \ldots \psi_+(z_N)}
\\ \\
\displaystyle{\times \,\, {\sf P}^-
\psistar_+ (1/\bar z_N)\ldots \psistar_+ (1/\bar z_1)
\psi_0 (-J_-)\ldots \psi_{N-1}(-J_-)  \rvac}.
\end{array}
$$
Since there are as many annihilation operators
to the right of ${\sf P}^-$ as creation ones, the state
that they produce from the vacuum is proportional to
the vacuum state itself, i.e.,
$$
\psistar_+ (1/\bar z_N)\ldots
\psistar_+ (1/\bar z_1)
\psi_0 (-J_-)\ldots \psi_{N-1}(-J_-)  \rvac =\rvac C_N,
$$
where the constant $C_N$ is
$$
\begin{array}{lll}
C_N&=&\displaystyle{
\lvac  \psistar_+ (1/\bar z_N)\ldots
\psistar_+ (1/\bar z_1)
\psi_0 (-J_-)\ldots \psi_{N-1}(-J_-)  \rvac}
\\ && \\
&=&\displaystyle{
\det_{1\leq j,k \leq N}\lvac \psistar_{+}(1/\bar z_j)\psi_{k-1} (-J_-)\rvac }.
\end{array}
$$
Because
$$
\begin{array}{c}
\displaystyle{
\lvac \psistar_{+}(1/\bar z_j)\psi_{k-1} (-J_-)\rvac =
\sum_{a,l\geq 0}\bar z_j^l \,  h_a (-{\bf t}_-) 
\lvac \psistar_{l}\psi_{k+a-1}\rvac}
\\ \\
\displaystyle{ \phantom{aaaaaaaaaa}=\,
\sum_{a\geq 0} \bar z_j^{k+a-1} h_a (-{\bf t}_-)\, = \,
\bar z_j^{k-1}e^{-\xi ({\bf t}_- , \bar z_j)}},
\end{array}
$$
the constant $C_N$ is explicitly given by
\beq\label{CN}
C_N = \Delta_N (\bar z_i)\prod_{l=1}^{N}
e^{-\xi ({\bf t}_- , \bar z_l)},
\eeq
where we use the convenient short-hand notation for the
Vandermonde determinant:
$$
\Delta_N (z_i) =\det_{1\leq j,k \leq N}
\left ( z_{k}^{j-1}\right )=
\prod_{i>j} (z_i -z_j).
$$
Now, it remains to calculate
$$
\begin{array}{c}
\displaystyle{
\lvac \psistar_{N-1}(-J_+)
\ldots \psistar_0 (-J_+)
\psi_+ (z_1) \ldots \psi_+(z_N)\rvac
=\det_{1\leq j,k\leq N}\lvac \psistar_{j-1}(-J_+)\psi_+(z_k)\rvac}
\\ \\
\displaystyle{
=\, \det_{1\leq j,k\leq N} z_{k}^{j-1}e^{\xi ({\bf t}_+ , z_k)}\, =\,
\Delta_N (z_i)\prod _{l=1}^{N}
e^{\xi ({\bf t}_+ , z_l)}},
\end{array}
$$
which can be done in a completely similar manner. Collecting everything
together, we obtain the result:
\beq\label{can4}
\tau_N ({\bf t}_+, {\bf t}_-)= \frac{1}{N!}\int_{\CCC ^N}|\Delta_N
(z_i)|^2 \prod_{l=1}^{N}
e^{\xi ({\bf t}_+ , z_l )-\xi ({\bf t}_- , \bar z_l)} d\mu (z_l).
\eeq

Assume that $d\mu (z)=e^{-U(z, \bar z)}d^2z$
is a smooth measure in the plane and $t_{-k}=-\bar t_k$, then
the expression $\xi ({\bf t}_+, z)-\xi ({\bf t}_- , \bar z)$ 
is purely real
and the integral (\ref{can4}) has a physical interpretation as
the partition function
of a canonical ensemble of $N$ identical 2D Coulomb particles
in the plane in the background
potential $W(z, \bar z)=-U(z, \bar z)+2{\cal R}e \sum_{k}t_k z^k$:
\beq\label{ZN}
\begin{array}{l}\displaystyle{
Z_N =\frac{1}{N!}\int_{\CCC ^N}|\Delta_N
(z_i)|^2 \prod_{l=1}^{N}
e^{W(z_l, \bar z_l)}d^2 z_l}
\\ \\
\displaystyle{\,\,\, \quad \,\, =\det_{1\leq i,j\leq N}
\int_{\CCC}z^{i-1}\bar z^{j-1}e^{W(z, \bar z)}d^2z.}
\end{array}
\eeq
It is proportional to the partition function
of the ensemble of normal random $N\times N$ matrices $\Phi$:
$Z_N \propto \int D\Phi \, e^{\mbox{tr}\, W(\Phi, \Phi^{\dag})}$,
with $z_i$ being their eigenvalues.

If the measure $d\mu$ is concentrated on a curve $\Gamma \subset \CC$,
then the 2D integrals $\int_{\CCC}(\ldots )d^2z$ are reduced to
1D integrals $\int_{\Gamma}(\ldots )|dz|$ along $\Gamma$. This means
that the 2D Coulomb particles are confined to the curve $\Gamma$.
For particular choices of $\Gamma$ the integral (\ref{can4}) yields
the partition functions of random matrix models of certain types
in terms of eigenvalues.
For example, if $\Gamma$ is the real line, one obtains the
partition function of Hermitian random matrices:
\beq\label{herm}
\tau_N ({\bf t}_+, {\bf t}_-)= \frac{1}{N!}\int_{\RRR ^N}(\Delta_N
(x_i))^2 \prod_{l=1}^{N}
e^{\xi ({\bf t}_+ -{\bf t}_-, \, x_l )} d\mu (x_l).
\eeq
It is the $\tau$-function of the 1D Toda chain because it depends 
on the differences $t_k -t_{-k}$ only. 
If $\Gamma$ is the
unit circle, the integral (\ref{can4}) becomes identical to (\ref{unitary})
which is the partition function of unitary random matrices.

\paragraph{The Hermitean two-matrix model.} 
An expression (\ref{can1}) for the element $G_0$ admits a generalization 
with a double integral with an arbitrary contour (see \cite{KMMM,HO06} for details). For example
\beq\label{twomm}
G_0=\normord \exp \left (\int_{\RRR}\!\! \int_{\RRR}
e^{xy}\psi_{+}(x)\psistar _{+}(1/y) d\mu_1 (x)d\mu_2 (y)-
\sum_{j\geq 0}\psi_j \psistar_j\right )\normord ,
\eeq
gives, through the same formula (\ref{can3}),
the partition function of the Hermitean two-matrix model:
\beq\label{2m}
\tau_N ({\bf t}_+, {\bf t}_-)= \frac{1}{N!}\int_{\RRR ^{2N}}\Delta_N (x_i) 
\Delta_N (y_i)
\prod_{l=1}^{N}
e^{x_ly_l+
\xi ({\bf t}_+ , \, x_l )-\xi ({\bf t}_- ,  \, y_l)} d\mu_1 (x_l)
d\mu_2 (y_l).
\eeq

\paragraph{The Harish-Chandra-Itzykson-Zuber (HCIZ) matrix model.} 

For an element $G_0$ similar to (\ref{twomm}) but with 
different choice of the integration contour and the interaction term,
\beq\label{IZelem}
G_0=\normord \exp \left (\oint\!\! \oint
e^{c(vw)^{-1}}\!
\psi_{+}(v)\psistar _{+}(1/w) \frac{dv}{2\pi i v} \frac{dw}{2\pi i w}-
\sum_{j\geq 0}\psi_j \psistar_j\right )\normord ,
\eeq
one obtains the partition function for the model of two coupled 
unitary random matrices: 
\beq\label{HCIZc}
\tau_N ({\bf t}_+, {\bf t}_-)= \frac{1}{N!}\oint \! \ldots \! \oint \Delta_N(v_i)\Delta_N(w_i)\prod_{k=1}^N
e^{c(v_k w_k)^{-1}+\xi ({\bf t}_+ , \, v_k )-\xi ({\bf t}_- ,  \, w_k)}\,\! \frac{dv_k}{2\pi i v_k} \frac{dw_k}{2\pi i w_k}.
\eeq
After the change of variables 
$$
t_k(a)=\frac{1}{k} \sum_{m=1}^N a_m^k, \,\,\,\,\,\, 
t_{-k}(b)=-\frac{1}{k} \sum_{m=1}^N b_m^k, \,\,\,\,\,\,\,\,\,\,\,\,\,k>0 
$$ 
(the Miwa transform),
this integral becomes equal to the partition function of the 
HCIZ matrix model:
\beq\label{IZ}
\tau_N({\bf t}_+(a), {\bf t}_-(b))=c^{\frac{N(N-1)}{2}} 
\prod_{k=1}^{N}\frac{1}{\Gamma(k)}
\int_{N\times N} \left[D U\right] \exp (c\,\, \tr U A U^{\dagger} B) .
\eeq
This integral is known as the Harish-Chandra-Itzykson-Zuber  (HCIZ)
integral \cite{HCIZ1,HCIZ2}.
It is a $N\times N$ unitary matrix integral with the Haar measure 
normalized by the condition $ \int_{N\times N} \left[D U\right]=1$, $A$ 
and $B$ are two diagonal matrices: $A=\diag(a_1,a_2,\ldots,a_n)$, $B=\diag(b_1,b_2,\ldots,b_n)$
(see \cite{Morozov} for details).
The integral (\ref{IZ}) is known to be a solution to the 2DTL 
\cite{KMMM93,HCIZ}. 

\subsubsection{Matrix models represented by 
diagonal group-like elements}

The simplest example of matrix models with diagonal group-like 
element $G_0$ is given by the model of unitary matrices 
(\ref{unitary}), it corresponds to trivial $G_0=1$. 
The operator representation (\ref{can3}) of such models 
is equivalent to the one suggested by Orlov et al
\cite{OSch00}. In fact all models of normal matrices with
axially symmetric measures $d\mu$, i.e., such 
that $d\mu (z, \bar z)=d\mu (|z|^2)$ belong to this class. 
For an axially symmetric measure,
the bilinear form in the fermion operators in (\ref{can1}) becomes
diagonal:
$$
\int_{\CCC}\psi_+(z)\psistar _{+}(1/\bar z)
d\mu (z)=\sum_{m,n\geq 0}\psi_n \psistar_m
\int_{\CCC}z^n \bar z^m e^{-U(|z|^2)}d^2z =
\sum_{n\geq 0} g_n \psi_n \psistar_n,
$$
where
$$
g_n = \int_{\CCC} |z|^{2n}e^{-U(|z|^2)}d^2z =
\pi \int_{0}^{\infty}x^n e^{-U(x)}dx
$$
(we assume that the measure is smooth with
$U(z, \bar z)=U(|z|^2)$),
so in this case
\beq\label{can5}
G_0 =\normord \exp \left (\sum_{n\geq 0} (g_n \! -\! 1) \psi_n \psistar_n
\right ) \normord
= \, \exp \left (\sum_{n\geq 0} \log g_n \, \psi_n \psistar_n \right )
\eeq
and the $\tau$-function (\ref{can3}) does have the form
$\lvacN e^{J_+}e^X e^{-J_-}\rvacN$ with
$\displaystyle{X \! = \!\! \sum_{j\in \z}  X_j \normord
\psi_j \psistar_j \normord }$.
In this case expansion (\ref{tau1b}) is diagonal, i.e., 
$c_{\lambda \mu}(n)=0$ unless $\lambda =\mu$ \cite{OSch00}.
More precisely, in our case we deal with a singular
limit of the vacuum expectation value 
with $X_j =\log g_j$ for $j\geq 0$ and
$X_j \to +\infty$ for all $j<0$. Indeed, writing
$$
e^X = \prod_{j\geq 0}\left ( 1+( e^{X_j}-1)\psi_j \psistar_j\right )\cdot
\prod_{j< 0}\left ( 1+( e^{-X_j}-1)\psistar_j \psi_j \right ),
$$
we see that the first product is equal to
$\prod\limits_{j\geq 0} (1+( g_j -1)\psi_j \psistar_j)=G_0$ while the
limit of the second one is the singular operator ${\sf P}^+$.

Below we give three examples.
One important example is 
$
U(z, \bar z)=c|z|^2,
$
then 
\beq\label{gelnorm}
g_n =\pi c^{-n-1} n!
=\pi c^{-n-1}\Gamma (n+1)
\eeq
and $X_n = -(n+1)\log c +
\log \Gamma (n+1)$ (the common constant $\log \pi$ is irrelevant).
Note that the analytic continuation of this
formula to negative values of $n$ with the help of the
gamma-function automatically implies the required
singular limit
$X_n = +\infty$ at $n<0$. The $\tau$-function (\ref{can4})
for this case has the following expansion in Schur functions
\cite{OSch00,OShi05}:
\beq\label{can4ex}
\tau_N ({\bf t}_+ , {\bf t}_-)=\pi^N c^{-N(N+1)/2}
\prod_{k=1}^{N}\Gamma (k) \cdot
\sum_{\ell (\lambda)\leq N}c^{-|\lambda |}(N)_{\lambda}\,
s_{\lambda}({\bf t}_+)s_{\lambda}(-{\bf t}_-).
\eeq
where the factor $(N)_{\lambda}$ is
$\displaystyle{
(N)_{\lambda}:=\prod_{i=1}^{\ell (\lambda )}
(N+1 -i)(N+2-i) \ldots (N+\lambda _i -i)}
$
(see also (\ref{pochh1}) and Fig.~2 in Appendix A).

The second example is the HCIZ model, where the element 
$G_0$ given by (\ref{IZelem}) has the same structure with
$
g_n=c^n/n!
$
which, up to an 
inessential $n$-independent factor, is inverse 
to the one considered in the previous example (\ref{gelnorm}), 
so that the expansion of (\ref{HCIZc}) 
in Schur functions is as follows \cite{Orlov02,Orlov-IZ}:
\beq
\tau_N ({\bf t}_+ , {\bf t}_-)=c^{N(N-1)/2}
\prod_{k=1}^{N}\frac{1}{\Gamma (k)} \cdot
\sum_{\ell (\lambda)\leq N}\frac{c^{|\lambda |}}{(N)_{\lambda}}\,
s_{\lambda}({\bf t}_+)s_{\lambda}(-{\bf t}_-).
\eeq

Our third example is the model of normal matrices with
$$
U(z, \bar z)=\frac{1}{2\beta}
\left ( \log \left |z/r\right |\right )^2.
$$
Here $\beta , r$ are parameters ($r$ plays the role of a scale
in the $z$-plane). The corresponding matrix model 
was introduced in \cite{Al12} (see also \cite{AMMN11}).
In this case we find:
$$
g_n=\pi \int_{0}^{\infty} x^n e^{-\frac{1}{2\beta}
\left ( \log (x/r^2)\right )^2}dx =
\pi r^{2(n+1)}\!\! 
\int_{-\infty}^{+\infty}\!\!\! e^{-\frac{t^2}{2\beta}+
(n+1)t}dt = \pi \sqrt{2\pi\beta} \, r^{2(n+1)} 
e^{\beta (n+1)^2/2},
$$
so, up to an inessential common factor,
\beq\label{gohur}
G_0=\exp \left (\sum_{n\geq 0}\Bigl ( (\beta +2\log r )\, n+
\frac{\beta}{2}\, n^2 \Bigr ) \psi_n \psistar_n \right )
\eeq
(see (\ref{can5})). The Schur function expansion of the 
$\tau$-function (\ref{can3}) with this $G_0$ reads:
\beq\label{can7}
\tau_N ({\bf t}_+ , {\bf t}_- )=
e^{\frac{\beta}{12}N(N-1)(2N-1)}(re^{\beta /2})^{N(N-1)}
\sum_{\ell (\lambda)\leq N}e^{\beta C_{\lambda}/2}
\Bigl ( r^2 e^{\beta(N+\frac{1}{2})}\Bigr )^{|\lambda |}
s_{\lambda}({\bf t}_+)s_{\lambda}(-{\bf t}_-),
\eeq
where
$$
C_{\lambda}=\sum_{i=1}^{\ell (\lambda )}
\left ( \Bigl ( \lambda_i -i+\frac{1}{2}\Bigr )^2 -
\Bigl (-i +\frac{1}{2}\Bigr )^2 \right ).
$$

\subsubsection{Matrix models, cut-and-join-like 
operators and bosonization}

Sometimes for description of matrix models and related 
$\tau$-functions instead of using the group-like 
elements (\ref{can1})  it is more convenient to use 
other group-like elements. The reason is that the truncated 
Fourier series is not very convenient for bosonization, 
see section \ref{Boson}.

For example, according to Morozov and Shakirov \cite{Wop}, 
the partition function of the Gaussian branch of the 
Hermitian matrix model (\ref{herm}) (here we consider 
only one semi-infinite family of times ${\bf t =t_+-t_-}$):
\beq\label{hmm}
\tau_N ({\bf t})= \frac{1}{N!}\int_{\RRR ^N}(\Delta_N
(x_i))^2 \prod_{l=1}^{N}
e^{-\frac{x_l^2}{2}+\xi ({\bf t}, \, x_l )} d x_l
\eeq
can be represented in terms of action of the 
cut-and-join-like operator on the trivial $\tau$-function:
\beq\label{wopmsh}
\widetilde{\tau}_N({\bf t})=w^{-N}e^{\frac{1}{2}W_{-2}}w^N,
\eeq
where the operator
\beq
W_{-2}=\sum_{a,b=0}^\infty\left(abt_at_b \frac{\p}{\p t_{a+b-2}}+(a+b+2)t_{a+b+2}\frac{\p^2}{\p t_a \p t_b}\right)
\eeq
belongs to the $W^{(3)}$ algebra and, in accordance with (\ref{bf102}), $\frac{\p}{\p t_0} :=w \p_w$ .
More precisely, 
$$
\tau_N({\bf t}) = (2 \pi)^{\frac{N}{2}}\prod_{k=1}^{N}\Gamma (k) \cdot \widetilde{\tau}_N({\bf t})
$$
so that $\tau_N({\bf t})=0$ for negative $N$.

The boson-fermion 
correspondence allows us to rewrite
(\ref{wopmsh}) as a fermionic vacuum expectation value
$$
\widetilde{\tau}_N({\bf t})=\left<N\right| e^{J_+({\bf t})} 
\exp\Bigl (\frac{1}{2}W_{-2}^F\Bigr ) \left| N \right>
$$ 
with the group element given by exponential of bilinear fermionic operator
\beq
W_{-2}^F=\frac{1}{6\pi i }\oint \normordboson \left(\p \phi(z)\right)^3\! 
\normordboson \, dz=-
\mbox{res}_z \Bigl (z^{-1} \normord \psistar (z)\p_z^2 \psi (z)
\normord  \Bigr )=\sum_{k \in \z} k(k-1)\psi_k\psistar_{k-2}
\eeq
where we used, in particular, expansion (\ref{bosdecomp}). 
The matrix model $\tau$-function
(\ref{hmm}) can be obtained with the help of the projection operator:
$$
\tau_N ({\bf t})=\left<N\right| e^{J_+({\bf t})} \exp \Bigl (\frac{1}{2}W_{-2}^F\Bigr )
G_0{\sf P}^+ \left| N \right>,
$$ 
where $G_0$ is a diagonal operator of the 
form (\ref{can5}) with coefficients up to inessential factor
coinciding with (\ref{gelnorm}), namely $g_n=\sqrt{2\pi}\,\Gamma(n+1)$.

Similarly, the group-like element for the $\tau$-function (\ref{can7}) can be rewritten in a more symmetric way: instead of (\ref{gohur}) one can use $G=\exp(W_{C}^F)$
\beq
\begin{array}{l}\displaystyle{
W_C^F=\sum_{n \in \z}\Bigl ( (\beta +2\log r )\, n+
\frac{\beta}{2}\, n^2 \Bigr ) \normord \psi_n \psistar_n \normord}
\\ \\ 
\displaystyle{=
-\mbox{res}_z \Bigl (z^{-1} \normord \psistar 
(z)\left((\beta +2\log r )\, z\p_z+ 
\frac{\beta}{2}(z\p_z)^2\right) \psi (z)
\normord  \Bigr )}
\end{array}
\eeq
since, as it follows from (\ref{ferm9}), $G{\sf P}^+ =G_0{\sf P}^+$. 
For $Q=2\log(r)+\frac{\beta}{2}$ this group-like element gives
the $\tau$-function of the two-component KP hierarchy, namely
\beq\label{cajapp}
\tau ({\bf t}_+ , {\bf t}_- )= 
\left<0\right| e^{J_+({\bf t}_+)} \exp\left(\frac{\beta}{2}\,W_0^F\right) Q^{L_0^F}e^{J_-(-{\bf t}_-)} \left| 0 \right>
\eeq
where 
\beq\label{caj}
W_0^F=\frac{1}{6\pi i }\oint \normordboson \left(\p \phi(z)\right)^3\normordboson z^2 dz=-\mbox{res}_z \Bigl (z^{-1} \normord \psistar 
(z)\left( (z\p_z)^2+z\p_z+\frac{1}{3} \right) \psi (z)
\normord  \Bigr )
\eeq
and
\beq
L_0^F=\frac{1}{4\pi i }\oint \normordboson \left(\p \phi(z)\right)^2\normordboson z dz=-\mbox{res}_z \Bigl (z^{-1} \normord \psistar 
(z)\left( z\p_z+\frac{1}{2} \right) \psi (z)
\normord  \Bigr )
\eeq
is known to be equal to the generating function for double Hurwitz numbers
\cite{OSch00,Okounkov00,Takasaki12}.  
The operator (\ref{caj}) is the fermionic counterpart 
of the famous cut-and-join operator \cite{GJ}: 
$$
W_0=\sum_{a,b=0}^\infty\left(a\,b\, t_a\, t_b \frac{\p}{\p t_{a+b}}+(a+b)t_{a+b}\frac{\p^2}{\p t_a \p t_b}\right),
$$
(more precisely, the cut-and-join operator does not include derivatives with respect to $t_0$, which do not affect the result) so that (\ref{cajapp}) is equal to application of this operator to the simple tau-function (\ref{id}):
\beq
\tau ({\bf t}_+ , {\bf t}_- )=\exp\left(\frac{\beta}{2}W_0\right)Q^{L_0}\, \exp\left(-\sum_{k>0}k t_k t_{-k}\right)=\sum_{\lambda}e^{\beta C_{\lambda}/2}  Q^{|\lambda |}
s_{\lambda}({\bf t}_+)s_{\lambda}(-{\bf t}_-).
\eeq

A generalization of the Hurwitz $\tau$-function is given by 
the vacuum expectation value with the group element
$$
G=\exp \left(\sum_{k=0} \frac{x_k}{(k+1)!} {\mathcal P}_k \right),
$$
where 
$$
{\mathcal P}_k=\sum_{n \in \z} \left(n+\frac{1}{2}\right)^k \normord\psi_n \psistar_n\normord .
$$
It describes the stationary Gromov-Witten theory on $\CC {\rm P}^1$ 
relative to two poles \cite{OP02}.
2DTL $\tau$-functions of this type, their representations in terms of 
free fermions and their Schur function expansions 
were considered already in \cite{KMMM93}.

Integrable properties of those vacuum expectation values with respect to variables $x_k$ are considered in the next section.

\subsection{Another type of time evolution}

Here we consider another type of multi-time evolution
of group-like elements and associated $\tau$-functions.
This class of $\tau$-functions has appeared naturally in
such diverse physical and mathematical 
contexts as Gromov-Witten theory on $\CC {\rm P}^1$ \cite{OP02},
supersymmetric gauge theories
\cite{NO,MN07}, the melting crystal model of topological strings
\cite{ORV,meltingcrystal} and
hydrodynamic description of Fermi gas in one spatial dimension
\cite{BAW06,BAW08}.
Their natural fermionic realization requires introducing 
evolution operators other than $e^{J_{\pm}({\bf t})}$.

\subsubsection{The Hamiltonian evolution operator}

Let us introduce the bilinear operators 
\beq\label{ham1}
H_k=\sum_{n\in \z}n^k \normord \psi_n \psistar_n\normord =
-\mbox{res}_z \Bigl (z^{-1} \normord \psistar (z)(z\p_z)^k \psi (z)
\normord  \Bigr ), \quad k\geq 1,
\eeq
which we call {\it Hamiltonians} in contrast to the 
{\it currents} (\ref{Jk})
\beq\label{ham2}
J_k=\sum_{j\in \z} \normord \psi_j \psistar _{j+k}\normord
=-\mbox{res}_z \Bigl (z^{-1} \normord \psistar (z)z^k \psi (z)
\normord  \Bigr ), \quad k \in \ZZ.
\eeq
The operators $H_k$ commute for all $k\geq 1$. 
Introducing an
infinite set of evolution parameters 
${\bf T}=(T_1, T_2, \ldots )$, 
we unify the Hamiltonians
in the generating operator
\beq\label{ham3}
H({\bf T})=\sum_{k\geq 1}T_k H_k.
\eeq
In the physical interpretation,
$H_1$ is momentum of the system of
free non-relativistic fermions and $H_2$ is the energy operator.
Accordingly, for purely imaginary $T_i$'s
$X=iT_1$ and $T=iT_2$ are physical 
space and time variables \cite{BAW06}.
One may also consider 
$$
H_0 = \sum_{n\in \z} \normord \psi_n \psistar_n\normord
$$
which is the charge operator $Q$ (\ref{chargeoper}).

With the help of the simple identity
$
e^{\beta \psi_n \psistar_n }=1+(e^\beta -1)\psi_n \psistar_n 
$,
an easy calculation shows that
$$
e^{\beta \normordsmall \psi_n \psistar_n \normordsmall }
\, \psi_n \, e^{-\beta \normordsmall \psi_n \psistar_n \normordsmall }=
e^{\beta }\psi_n
$$
$$
e^{\beta \normordsmall \psi_n \psistar_n \normordsmall }
\, \psistar_n \, e^{-\beta \normordsmall \psi_n \psistar_n \normordsmall }=
e^{-\beta }\psistar_n
$$
and, therefore,
\beq\label{ham4}
\begin{array}{l}
e^{H({\bf T})}\psi_n e^{-H({\bf T})}=e^{\sum_{k\geq 1}T_k n^k}\psi_n
\\ \\
e^{H({\bf T})}\psistar_n e^{-H({\bf T})}=
e^{-\sum_{k\geq 1}T_k n^k}\psistar_n
\end{array}
\eeq
(cf. (\ref{ferm3})).
We call $e^{H({\bf T})}$ the Hamiltonian evolution operator.
Its adjoint action is thus diagonal on the fermionic modes 
$\psi_n$, $\psistar_n$. The action on $\psi (z)$ and $\psistar (z)$ 
is more complicated.
However, the action of the special 
operator $e^{T_1 H_1}$ is simple.
It shifts the argument of the series $\psi (z)$, $\psistar (z)$:
\beq\label{ham5}
\begin{array}{l}
e^{T_1H_1}\psi (z)  \, e^{-T_1H_1}=\psi (e^{T_1}z)
\\ \\
e^{T_1H_1}\psistar (z)  \, e^{-T_1H_1}=\psistar (e^{-T_1}z)
\end{array}
\eeq
which is in agreement with interpretation of $X=iT_1$ as
the physical space variable for the system on a circle.

From (\ref{ham5}) it is clear that the basis states 
(\ref{lambda1}) are eigenstates for the Hamiltonian evolution operator.
Namely, we have:
\beq\label{ham6}
\prod_{j\in \z}e^{b_j \normordsmall \psi_j \psistar_j  \normordsmall}
\left |\lambda , n\right >=c_n e^{B_{\lambda}(n)}
\left |\lambda , n\right >,
\eeq
where
\beq\label{ham7}
c_n =\left \{ \begin{array}{l}
e^{-b _{-1}-b _{-2}-\ldots -b _n}, \quad n<0
\\
1,  \quad \quad \quad \quad \quad \quad \quad n=0
\\
e^{b _{0}+b _{1}+\ldots + b _{n-1}}, \quad \quad \! n>0
\end{array}
\right.
\eeq
is the solution to the difference equation 
$c_{n+1}/c_{n}=e^{b _n}$ and
\beq\label{ham8}
B_{\lambda}(n)=\sum_{i=1}^{d(\lambda )}(b _{n+\alpha_i} 
- b_{n-\beta_i-1})=
\sum_{j\geq 1}(b_{n+\lambda_j -j} - b_{n-j}).
\eeq
Similar formulas hold for the left action to the 
dual basis vectors.

{\bf Remark.} The construction of the Hamiltonian evolution
operator admits a ``$q$-deformation'' related to the quantum torus
algebra and to the integrable structure of the 
melting crystal model \cite{meltingcrystal}.

\subsubsection{The $\tau$-function in ${\bf T}$-variables}

Given any group-like element $G$ (with zero charge), 
let us consider the expectation values
\beq\label{TT1}
\lvacn Ge^{H({\bf T})}
e^{J_{-}({\bf t})}\rvacn \quad \mbox{or} \quad 
\lvacn e^{J_{+}({\bf t})}e^{H({\bf T})}G\rvacn .
\eeq
At fixed ${\bf T}$, they are KP $\tau$-functions 
in the variables ${\bf t}$ (and MKP $\tau$-functions in 
${\bf t}, n$).
We are interested in the cases when they are $\tau$-functions 
in the variables ${\bf T}$. First of all,
it follows from formulas (\ref{ham6})
and their duals 
that at ${\bf t}=0$ or $G=1$ these expectation values are,
at each fixed $n$, exponents of a linear form in the variables 
${\bf T}$ and they thus represent the trivial $\tau$-function. 

In what follows we set $n=0$ and consider the expectation value
$\lvac e^{J_{+}({\bf t})}e^{H({\bf T})}G\rvac$. Using 
(\ref{tau401}) and (\ref{ham6}), (\ref{ham8}), we have:
$$
G\rvac =\sum_{\lambda}(-1)^{b(\lambda)}c_{\lambda}\left |
\lambda , 0\right >=
\sum_{d\geq 0}c_{\emptyset}^{1-d}
\!\!\!\!\sum_{{\alpha_1 >\alpha_2 >\ldots >\alpha_d\geq 0\atop
\beta_1 >\beta_2 >\ldots >\beta_d\geq 0}} (-1)^{\sum_i (\beta_i+1)}
\det_{1\leq r,s\leq d} c_{(\alpha_r |\beta_s)}
\left |\lambda , 0\right >,
$$
where $\lambda =(\alpha_1 , \ldots , \alpha_d|
\beta_1 ,\ldots , \beta_d )$, $c_{\emptyset}=\lvac G\rvac$ and
$$
e^{H({\bf T})}\left |\lambda , 0\right >=
\exp \Bigl (\sum_{k\geq 1}T_k
\sum_{l=1}^{d(\lambda )}
\bigl (\alpha_l^k -(-\! \beta_l \! -\! 1)^k\bigr )\Bigr )
\left |\lambda , 0\right >.
$$
Combining this with (\ref{lambda3}) and using the 
Giambelli determinant 
formula for $s_{\lambda}({\bf t})$, we can compute the 
expectation value:
$$\begin{array}{ll}
\lvac  e^{J_{+}({\bf t})}e^{H({\bf T})}G\rvac &=\,\,
\displaystyle{
\sum_{d\geq 0}c_{\emptyset}^{1-d}
\!\!\!\!\!\!\sum_{{\alpha_1 >\alpha_2 >\ldots >\alpha_d\geq 0\atop
\beta_1 >\beta_2 >\ldots >\beta_d\geq 0}}
\det_{1\leq r,s\leq d} c_{(\alpha_r |\beta_s)}
\det_{1\leq i,j\leq d}s_{(\alpha_i|\beta_j)}({\bf t})}
\\ &\\
&\displaystyle{\,\,\,\,\,\, \times 
\prod_{l=1}^{d}e^{\sum_{k\geq 1}T_k(\bigl 
(\alpha_l^k -(-\! \beta_l \! -\! 1)^k\bigr ).}}
\end{array}
$$
Now, following Orlov's idea \cite{Orlov03}, 
we are going to put ${\bf t}= a[w^{-1}]=
(aw^{-1}, \frac{a}{2}\, w^{-2}, \frac{a}{3}\, w^{-3}, \ldots )$
for any complex $a,w$
and use equation (\ref{hook6}) from Appendix A for the explicit 
value of
the Schur function $s_{(\vec \alpha |\vec \beta )}(a[w^{-1}])$.
We get:
$$\begin{array}{ll}
\lvac  e^{aJ_{+}([w^{-1}])}e^{H({\bf T})}G\rvac &=\,\,
\displaystyle{
\sum_{d\geq 0}c_{\emptyset}^{1-d}
\!\!\!\!\!\!\sum_{{\alpha_1 >\alpha_2 >\ldots >\alpha_d\geq 0\atop
\beta_1 >\beta_2 >\ldots >\beta_d\geq 0}}
\det_{1\leq r,s\leq d} c_{(\alpha_r |\beta_s)}
\det_{1\leq i,j\leq d}\Bigl ( \frac{\beta_j+1}{\alpha_i +
\beta_j +1}\Bigr )}
\\ &\\
&\displaystyle{\,\,\,\, \times 
\prod_{l=1}^{d}(-1)^{\beta_l+1}\frac{\Gamma (a+\alpha_l +1)
\Gamma (1-a+\beta_l )e^{\sum_{k\geq 1}T_k \bigl 
(\alpha_l^k -(-\! \beta_l 
\! -\! 1)^k\bigr )}}{aw^{\alpha_l+\beta_l+1}
\Gamma (\alpha_l+1)\Gamma (\beta_l+2)
\, \Gamma (a) \Gamma (-a)}}\,,
\end{array}
$$
where $\Gamma (x)$ is the gamma-function.
Let us compare the r.h.s. with the multi-soliton 
$\tau$-function (\ref{sol2}). In terms of the matrix
$$
A_{ik}= (-1)^k\frac{c_{(i-1, k-1)}}{\lvac G\rvac}\,
\frac{w^{-i-k+1}\Gamma (i+a)\Gamma (k-a)}{a\Gamma (i)\Gamma(k+1)
\, \Gamma (a) \Gamma (-a)}
$$
the former can be written as 
$$
\begin{array}{ll}\displaystyle{
\frac{\lvac  
e^{aJ_{+}([w^{-1}])}e^{H({\bf T})}G\rvac}{\lvac G\rvac}} &=\,\,
\displaystyle{
\sum_{d=0}^{\infty}
\!\!\sum_{{1\leq i_1 <i_2 <\ldots <i_d\atop
1\leq k_1 <k_2 <\ldots <k_d}}
\det_{1\leq r,s\leq d} A_{i_rk_s}\det_{1\leq r,s\leq d}
\Bigl (\frac{k_s}{i_r\! -\! 1\! +\! k_s}\Bigr )}
\\&\\
&\displaystyle{\,\, \times 
\prod_{m=1}^{d} \exp \Bigl (\sum_{j\geq 1}T_j \bigl (
(i_m \! -\! 1)^j -(-k_m)^j\bigr )\Bigr ).
}
\end{array}
$$
After the identification $p_i= i-1$, $q_i=-i$, $i\geq 1$,
this expression formally looks like the $\infty$-soliton 
$\tau$-function (\ref{sol2}). More precisely, we have:
$$
\frac{\lvac  
e^{aJ_{+}([w^{-1}])}e^{H({\bf T})}G\rvac}{\lvac G\rvac} =
\lvac e^{J_+({\bf T})}
\normordbare \exp \Bigl ( \sum_{i,k\geq 1} A_{ik}
\psistar (-k)\psi (i\! -\! 1) \Bigr ) \normordbare \rvac .
$$
The form of the r.h.s. explicitly proves that 
the l.h.s. is the $\tau$-function of the KP
hierarchy in the ${\bf T}$-variables. We do not know what is
the most general choice of the times ${\bf t}$ such that 
the vacuum expectation value $\lvac e^{J_+({\bf t})}
e^{H({\bf T})} G \rvac$ is a $\tau$-function in ${\bf T}$-variables
for any group-like $G$.

{\bf Remark.} As is argued in \cite{BAW08}, a more general form
of the $\tau$-function in ${\bf T}$-variables is
$$
\tau ({\bf T}, {\bf T'})=\lvac G' V_{-a}(w,{\bf T})
V_{a}(z,{\bf T'})G\rvac ,
$$
where $G, G'$ are group-like elements and 
$$
V_{a}(z,{\bf T})=e^{-H([{\bf T}])}\, \normordboson
e^{a\phi (z)}\normordboson \,\, e^{H([{\bf T}])}.
$$
This fact was established in \cite{BAW08} using some 
field-theoretical arguments. 
Note that at $G'=1$ and ${\bf T'}=0$
the statement reduces to the example above after the redefinition
$G\to \normordboson
e^{a\phi (z)}\normordboson G$. However, we do not know how to prove
the more general statement by the same method or by operator 
methods developed in \cite{meltingcrystal}.


\newpage
\section*{Appendix A: Young diagrams and Schur functions}
\label{schur}
\addcontentsline{toc}{section}{Appendix A}
\def\theequation{A\arabic{equation}}
\def\theHequation{\theequation}
\setcounter{equation}{0}

\begin{figure}
\begin{center}
\begin{tikzpicture}[scale=0.8,line width=1pt]
\fill[brown!60!] (0,0) -- (1,0) -- (1,-1) -- (0,-1);
\fill[brown!60!] (1,-1) -- (2,-1) -- (2,-2) -- (1,-2);
\fill[brown!60!] (2,-2) -- (3,-2) -- (3,-3) -- (2,-3);
\draw (0,0) -- (9,0) -- (9,-1) -- (7,-1) -- (7,-2) -- (6,-2) -- (6,-3) -- (3,-3) -- (3,-4)  -- (2,-4) -- (2,-5) -- (1,-5) --
(1,-7) -- (0,-7) -- cycle;
\draw (0,-1) -- (9,-1);
\draw (0,-2) -- (7,-2);
\draw (0,-3) -- (6,-3);
\draw (0,-4) -- (3,-4);
\draw (0,-5) -- (2,-5);
\draw (0,-6) -- (1,-6);
\draw (0,-7) -- (1,-7);
\draw (0,0) -- (0,-7);
\draw (1,0) -- (1,-7);
\draw (2,0) -- (2,-5);
\draw (3,0) -- (3,-4);
\draw (4,0) -- (4,-3);
\draw (5,0) -- (5,-3);
\draw (6,0) -- (6,-3);
\draw (7,0) -- (7,-2);
\draw (8,0) -- (8,-1);
\draw (9,0) -- (9,-1);
\draw (10.2,-0.5) node {$\alpha_1=8$};
\draw (8.2,-1.5) node {$\alpha_2=5$};
\draw (7.2,-2.5) node {$\alpha_3=3$};
\draw (0.5,-7.5) node {$\beta_1=6$};
\draw (1.8,-5.5) node {$\beta_2=3$};
\draw (2.8,-4.5) node {$\beta_3=1$};
\end{tikzpicture}
\end{center}
\caption{The Frobenius notation. The Young diagram $\lambda=(9,7,6,3,2,1,1)=(8,5,3|6,3,1)$.}
\end{figure}
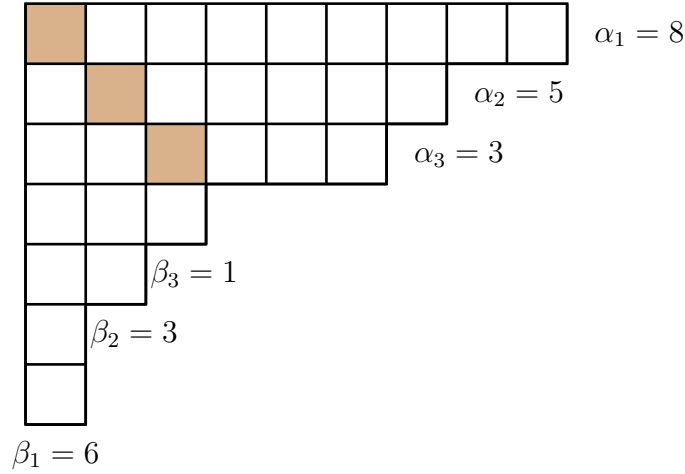

We use the notation of \cite{Macdonald}.

\paragraph{Partitions and diagrams.}
A partition $\lambda =
(\lambda_1 , \lambda_2, \ldots , \lambda_{\ell})$ 
is a sequence of positive integer numbers 
$\lambda_i$ such that $\lambda_1 
\geq \lambda_2 \geq \ldots \geq \lambda_{\ell}$.
The number $\ell =\ell (\lambda )$ is called the length 
of the partition.
The partitions are naturally represented by Young diagrams.
The Young diagram of $\lambda$ 
is a table whose $j$-th row (counting from 
the top) consists of $\lambda_j$ boxes. 
We will identify partitions with diagrams and 
will denote the diagram of $\lambda$ by the same symbol $\lambda$.
The total number of
boxes in the diagram $\lambda$ is
$\displaystyle{|\lambda |=\sum_{i=1}^{\ell}\lambda_i}$.
The empty diagram is denoted by $\emptyset$.

By $\lambda '$ we denote the transposed Young diagram 
which is obtained from $\lambda$ by reflection in the main diagonal.
Namely, $\lambda'_j$ is the height of the $j$-th column 
of $\lambda$.

Given a Young diagram $\lambda =
(\lambda_1 , \ldots , \lambda_{\ell})$ with $\ell =\ell (\lambda )$
nonzero rows, let
$(\vec \alpha |\vec \beta )=(\alpha_1, \ldots , \alpha_{d}|
\beta_1 , \ldots , \beta_{d})$ be the Frobenius notation
for the diagram $\lambda$. Here $d=d(\lambda )$ is the number of
boxes in the main diagonal and $\alpha_i =\lambda_i -i$,
$\beta_i =\lambda'_i -i$.  In other words,
$\alpha_i$ is the lenght of the part of the
$i$-th row to the right from the
main diagonal and $\beta_i$ is the length of the part of the $i$-th column
under the main diagonal (not counting the diagonal box).
Clearly, $\alpha_1 > \alpha_2 >\ldots > \alpha _d\geq 0$,
$\beta_1 > \beta_2 >\ldots > \beta _d\geq 0$.
If $\lambda =(\vec \alpha |\vec \beta )$, then 
$\lambda '=(\vec \beta |\vec \alpha )$.
Note that 
$$
\sum_{i=1}^{d}(\alpha_i +\beta_i ) +d =|\lambda |.
$$

A box $x\in \lambda$ has coordinates 
$(i,j)$ if it is in the $i$-th line (from the top) and $j$-th 
column (from the left). Let $u\in \CC$. 
One can define the generalized 
Pochhammer symbol $(u)_{\lambda}$ as
\beq\label{pochh1}
(u)_{\lambda}=\prod_{(i,j)\in \lambda}
(u+j-i)=
(u)_{\lambda_1}(u-1)_{\lambda_2}\ldots 
(u-\ell (\lambda )+1)_{\lambda_{\ell (\lambda )}},
\eeq
where $(u)_k =u(u+1)\ldots (u+k-1)$ is the usual 
Pochhammer symbol (see Fig.~1). 
In the Frobenius notation, we have:
\beq\label{pochh2}
(u)_{(\vec \alpha |\vec \beta )}=\prod_{i=1}^{d(\lambda )}
(-1)^{\beta_i}(u)_{\alpha_i +1}(1-u)_{\beta_i}.
\eeq

\begin{figure}
\begin{center}
\begin{tikzpicture}[scale=1,line width=1pt]
\draw (0,0) -- (5,0) -- (5,-1) -- (4,-1) -- (4,-2) -- (1,-2) -- (1,-4) -- (0,-4) -- cycle;
\draw (0,-1) -- (5,-1);
\draw (0,-2) -- (4,-2);
\draw (0,-3) -- (1,-3);
\draw (0,-4) -- (1,-4);
\draw (1,0) -- (1,-4);
\draw (2,0) -- (2,-2);
\draw (3,0) -- (3,-2);
\draw (4,0) -- (4,-2);
\draw (5,0) -- (5,-1);
\draw (0.5,-0.5) node {$u$};
\draw (1.5,-0.5) node {$u\! +\! 1$};
\draw (2.5,-0.5) node {$u\! +\! 2$};
\draw (3.5,-0.5) node {$u\! +\! 3$};
\draw (4.5,-0.5) node {$u\! +\! 4$};
\draw (0.5,-1.5) node {$u\! -\! 1$};
\draw (1.5,-1.5) node {$u$};
\draw (2.5,-1.5) node {$u\! +\!1$};
\draw (3.5,-1.5) node {$u\! +\! 2$};
\draw (0.5,-2.5) node {$u\! -\! 2$};
\draw (0.5,-3.5) node {$u\! -\! 3$};
\end{tikzpicture}
\end{center}
\caption{The Young diagram $\lambda=(5,4,1,1)=(4,2|3,0)$.
The generalized Pochhammer symbol $(u)_{\lambda}$ is 
the product of the contents of all boxes.}
\end{figure}

\paragraph{Schur functions.}
Let ${\bf t}=\{t_1, t_2, t_3, \ldots \}$ be an infinite set
of parameters.
The Schur polynomials $s_{\lambda}({\bf t})$
labeled by Young diagrams $\lambda$ can be defined
by the determinant formula
\beq\label{T1}
s_{\lambda}({\bf t})=\det_{i,j=1, \ldots , \lambda_1'}
h_{\lambda_i -i +j}({\bf t}),
\eeq
where the polynomials $h_j$ are defined with the help of
the generating series
$$
\exp \Bigl ( \sum_{k\geq 1}t_k z^k\Bigr )=
1+ h_1({\bf t})z +h_2({\bf t})z^2 +\ldots
$$
or, explicitly,
$$
h_k ({\bf t})=\sum_{k_1+2k_2+\ldots =k}
\frac{t_1^{k_1}}{k_1!}\, \frac{t_2^{k_2}}{k_2!}\ldots=\sum_{m=1}^{k} \frac{1}{m!} \sum_{{k_1, \ldots , k_m\geq 1}\atop {k_1+\ldots +k_m=k}}
t_{k_1}t_{k_2}\ldots t_{k_m},
\quad k>0.
$$
For example,
$$
\begin{array}{l}
h_1({\bf t})=t_1\,,
\\ \\
h_2({\bf t})=\frac{1}{2}\, t_1^2 +t_2\,,
\\ \\
h_3({\bf t})=\frac{1}{6}t_1^3 +t_1t_2 +t_3\,,
\\ \\
h_4({\bf t})=\frac{1}{24}t_1^4 +\frac{1}{2}\, t_2^2 +
\frac{1}{2}\, t_1^2 t_2 +t_1t_3 +t_4\,.
\end{array}
$$
It is convenient to put $h_0({\bf t})=1$, $h_{k}({\bf t})=0$ for $k<0$
and $s_{\emptyset}({\bf t})=1$. 
From the definition it follows that $\p_{t_n}h_k({\bf t})=
h_{k-n}({\bf t})$.
The functions $h_k$ are elementary Schur polynomials in the sense that
for 1-row diagrams $\lambda =(j)$ with $j$ boxes
$s_{(j)}({\bf t})=h_j ({\bf t})$. Equivalently, one can define
\beq\label{T2}
s_{\lambda}({\bf t})=\det_{i,j=1, \ldots , \lambda_1}
e_{\lambda'_i -i +j}({\bf t}),
\eeq
where the polynomials $e_j$ are defined with the help of
the generating series
$$
\exp \Bigl (\sum_{k\geq 1}(-1)^{k-1}t_k z^k\Bigr )=
1+ e_1({\bf t})z +e_2({\bf t})z^2 +\ldots
$$
It is clear that
$$
e_j({\bf t})=(-1)^{j}h_j(-{\bf t}).
$$
For 1-column diagrams $\lambda =(1^j)$ with $j$ boxes
$s_{(1^j)}({\bf t})=e_j ({\bf t})$. Equations (\ref{T1}),
(\ref{T2}) are known as Jacobi-Trudi formulas.
It can be proved \cite{Macdonald} that the Schur
polynomials form a basis in the space of symmetric functions
of the variables $x_i$ defined by $kt_k =\sum_i x_i^k$.
In terms of the variables $x_i$ ($i=1,\ldots , N$),
$$
s_{\lambda}({\bf t})=\frac{\det _{1\leq i,j\leq N}
\Bigl ( x_{i}^{N+\lambda_j -j}\Bigr )}{\det _{1\leq i,j\leq N}
\Bigl ( x_{i}^{N -j}\Bigr )}\,.
$$
We note that
$$
s_{\lambda}({\bf t})=(-1)^{|\lambda |}s_{\lambda '}(-{\bf t}).
$$
It immediately follows form the definition that if 
$t_k \rightarrow a^k t_k$, then $h_k \rightarrow a^kh_k$ and
$s_{\lambda}\rightarrow a^{|\lambda |}s_{\lambda}$ (the quasihomogeneity
property).

The following formulas for Schur functions for hook diagrams
$\lambda = (\alpha |\beta )=(\alpha +1, 1^{\beta})$ directly follow
from the Jacobi-Trudi formulas:
\beq\label{hook1}
\begin{array}{lll}
s_{(\alpha |\beta )}({\bf t})&=&
\displaystyle{(-1)^{\beta +1}
\sum_{m=0}^{\alpha}h_{\alpha -m}({\bf t})h_{\beta +m+1}(-{\bf t})}
\\ &&\\
&=&\displaystyle{(-1)^{\beta}
\sum_{m=0}^{\beta}h_{\beta -m}(-{\bf t})h_{\alpha +m+1}({\bf t})}.
\end{array}
\eeq
The Giambelli formulas represent the Schur function for 
an arbitrary diagram $\lambda =(\vec \alpha |\vec \beta )$ 
as a determinant of the hook ones:
\beq\label{hook2}
s_{\lambda}({\bf t})=\det _{1\leq i,j\leq d(\lambda )}
s_{(\alpha _i |\beta_j)}({\bf t}).
\eeq

\paragraph{The Cauchy-Littlewood identity.}
We note the Cauchy-Littlewood
identity
\beq\label{id}
\sum_{\lambda}s_{\lambda}({\bf t})s_{\lambda}({\bf t}')=
\exp \Bigl ( \sum_{k\geq 1}kt_k t'_k\Bigr ),
\eeq
where the sum is over all Young diagrams including the empty one.
Writing it in the form
$$
\sum_{\lambda}s_{\lambda}({\bf y})s_{\lambda}(\tilde \p )=
\exp \Bigl ( \sum_{k\geq 1}y_k \p_{t_k}\Bigr ),
$$
where $\tilde \p =\{\p_{t_1} , \frac{1}{2}\p_{t_2}, \frac{1}{3}\p_{t_3},
\ldots \, \}$
and applying to $s_{\mu}({\bf t})$, we get the relation
\beq\label{id1}
\left. \phantom{\int}
s_{\lambda}(\tilde \p )s_{\mu}({\bf t})\right |_{{\bf t}=0}=
\delta_{\lambda \mu}
\eeq
which reflects the orthonormality of the Schur functions.

\paragraph{Skew Schur functions.}
Let $\mu \subset \lambda$ be two Young diagrams.
The skew Schur functions (or the Schur functions for the 
skew diagram $\lambda \setminus \mu$) are defined as
\beq\label{skew1}
s_{\lambda \setminus \mu} ({\bf t})=
\sum_{\nu}c_{\mu \nu}^{\lambda}s_{\nu}({\bf t}),
\eeq
where the Littlewood-Richardson
coefficients $c_{\mu \nu}^{\lambda}$ are determined by
$$
s_{\mu}({\bf t})s_{\nu}({\bf t})=\sum_{\lambda}
c_{\mu \nu}^{\lambda}s_{\lambda}({\bf t}).
$$
One also has \cite{Macdonald}
$$
s_{\lambda}({\bf t}+{\bf t}')=\sum_{\mu}
s_{\lambda \setminus \mu} ({\bf t})s_{\mu}({\bf t}')=
\sum_{\mu , \nu}c_{\mu \nu}^{\lambda}s_{\mu}({\bf t}')
s_{\nu}({\bf t})
$$
and, as an easy consequence,
$$
s_{\lambda \setminus \mu} ({\bf t})=
s_{\mu}(\tilde \p )s_{\lambda}({\bf t}).
$$
There are generalized Jacobi-Trudi formulas for the 
skew Schur functions \cite{Macdonald}:
\beq\label{skew2}
s_{\lambda \setminus \mu} ({\bf t})=
\det_{1\leq i,j\leq \ell (\lambda )}
h_{\lambda_i -\mu_j -i+j} ({\bf t}) =
\det_{1\leq i,j\leq \ell (\lambda ' )}
e_{\lambda'_i -\mu'_j -i+j}({\bf t}).
\eeq

\paragraph{Hook lengths.} 
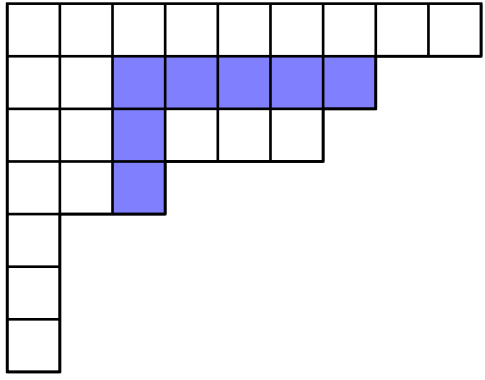
\begin{figure}
\begin{center}
\begin{tikzpicture}[scale=0.7,line width=1pt]
\fill[blue!50!] (2,-1) -- (7,-1) -- (7,-2) -- (2,-2);
\fill[blue!50!] (2,-2) -- (3,-2) -- (3,-4) -- (2,-4);
\draw (0,0) -- (9,0) -- (9,-1) -- (7,-1) -- (7,-2) -- (6,-2) -- (6,-3) -- (3,-3) -- (3,-4) -- (1,-4) --
(1,-7) -- (0,-7) -- cycle;
\draw (0,-1) -- (9,-1);
\draw (0,-2) -- (7,-2);
\draw (0,-3) -- (6,-3);
\draw (0,-4) -- (3,-4);
\draw (0,-5) -- (1,-5);
\draw (0,-6) -- (1,-6);
\draw (0,-7) -- (1,-7);
\draw (0,0) -- (0,-7);
\draw (1,0) -- (1,-7);
\draw (2,0) -- (2,-4);
\draw (3,0) -- (3,-4);
\draw (4,0) -- (4,-3);
\draw (5,0) -- (5,-3);
\draw (6,0) -- (6,-3);
\draw (7,0) -- (7,-2);
\draw (8,0) -- (8,-1);
\draw (9,0) -- (9,-1);
\end{tikzpicture}
\end{center}
\caption{The hook length $h(2,3)=7$.}
\end{figure}

The hook length at $x=(i,j)$ is defined as
$$
h(x)=h(i,j)=\lambda_i +\lambda '_j -i-j+1
$$
(see Fig.~2).
An important quantity is the product of all hook lengths
of a diagram
\beq\label{hook3}
H_{\lambda}=\prod_{x\in \lambda}h(x).
\eeq
In particular, for a hook $(\alpha |\beta )$ we have
\beq\label{hook4}
H_{(\alpha |\beta )}=\alpha ! \, \beta ! \, (\alpha \! +\! \beta \! +\! 1).
\eeq

Set ${\bf t}_{*}=(1, 0, 0, \ldots )$, $[1]=
(1, \frac{1}{2}, \frac{1}{3}, \ldots )$, then 
\beq\label{hook5}
s_{\lambda}({\bf t}_{*})=\frac{1}{H_{\lambda}}\,, 
\quad \quad
s_{\lambda}(u[1])=
\frac{(u)_{\lambda}}{H_{\lambda}}
\eeq
and, more generally,
\beq\label{hook7}
s_{\lambda}(u[w^{-1}])=
\frac{w^{-|\lambda |}(u)_{\lambda}}{H_{\lambda}}
\eeq
(see \cite{Macdonald} for the proofs). 
Using (\ref{hook7}), (\ref{hook4}) and the Giambelli formula
together with the Cauchy determinant formula
$$
\det _{1\leq i,j\leq d}\frac{1}{\alpha_i +\beta _j+1}
=\frac{\prod_{j<j'}^d (\alpha_j -\alpha_{j'})
(\beta_j-\beta_{j'})}{\prod_{k,l=1}^{d}
(\alpha_k +\beta_l +1)},
$$
we get
\beq\label{hook6}
s_{(\vec \alpha |\vec \beta )}(u[w^{-1}])=
\prod_{i=1}^{d}\left (\frac{ (-1)^{\beta_i} \Gamma (u+\alpha_i +1)
\Gamma (1-u+\beta_i )}{\Gamma (\alpha_i +1)\Gamma (\beta_i +1)
\Gamma (u)\Gamma (1-u)}\right ) \cdot 
\frac{\prod_{j<j'}^d (\alpha_j -\alpha_{j'})
(\beta_j-\beta_{j'})}{w^{|\lambda |}\prod_{k,l=1}^{d}
(\alpha_k +\beta_l +1)}\,.
\eeq

Following \cite{BAW06}, we show how to prove (\ref{hook7}) for 
hook diagrams $(\alpha |\beta )$ using the fermionic operators.
We start with
$$
\begin{array}{c}
s_{(\alpha |\beta )}({\bf t}) =(-1)^{\beta +1}
\lvac e^{J_+({\bf t})}\psistar_{-\beta -1}\psi_{\alpha}\rvac
\\ \\ \displaystyle{
=\frac{(-1)^{\beta +1}}{(2\pi i)^2}
\oint_{0}\frac{dz}{z}\oint_{0}\frac{d\zeta}{\zeta}
z^{-\alpha}\zeta^{-\beta -1}\lvac e^{J_+({\bf t})} \psistar (\zeta )
\psi (z)\rvac },
\end{array}
$$
where the integrals are over small contours around $0$.
From this it is clear that
$$
(\alpha +\beta +1)s_{(\alpha |\beta )}({\bf t}) 
=\frac{(-1)^{\beta +1}}{(2\pi i)^2}
\oint_{0}\frac{dz}{z}\oint_{0}\frac{d\zeta}{\zeta}
z^{-\alpha}\zeta^{-\beta -1}
(z\p_z +\zeta \p_{\zeta})\lvac e^{J_+({\bf t})} \psistar (\zeta )
\psi (z)\rvac .
$$
The vacuum expectation value under the integral 
at ${\bf t}=u[w^{-1}]$, i.e., $t_k=uw^{-k}/k$ is
$$
\lvac e^{J_+(u[w^{-1}]) } \psistar (\zeta )
\psi (z)\rvac =\left (\frac{w-\zeta }{w-z}\right )^u 
\frac{\zeta}{\zeta -z}\,.
$$
Note that 
$$
(z\p_z +\zeta \p_{\zeta})\left [\left (\frac{w-\zeta }{w-z}\right )^u 
\frac{\zeta}{\zeta -z}\right ]=-uw\zeta
(w-z)^{-u-1} (w-\zeta )^{u-1}
$$
factorizes, so the double integral becomes a product of two
ordinary integrals:
$$
\begin{array}{c}
(\alpha +\beta +1)s_{(\alpha |\beta )}(u[w^{-1}]) 
\\ \\
=\, \displaystyle{
\frac{(-1)^{\beta}u}{w}
\oint_{0}\frac{dz}{2\pi i z}\, 
z^{-\alpha}\left (1-\frac{z}{w}\right )^{-u-1}
\oint_{0}\frac{d\zeta}{2\pi i \zeta}\,\zeta^{-\beta}
\left (1-\frac{\zeta}{w}\right )^{u-1}}
\\ \\
=\, \displaystyle{
\frac{(-1)^{\beta}(u)_{\alpha +1}(1-u)_{\beta}}{\alpha ! \,
\beta ! \, w^{\alpha +\beta +1}}}
\end{array}
$$
which is (\ref{hook7}) for the hook $\lambda =(\alpha |\beta )$.

\section*{Appendix B: Proof of Proposition \ref{dcharge}}
\label{chargepr}
\addcontentsline{toc}{section}{Appendix B}
\def\theequation{B\arabic{equation}}
\def\theHequation{\theequation}
\setcounter{equation}{0}

Let $G$ be a solution of (\ref{commute}) consisting of parts with various charges: 
$$G=\sum_{k} G_k,$$ 
where the charge of $G_k$ is $k$. Assume that for some $p$ and $q$ 
with $p-q=a>0$ both $G_p$ and $G_q$ are non-trivial. 
Then from (\ref{commute}) it follows that
\begin{equation}\label{commute1}
\displaystyle{\sum_{k \in {\mathbb Z}} \psi_{k} G_q 
\otimes  \psi_{k}^{*} G_p =
\sum_{k \in {\mathbb Z}}G_q\psi_{k} \otimes G_p \psi_{k}^{*}}.
\end{equation}
Let us define two functions:
\begin{equation}\label{tauanalogs}
\begin{array}{c}
\displaystyle{Z_n({\bf t_+},{\bf t_-})=\left < n\! +\! p \right| e^{J_+({\bf t_+})} G_p e^{J_-({\bf t_-})} \left | n \right >},
\\ \\
\displaystyle{W_n({\bf t_+},{\bf t_-})=\left < n\! +\! q \right| e^{J_+({\bf t_+})} G_q e^{J_-({\bf t_-})} \left | n \right >}.
\end{array}
\end{equation}
To prove that at least one of $G_p$ and $G_q$ is 
identically equal to zero it is enough to show that for all $m$ and $m'$ (see Proposition \ref{unidec}):
\beq\label{contr}
W_m ({\bf t}_+ , {\bf t}_-)\,
Z_{m'} ({\bf t}_+ , {\bf t}_-)\, = 0.
\eeq

From (\ref{commute1}) in full analogy with (\ref{2D1}) 
for all $n$ and $n'$ we have
\beq\label{2Dm}
\begin{array}{c}
\displaystyle{
\oint_{{\infty}}
z^{n-n'-a}e^{\xi ({\bf t}_+ -{\bf t}'_+,z)}
W_n ({\bf t}_+ \!-\! [z^{-1}], {\bf t}_-)\,
Z_{n'} ({\bf t}'_+ \!+\! [z^{-1}], {\bf t}'_-)\, dz}
\\ \\
\displaystyle{
=\, \oint_{{0}}
z^{n-n'}e^{\xi ({\bf t}_- -{\bf t}'_-,z^{-1})}
W_{n+1} ({\bf t}_+, {\bf t}_- \!-\! [z])\,
Z_{n'-1} ({\bf t}'_+ , {\bf t}'_- \!+\! [z])\, dz}.
\end{array}
\eeq
We denote $b=n-n'$. 
Let us consider different values of $b$ assuming that $n$ is arbitrary.
For $b=a-1$ we 
put ${\bf t}'_+={\bf t}_+$ and ${\bf t}_-'={\bf t}_-$. 
Then only the left hand side of ({\ref{2Dm}}) survives, and we get:
\beq\label{2Dmm}
W_n ({\bf t}_+ , {\bf t}_-)\,
Z_{n-a+1} ({\bf t}_+ , {\bf t}_-)\, =0.
\eeq
If $-1<b<a-1$ 
we first differentiate ({\ref{2Dm}}) with respect to $t_{a-b-1}$ and then put  ${\bf t}_+'={\bf t}_+$ and ${\bf t}_-'={\bf t}_-$. 
Again, only the left hand side survives, so we get:
\beq\label{2Dm1}
W_n ({\bf t}_+ , {\bf t}_-)\,
Z_{n-b} ({\bf t}_+ , {\bf t}_-)\, =0 \,\,\,\,\,\,\,\,  \quad 
-1<b<a-1.
\eeq
For $b=-1$ we put ${\bf t}_+'={\bf t}_+$, ${\bf t}_-'={\bf t}_-$ 
so that only the right hand side of ({\ref{2Dm}}) survives:
\beq\label{2Dm2}
W_{n} ({\bf t}_+ , {\bf t}_-)\,
Z_{n-1} ({\bf t}_+ , {\bf t}_-)\, =0.
\eeq
For $-1<b<a-1$ 
we first differentiate ({\ref{2Dm}}) 
with respect to $t_{-b-1}$ and then put  ${\bf t}_+'={\bf t}_+$ and ${\bf t}_-'={\bf t}_-$. Again, only the right hand side survives, so we get:
\beq\label{2Dm3}
W_n({\bf t}_+ , {\bf t}_-)\, Z_{n-b-2}({\bf t}_+ , 
{\bf t}_-)\,=0, \,\,\,\,\,\,\,\, \quad -1<b<a-1.
\eeq
Therefore, combining (\ref{2Dmm}), (\ref{2Dm1}), 
(\ref{2Dm2}) and (\ref{2Dm3}), 
we get 
\beq\label{2Dm4}
W_n ({\bf t}_+ , {\bf t}_-)\,
Z_{n-k} ({\bf t}_+ , {\bf t}_-)\, =0, \,\,\,\,\,\,\,\,  
\quad   0\leq k \leq a.
\eeq
Now assume that for some fixed $n$ and $k<0$ 
$$
W_n ({\bf t}_+ , {\bf t}_-)\,
Z_{n-k} ({\bf t}_+ , {\bf t}_-)\, \neq 0.
$$
Then we take a derivative of (\ref{2Dm}) with respect to $t_{a-k-1}$ and put ${\bf t}_+={\bf t}_+'$ 
and ${\bf t}_-={\bf t}_-'$:
$$
W_n ({\bf t}_+, {\bf t}_-)\,
Z_{n-k} ({\bf t}_+, {\bf t}_-)\, 
=\, \oint_{{0}}
z^{k}
\left(\frac{\p}{\p t_{a-k-1}}W_{n+1} ({\bf t}_+, {\bf t}_- \!-\! [z])\right)\,
Z_{n-k-1} ({\bf t}_+ , {\bf t}_- \!+\! [z])\, dz
$$
so that $W_{n+1} ({\bf t}_+, {\bf t}_-)\,
Z_{n-k-1} ({\bf t}_+, {\bf t}_-)\, \neq 0$. Thus, if $-2\leq k<0$, we 
have got a contradiction with (\ref{2Dm3}), 
if $k<-2$, then we come to
a contradiction after a finite number of steps. 
For $k>a$ the argument is similar.

\end{document}